\documentclass[a4paper,11pt]{article}
\pdfoutput=1 % if your are submitting a pdflatex (i.e. if you have
\usepackage{jheppub} % for details on the use of the package, please
\usepackage[T1]{fontenc} % if needed
\usepackage[toc,page]{appendix}
\usepackage{amsmath,amssymb,amsfonts} %equations
\usepackage{mathrsfs} %equations
\usepackage{mathtools}
\usepackage{graphicx} %scaling
\usepackage{enumerate}
\usepackage{multirow}
\usepackage{xcolor}
\usepackage{bbold}
\usepackage{tikz}
\usetikzlibrary{shapes.geometric, arrows}

\title{\boldmath{Extended DBI and its generalizations from\\ graded soft theorems }}

\author{Karol Kampf,}
\author{Ji\v{r}\'{i} Novotn\'{y}}
\author{and Petr Va\v{s}ko}
\affiliation{Institute of Particle and Nuclear Physics, Charles University\\V Hole\v{s}ovi\v{c}k\'{a}ch 2, 180 00 Prague 8, Czech Republic}

\emailAdd{kampf@ipnp.mff.cuni.cz}
\emailAdd{novotny@ipnp.mff.cuni.cz}
\emailAdd{vasko@ipnp.mff.cuni.cz}

\abstract{We analyze a theory known as
\emph{extended DBI}, which interpolates between
DBI and the $U(N)\times U(N)/U(N)$ non-linear sigma model and represents a nontrivial example of theories with mixed power counting. We discuss symmetries of the action and their geometrical origin; the special case of
$SU(2)$ extended DBI theory is treated in great detail.
The revealed symmetries lead to a new type of \emph{graded soft theorem} that allows us to prove on-shell
constructibility of the tree-level S-matrix. It turns out that the on-shell constructibility of the full  extended DBI remains valid, even if its DBI sub-theory is modified in such a way to preserve its own on-shell constructibility. We thus propose a slight generalization of the DBI sub-theory, which we call \emph{2-scale DBI theory}. Gluing it back to the rest of the extended DBI theory gives a new set of on-shell reconstructible theories -- the \emph{2-scale extended DBI theory} and its descendants.  The uniqueness of the parent theory is confirmed by the bottom-up approach that uses on-shell amplitude methods exclusively.}

\begin{document} 
\maketitle
\flushbottom

%%%%%%%%%%%%%%%%%%%%%%%%%%%%%%%%%%%%%%%%%%%%%%%%%%%%%%%%%%%%
\section{Introduction}

In past few decades, the calculational techniques for perturbative on-shell
scattering amplitudes in weakly coupled quantum field theories experienced
rapid development. The textbook approach based on Feynman diagrams has been
supplemented by new methods which either allow to calculate the amplitudes
more effectively or bring about completely new insight in the very structure
of the perturbative quantum field theory. The former are based on the most
general properties such as locality, unitarity and analytical structure of
the amplitudes and represent a modern reincarnation of the bootstrap methods
originally developed in the sixties of past century - typical examples are those
based on generalized unitarity \cite{Bern:1994cg,Bern:1994zx} and various sorts of recursion relations
\cite{Berends:1987me, Britto:2004ap,Britto:2005fq,Cohen:2010mi,Cheung:2015cba,Cheung:2015ota,ArkaniHamed:2010kv,Baadsgaard:2015twa}. The latter methods have the ambition to completely reformulate the very
paradigm of quantum field theory and reveal new mathematical structures
behind it - let us mention e.g. the color-kinematics duality \cite{Bern:2008qj,Bern:2010ue,Bern:2019prr} and the
geometrical approach based on positive geometry, namely the amplituhedron
for planar $\mathcal{N}=4$ SYM \cite{Arkani-Hamed:2016byb,Arkani-Hamed:2013kca,Arkani-Hamed:2013jha,Arkani-Hamed:2014dca,Arkani-Hamed:2017vfh,Arkani-Hamed:2018rsk,Damgaard:2019ztj,Herrmann:2020qlt} or the associahedron \cite{Arkani-Hamed:2017mur,Arkani-Hamed:2019vag,Arkani-Hamed:2020cig}. Another
interesting representation of the scattering amplitudes of a particular set
of theories is provided by the CHY formula \cite{Cachazo:2013hca,Cachazo:2014nsa} which calculates them as an
integral over a punctured sphere which can be transformed into a sum over
solutions of the scattering equations. The CHY representation reveals among
others deep interrelations between amplitudes of different theories which
belong to the set known as the web of theories \cite{Cachazo:2014xea}, and also manifestly
incorporates various sorts of soft theorems \cite{Cachazo:2015ksa,Cachazo:2016njl,Schwab:2014xua}.

Remarkably, this progress is not limited to the case of well-behaved
renormalizable theories. Many of the above new methods are applicable also
to the power-counting non-renormalizable low-energy effective ones. This
usually happens provided they posses some special properties, namely
particular symmetries, which are strong enough to define the theory
uniquely. Such symmetries are in close relation to the properties of the
scattering amplitudes. Typically they are responsible for some sort of soft
theorems and in many cases just these soft theorems can be used to define
the theory \cite{Kampf:2012fn,Kampf:2013vha,Cheung:2015ota,Cheung:2016drk,Cheung:2018oki,Kampf:2020tne,Elvang:2018dco,Padilla:2016mno}. This applies e.g. to the case of exceptional scalar field
theories (the nonlinear sigma model (NLSM), the DBI scalar, the Special
Galileon), vector theories (Born-Infeld electrodynamics) or
scalar-vector theories (the Special scalar-vector Galileon \cite{Kampf:2021bet}), all of which
can be uniquely reconstructed from the corresponding soft limits \cite{Cheung:2016drk}.

Some of
these theories were discovered first just by constructing their tree-level
amplitudes and only then identified with a particular Lagrangian. And even
when the Lagrangian was known, the symmetry responsible for the soft
behavior of the amplitudes was far from being manifest.
In this paper we discuss in more detail one of these  cases, namely the
theory proposed by Cachazo et al. \cite{Cachazo:2014xea} which is referred to as \emph{extended
Dirac-Born-Infeld theory}. The CHY representation of this theory was also mentioned in \cite{Low:2020ubn}, where the double copy structure was established, and in \cite{Rodina:2021isd}, where the soft behavior was discussed. 

Originally, the amplitudes of the extended Dirac-Born-Infeld theory were
constructed using a particular procedure of squeezing and dimensional
reduction applied to the CHY representation of the General relativity
amplitudes and subsequently, its Lagrangian was conjectured in a closed
form. The latter has been recognized as a theory interpolating between NLSM
and DBI theory. 
In this context, a natural question arises, whether this Lagrangian is
unique, what are the symmetries and what is their geometrical origin, and
which properties of the amplitudes are the key ones for the possible
amplitude bootstrap. Note also that this theory differs from the above
mentioned exceptional theories: its amplitudes are not homogeneous functions
of the momenta, while almost all the  theories discussed in the context of amplitude methods were limited to cases with unmixed powercounting\footnote{See however \cite{Padilla:2016mno}, where the DBI Galileon was studied as an example of application of amplitude methods to theories with mixed powercounting.}.  Also, it was so far not completely clear how to study soft limits of amplitudes without unique powercounting. Note that the two ``boundary theories'' of the extended DBI theory, namely NLSM and DBI, have
different soft behavior and their on shell reconstruction is based on different soft theorems. This indicates that provided the extended DBI theory is in some
sense uniquely reconstructible from its soft limits, some generalization of
the usual soft bootstrap is needed. In this paper we address all these
issues in detail.

Let us briefly summarize the main outcomes of this paper. First, we introduce the framework of effective field theories (EFTs) whose Lagrangian consists of operators with mixed power counting (so called multi-$\rho$ theories). Section~\ref{sec:Graded_soft} then generalizes the concept of soft theorems, which were so far mostly applied to theories built of operators with a fixed power counting (single $\rho$ theories), to the more general arena of multi-$\rho$ theories. In particular, formulas~\eqref{eq:soft_crit1}--\eqref{hierarchy_reconstructible} build towards a criterion for a multi-$\rho$ EFT, that would decide whether it has an on-shell constructible tree-level S-matrix. This criterion is formulated in~\eqref{graded_soft_theorem} as the \emph{graded soft theorem}. Section~\ref{section: simple example} introduces a simple multi-$\rho$ EFT consisting of scalars only, which allows us to illustrate the application of graded soft theorems to a not overly complicated model. It also sets the stage for the introduction of the main character of this paper -- the \emph{extended DBI theory} -- in Section~\ref{sec:eDBI}. There we analyze its symmetries and various subtheories resulting as limits in couplings of the Lagrangian. The discussion of implications of the symmetries for soft theorems is touched upon in this section, but is mostly deferred to Section~\ref{sec:reconstr}. Section~\ref{sec:SU2_theory} treats the special case of $\mathrm{SU}(2)$ extended DBI theory in detail. Its main purpose is to clarify the more abstract constructions from previous sections and provide the reader with completely explicit expressions. In Section~\ref{sec:reconstr} we extract the potential of symmetries previously established for the extended DBI theory and reformulate them in terms of  soft theorems. Those are then used  in the proof of on-shell constructibility of the tree-level S-matrix of the extended DBI theory. Section~\ref{sec:bottomup} serves two purposes. First, based on the bottom-up approach of recursive amplitude construction, it verifies that conclusions made before are consistent. On top of that, it also provides a strong hint that there might exist more general on-shell reconstructible theories (parametrized by more couplings) than the DBI theory or extended DBI theory, respectively. The hunt for these generalized theories is organized in Section~\ref{sec:generalization}. Their existence is indeed confirmed and the Lagrangian of the \emph{2-scale DBI theory} is presented in~\eqref{eq:2DBI} while the one for the \emph{2-scale extended DBI theory} in~\eqref{eq:2eDBI}. We briefly summarize and draw our conclusions in Section~\ref{sec:conclusions}. Technical results are collected in three appendices.     

%%%%%%%%%%%%%%%%%%%%%%%%%%%%%%%%%%%%%%%%%%%%%%%%%%%%%%%%%%%%
\section{Multi-$\rho$ effective field theories\label{section_multi_rho}}

The Lagrangian of low-energy effective field theory contains usually a
(possibly infinite) tower of elementary vertices $V$. These correspond to
monomials $\mathcal{L}_{V}$ in fields decorated with increasing number of
derivatives and accompanied by couplings $g_{V}$ with decreasing mass
dimension,
\begin{equation}
\mathcal{L}=\sum_{V}g_{V}\mathcal{L}_{V}.  \label{Lagrangian_general}
\end{equation}
Each vertex $V$ represented by $\mathcal{L}_{V}$ gives rise to a unique
tree-level contact term $g_{V}A_{V}^{ct}$ contributing to some scattering
amplitude. It is useful to characterize the individual vertices $V$ using
the power-counting parameters $\rho _{V}$ which are defined as follows
\begin{equation}
\rho _{V}\equiv \frac{D_{V}-2}{N_{V}-2}. 
\end{equation}
Here $D_{V}$ is the mass dimension of the contact term\footnote{With the coupling constant $g_V$ stripped off.} $A_{V}^{ct}$ and $%
N_{V}$ is the number of its external legs (i.e. the number of fields of the
corresponding term $\mathcal{L}_{V}\ $in the Lagrangian).

Assume now a contribution to some scattering amplitude given by a Feynman
graph $\Gamma $ with $V_{\Gamma }$ vertices, $I_{\Gamma }$ internal lines, $%
N_{\Gamma }$ external lines and $L_{\Gamma }$ loops. The mass dimension $%
D_{\Gamma }$ of such contribution (with the coupling constants stripped) is
then 
\begin{equation}
D_{\Gamma }=\sum_{V}D_{V}-2I_{\Gamma }+4L_{\Gamma } 
\end{equation}
and the number of external legs $N_{\Gamma }$ can be expressed as%
\begin{equation}
N_{\Gamma }=\sum_{V}N_{V}-2I_{\Gamma }. 
\end{equation}
Using the topological relation
\begin{equation}
L_{\Gamma }=I_{\Gamma }-V_{\Gamma }+1, 
\end{equation}
we get the generalization of the Weinberg formula 
\begin{eqnarray}
D_{\Gamma }-2 &=&\sum_{V}\left( D_{V}-2\right) +2L_{\Gamma }  \nonumber \\
N_{\Gamma }-2 &=&\sum_{V}\left( N_{V}-2\right) -2L_{\Gamma }.
\label{power counting}
\end{eqnarray}
For a given graph $\Gamma $ we can then define the power-counting parameter 
\begin{equation}
\rho _{\Gamma }\equiv \frac{D_{\Gamma }-2}{N_{\Gamma }-2}.  \label{rho_Gamma}
\end{equation}

The relations (\ref{power counting}) have a nice graphical interpretation in
the two-dimensional plane. Each graph can be represented in such a plane by
a point $P_{\Gamma }$ with coordinates $( N_{\Gamma }-2,$ $D_{\Gamma
}-2)$, while the elementary vertices correspond to the vectors $%
\mathbf{v}_{V}$ with components $(N_{V}-2, D_{V}-2)$. To get the
point $P_{\Gamma }$ representing the graph, we have to shift the starting
point $\left( -2L_{\Gamma },2L_{\Gamma }\right) $ by the sum of the vectors $%
\mathbf{v}_{V}$ corresponding to all the vertices of $\Gamma $,%
\begin{equation}
P_{\Gamma }=\left( -2L_{\Gamma },2L_{\Gamma }\right) +\sum_{V}\mathbf{v}_{V}. 
\end{equation}%
Of course, the same point can represent several different graphs (see Section \ref{section: simple example} and Fig.\ref{fig:1} for a particular example).

Note that the power-counting parameter $\rho _{V}$ has the meaning of the
slope of the vector $\mathbf{v}_{V}$. Provided $\rho _{V}$ is the same for
all the elementary vertices of the theory, than all the vectors $\mathbf{v}%
_{V}$ are parallel. Therefore the points corresponding to the contributions of all
the $L-$loop graphs sit on a single line in the $\left( N_{\Gamma
}-2,D_{\Gamma }-2\right) -$plane. For tree level graphs such a line goes
through the origin. We refer to such cases as single-$\rho $ theories.
Typical examples are the non-linear sigma model ($\rho _{V}=0$), the DBI
scalar ($\rho _{V}=1$), the Born-Infeld electrodynamics ($\rho _{V}=1$) and
the Galileon ($\rho _{V}=2$).

The opposite cases, which we denote as multi-$\rho $ theories, contain
elementary vertices with at least two different $\rho _{V}$'s. In such a
multi-$\rho $ theory, let us define two distinguished $\rho $'s, namely $%
\rho _{\min }=\min_{V}\rho _{V}$ and $\rho _{\max }=\max_{V}\rho _{V}$.
Clearly, the points $P_{\Gamma }$ representing tree-level contributions
of individual graphs to the scattering amplitudes are situated inside
a wedge whose vertex is at the origin of the $\left( N_{\Gamma }-2,D_{\Gamma
}-2\right) -$plane, and which is bounded by lines with slopes $\rho _{\min }$ and $\rho
_{\max }$. The points on the border of this wedge correspond to graphs built solely using vertices either with $\rho _{V}=\rho _{\min }$
or $\rho_{V} =\rho _{\max }$. They correspond therefore to scattering
amplitudes of two single-$\rho $ theories with Lagrangians
\begin{equation}
\mathcal{L}_{\rho _{\min }}=\sum_{V,~\rho _{V}=\rho _{\min }}g_{V}\mathcal{L}%
_{V},~~~~\mathcal{L}_{\rho _{\max }}=\sum_{V,~\rho _{V}=\rho _{\max }}g_{V}%
\mathcal{L}_{V}, 
\end{equation}
which can be treated as ``subtheories'' of the multi-$\rho $ theory. The complete theory then ``interpolates'' between these two. Typical example of such a
multi-$\rho $ theory is the DBI Galileon \cite{deRham:2010eu}, which interpolates between the DBI
scalar with $\rho =\rho _{\min }=1$ and the general Galileon with $\rho
=\rho _{\max }=2.$

The points $P_{\Gamma }$ in the interior of the above mentioned wedge
correspond to the graphs with vertices with different $\rho _{V}$'s. For a
general tree-level amplitude (with a fixed number of external legs $N_\Gamma$) we can then write%
\begin{equation}
A=\sum_{\rho _{\min }\leq \rho \leq \rho _{\max }}A^{(\rho )} \,,
\label{eq:graded_A}
\end{equation}
where the single-$\rho $ components $A^{(\rho )}$ are sums of contributions
of all graphs with the same $\rho _{\Gamma }=\rho $ (cf. (\ref{rho_Gamma})).
In our graphical language all the graphs which contribute to $A^{(\rho )}$
sit at the same point in the $\left( N_{\Gamma }-2,D_{\Gamma }-2\right) -$%
plane.

As mentioned above, the only graphs which contribute to $A^{\left( \rho
_{\min }\right) }$ and to $A^{\left( \rho _{\max }\right) }$ are those with
vertices from $\mathcal{L}_{\rho _{\min }}$ and $\mathcal{L}_{\rho _{\max }}$
respectively. Therefore the components $A^{\left( \rho _{\min }\right) }$
and $A^{\left( \rho _{\max }\right) }$correspond to the amplitudes of the
single-$\rho $ subtheories $\mathcal{L}_{\rho _{\min }}$ and $\mathcal{L}%
_{\rho _{\max }}$ with smallest and largest $\rho $. On the other hand, the
components $A^{\left( \rho \right) }$ with $\rho _{\min }<\rho <\rho _{\max
} $ cannot be treated independently and attributed to amplitudes of some
single-$\rho $ theory since in general the graphs with vertices with
different $\rho _{V}$ contribute to them.

The general properties of the single-$\rho $ components $A^{\left( \rho
\right) }$ can differ. Nevertheless, given some property based on
cancellation between different Feynman graphs which is valid for the
complete amplitude $A$, the same property has to be shared also by the
individual components $A^{\left( \rho \right) }$, since graphs with
different $\rho _{\Gamma }$ cannot communicate with each other.

\section{Graded soft theorems \label{sec:Graded_soft}}

The Lagrangians of the low-energy effective field theories (\ref%
{Lagrangian_general}) can have in principle an infinite number of elementary
vertices $\mathcal{L}_{V}$ and an infinite number of corresponding coupling
constants $g_{V}$. This general picture can be substantially changed in the
cases, when the theory is subject to some symmetry. Provided the symmetry
requirements are strong enough, the number of independent parameters can
be reduced to a finite set or even to just one independent coupling constant (representing then the so-called exceptional theory \cite{Cheung:2016drk}).
Such symmetric theories have usually many interesting properties which
manifest themselves at the level of the tree on-shell scattering
amplitudes. The most prominent such properties are related with soft
limits of the amplitudes and can be expressed in terms of soft theorems.
For instance, provided the low energy theory describes dynamics of Goldstone bosons of some spontaneously broken symmetry, the amplitudes
possess in many cases a so called Adler zero \cite{Adler:1964um}, i.e. they vanish in the limit
when one external Goldstone particle becomes soft. Provided the broken
symmetry manifests itself at the Lagrangian level as a generalized polynomial
shift symmetry, the Adler zero can be even enhanced \cite{Cheung:2014dqa}, that means the
amplitudes behave in the soft Goldstone boson limit $p\rightarrow 0$ as 
\begin{equation}
A(p) =O\left( p^{\sigma }\right) \,,
\end{equation}%
where the soft exponent $\sigma >1$. The enhanced Adler zero condition can
be often used to define the theory uniquely in terms of the soft BCFW
recursion \cite{Cheung:2015ota}. 
We call them the {\it on-shell reconstructible theories\/} and they can be characterized by the power counting parameter $\rho$ (provided they are single-$\rho$ theories) and by the soft exponent $\sigma$. 
The criterion of reconstructibility for scalar theories can be
expressed as (cf. \cite{Cheung:2015ota,Cheung:2016drk}) 
\begin{equation}
\rho \leq \sigma \,,
\end{equation}%
for $\left( \rho ,\sigma \right) \neq \left( 1,1\right) $. The well known
examples of the pure scalar reconstructible theories are the nonlinear sigma
model ($\rho =0$, $\sigma =1$), the DBI scalar ($\rho =1$, $\sigma =2$), the
general Galileon ($\rho =2$, $\sigma =2$) and the Special Galileon 
($\rho =2$, $\sigma =3$). All these theories allow to reconstruct all their tree-level
amplitudes recursively, using the seed amplitudes (four-point and/or
five-point) as the only free input.

The reconstructibility is not limited to theories of Goldstone
bosons only. Recently it has been shown \cite{Cheung:2018oki}, that Born-Infeld
electrodynamics is reconstructible using the soft behavior of its amplitudes
with respect to the \emph{multi-chiral soft limit} as its defining property.
The latter is defined as a special limit when all the photons with the same
helicity simultaneously become soft. Provided such a limit is applied to
helicity plus photons and writing their momenta in terms of the helicity
spinors, $p_{i}=[i|\sigma |i\rangle/2 $, then it is taken so that the
corresponding holomorphic spinors $|i\rangle $ are sent to zero (similarly
for helicity minus photons we send instead the antiholomorphic spinors $[i|$
to zero). In both cases of the multi-chiral soft limit, the tree-level
amplitudes of the Born-Infeld electrodynamics vanish. Let us note that
Born-Infeld theory is a single-$\rho $ theory with $\rho =1$.

Also the theories which couple photons (or massless vector particles) to
Goldstone bosons can be reconstructible from their soft limits. This is the
case for the Special scalar-vector Galileon and its generalizations discussed
recently in \cite{Kampf:2021bet}. In this theory, which is a single-$\rho $ theory with $\rho
=2$, the enhanced Adler zero condition for the soft scalars is combined with
the generalized soft theorem for the soft photons. In the latter case, the
amplitude does not vanish in the soft photon limit but it is rather related
to the lower point amplitudes, which enables the recursion.

All the above mentioned examples are single-$\rho $ theories. The problem of
reconstructibility of multi-$\rho $ theories has been addressed in \cite{Padilla:2016mno},
however, the systematical classification is still missing. As discussed in
\cite{Padilla:2016mno}, the criterion of reconstructibility for a scalar multi-$\rho $ theory
with enhanced soft limit with soft exponent $\sigma $ is similar to the case
of single-$\rho $ theories, namely for $\left( \rho _{\max },\sigma \right)
\neq \left( 1,1\right) $ we require 
\begin{equation}
\rho _{\max }\leq \sigma . 
\end{equation}%
The example of such \ a reconstructible theory is the DBI Galileon mentioned
above where the soft behavior with $\sigma =2$ is a consequence of the
non-linearly realized Lorentz symmetry in dimension $D=5$.

The most interesting case is the multi-$\rho $ theory which interpolates
between $\mathcal{L}_{\rho _{\min }}$ and $\mathcal{L}_{\rho _{\max }}$,
where both these two subtheories are reconstructible using the soft BCFW
recursion based on different sorts of soft theorems. For instance, the
soft exponent $\sigma $ of the enhanced Adler zeros can be different, i.e.
for the subamplitudes $A^{\left( \rho _{\min }\right) }$ and $A^{\left( \rho
_{\max }\right) }$ we can have
\begin{equation}\label{eq:soft_crit1}
A^{\left( \rho _{\min }\right) }=O\left( p^{\sigma _{\min }}\right)
,~~~~A^{\left( \rho _{\max }\right) }=O\left( p^{\sigma _{\max }}\right) 
\end{equation}
with $\sigma _{\min }<\sigma _{\max }$. Clearly, the complete amplitude
cannot then behave better than 
\begin{equation}\label{eq:soft_crit2}
A(p) =O\left( p^{\sigma _{\min }}\right) .
\end{equation}
Provided this is the case, only the amplitude components $A^{\left( \rho
\right) }$ with $\rho \leq \sigma _{\min }$ satisfy the reconstructibility
conditions. The reconstructibility of the complete amplitude depends then on
the value of the power-counting parameter $\rho _{\max }$. 
Assuming the hierarchy:
\begin{equation}\label{eq:soft_crit3}
\rho _{\min }\leq \sigma _{\min }<\rho _{\max }\leq \sigma _{\max }\,, 
\end{equation}
then there is a gap $\sigma _{\min }<\rho <\rho _{\max }$ for which the
amplitude components $A^{\left( \rho \right) }$ cannot be reconstructed using
the recursion based on the soft theorem (\ref{eq:soft_crit2}). However,
for $\sigma _{\min }=\rho _{\max }$ such a gap shrinks to an empty set and in
fact the complete amplitude is reconstructible, since the $A^{\left( \rho
_{\max }\right) }$ component can be reconstructed using the recursion based
on its own soft behavior $A^{\left( \rho _{\max }\right) }=O\left( p^{\sigma
_{\max }}\right) $. In summary, provided
\begin{equation}
\rho _{\min }\leq \sigma _{\min }=\rho _{\max }\leq \sigma _{\max }\,,
\label{hierarchy_reconstructible}
\end{equation}
we can construct the soft BCFW recursion leaning on the \emph{graded soft
theorem}
\begin{eqnarray}
A(p) -A^{\left( \rho _{\max }\right) } &=&O\left( p^{\sigma
_{\min }}\right)  \nonumber \\
A^{\left( \rho _{\max }\right) } &=&O\left( p^{\sigma _{\max }}\right) .
\label{graded_soft_theorem}
\end{eqnarray}
For a pure scalar theory, it is based on the analytical properties of the
function
\begin{equation}
f_{n}(z) =\frac{\widehat{A}_{n}(z) -\widehat{A}%
_{n}^{\left( \rho _{\max }\right) }(z) }{\prod\limits_{i=1}^{n}%
\left( 1-a_{i}z\right) ^{\sigma _{\min }}}+\frac{\widehat{A}_{n}^{\left(
\rho _{\max }\right) }(z) }{\prod\limits_{i=1}^{n}\left(
1-a_{i}z\right) ^{\sigma_{\max }}}\,,
\label{eq:f_n}
\end{equation}
where $\widehat{A}_{n}(z) $ and $\widehat{A}_{n}^{\left( \rho
_{\max }\right) }(z) $ are the deformed $n-$point amplitudes
depending on the complex parameter $z$ through all-line soft shift of the
original kinematic configuration (cf. \cite{Cheung:2015ota,Cheung:2016drk})
\begin{eqnarray}
\widehat{p}_{i}(z) &=&\left( 1-a_{i}z\right) p_{i}, \notag\\
\widehat{A}_{n}(z) &=&A_{n}|_{p_{i}\rightarrow \widehat{p}%
_{i}(z) }\,\,\,\,\,\,\,\,\widehat{A}_{n}^{\left( \rho _{\max
}\right) }(z) ={A}_{n}^{\left( \rho _{\max }\right)
}(z) |_{p_{i}\rightarrow \widehat{p}_{i}(z) }\,.
\label{eq:soft_shift_general}
\end{eqnarray}
Such a deformation is possible in $D$ dimensions provided $n>D+1$. Note that
as a consequence of (\ref{hierarchy_reconstructible}) 
\begin{equation}
\lim_{z\rightarrow \infty }f_{n}(z) =0 
\end{equation}
and the soft behavior of the amplitudes cancels the apparent poles of $%
f_{n}(z) $ at $\ z=1/a_{i}$. Therefore, the only singularities of 
$f_{n}(z) $ are the unitarity poles $z_{\mathcal{F}}^{\pm }$
related to the factorization channels $\mathcal{F}$ . The latter are determined
by vanishing of the corresponding propagator denominator 
\begin{equation}
p_{\mathcal{F}}^{2}(z) \equiv \left( \sum_{i\in \mathcal{F}} \widehat{p}_{i}(z) \right) ^{2}=0 
\end{equation}
and $z_{\mathcal{F}}^{\pm }$ are the two roots of this quadratic equation.

Applying now the residue theorem to the meromorphic function $f_{n}(z)/z$, i.e. the fact that the sum of the residues at all its poles
(including infinity) vanishes, we can write
\begin{equation}
A_{n}=\widehat{A}_{n}\left( 0\right) =\mathrm{res~}\left( f_{n}/z,0\right)
=-\sum_{\mathcal{F},I=\pm }\mathrm{res~}\left( f_{n}/z,z_{\mathcal{F}%
}^{I}\right) .  \label{recursion}
\end{equation}
The residue at the unitarity poles factorizes into products of lower-point amplitudes,
\begin{eqnarray}
-\mathrm{res~}\left( f_{n}/z,z_{\mathcal{F}}^{I}\right)  &=&\Biggl[ \frac{
\widehat{A}_{L}\left( z_{\mathcal{F}}^{I}\right) \widehat{A}_{R}\left( z_{
\mathcal{F}}^{I}\right) -\widehat{A}_{L}^{\left( \rho _{\max }\right)
}\left( z_{\mathcal{F}}^{I}\right) \widehat{A}_{R}^{\left( \rho _{\max
}\right) }\left( z_{\mathcal{F}}^{I}\right) }{\prod\limits_{i=1}^{n}\left(
1-a_{i}z_{\mathcal{F}}^{I}\right) ^{\sigma _{\min }}}\notag \\
&&+\frac{\widehat{A}_{L}^{\left( \rho _{\max }\right) }\left( 
z_{\mathcal{F}}^{I}\right) \widehat{A}_{R}^{\left( \rho _{\max }\right) }\left(
z_{\mathcal{F}}^{I}\right) }{\prod\limits_{i=1}^{n}\left( 1-a_{i}z_{\mathcal{F}}^{I}\right) ^{\sigma _{\max }}}\Biggr] \frac{1}{z_{\mathcal{F}}^{I}}
\mathrm{res~}\left( \frac{1}{p_{\mathcal{F}}^{2}(z) },
z_{\mathcal{F}}^{I}\right) 
\end{eqnarray}
and thus the formula (\ref{recursion}) gives the desired recursion.

\section{Simple example\label{section: simple example}}

Let us give a simple example of the above considerations. Assume a theory
in $D=4$ dimensions given by the Lagrangian\footnote{Here and in what follows  we use the mostly minus signature $\eta_{\mu\nu}=diag(1,-1,-1,-1)$.}
\begin{equation}
\mathcal{L}=-\Lambda ^{4}\sqrt{-\det \left( \eta _{\mu \nu }-\frac{1}{%
4\lambda ^{2}\Lambda ^{4}}\left\langle  \partial _{\mu }U^{\dagger}\partial
_{\nu }U \right\rangle \right) }+\Lambda ^{4}
\label{multi_rho_example}
\end{equation}
containing two dimensionful parameters $\Lambda $ and $\lambda $, with mass
dimensions $\left[ \Lambda \right] =1$, $\left[ \lambda \right] =-1$. The
Lagrangian describes the dynamics of a $U(N) $ multiplet of
scalars in the adjoint representation $\phi ^{a}$, $a=1,\ldots ,N^{2}$, arranged into a $U(N) $
matrix 
\begin{equation}
U=\frac{\mathbb{1}+\lambda \phi }{\mathbb{1}-\lambda \phi }\,. 
\end{equation}
Here $\phi =\phi ^{a}T^{a}$ and the $U(N) $ generators $T^{a}$
satisfy 
\begin{equation}
T^{a\dagger}=-T^{a}\,,\qquad \langle T^{a}T^{b}\rangle =-\delta^{ab}
\label{eq:generators}
\end{equation}
and we use the shorthand notation for the traces 
$
\langle \cdot \rangle \equiv \mathrm{Tr}\left( \cdot \right) 
$.
The Lagrangian is manifestly invariant with respect to the $U(N)
_{L}\times U(N) _{R}$ $\ni \left( V_{L},V_{R}\right) $ symmetry
acting on $U$ as
\begin{equation}
U^{\prime }=V_{L}UV_{R}^{\dagger}. 
\end{equation}
Its vectorial subgroup $U(N) _{V}\ni \left( V,V\right) $ is
realized linearly on the fields $\phi ^{a}$
\begin{equation}
\phi ^{\prime }=V\phi V^{\dagger}, 
\end{equation}
while the axial transformations of the form $\left( V,V^{\dagger}\right) $ are
spontaneously broken and their action on the fields $\phi ^{a}$ is
nonlinear. Explicitly for $V=\exp \left( \lambda \alpha \right) $, we get in
the first order in $\alpha =\alpha ^{a}T^{a}$ (cf. (\ref{eq:A16})-(\ref{eq:A18}) in Appendix \ref{sec:appA})
\begin{equation}
\phi ^{\prime }=\phi +\alpha -\lambda ^{2}\phi \alpha \phi +O\left( \alpha
^{2}\right) .  \label{chiral_shift_symmetry}
\end{equation}
This second symmetry is of the form of a generalized shift symmetry and
therefore the amplitudes of the theory are guaranteed to have Adler zero
with soft index $\sigma =1$.

\begin{figure}[t]
\centering
\includegraphics[width=12cm]{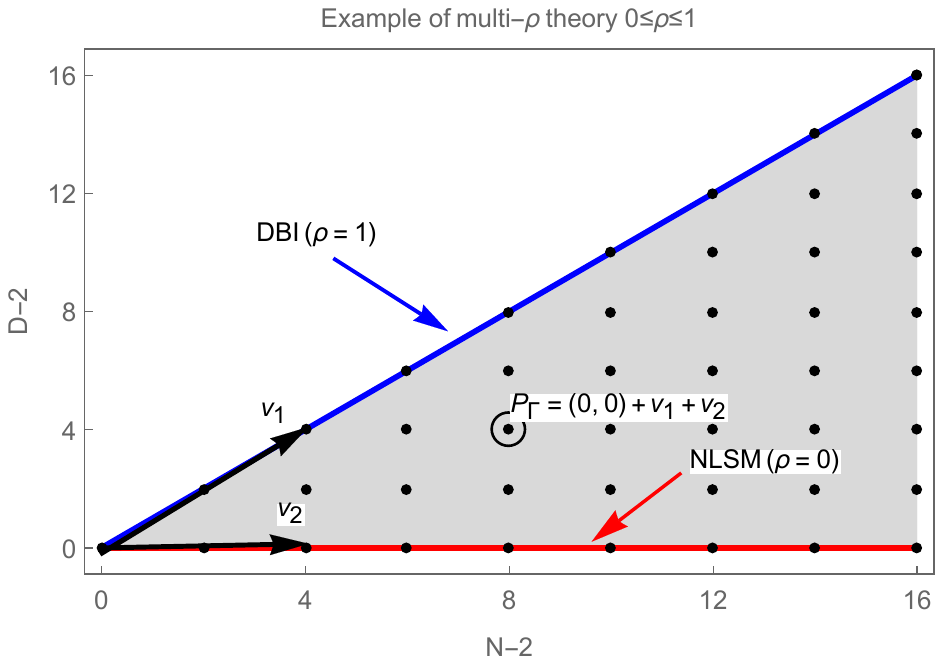}
\caption{Graphical representation of the model with Lagrangian (\ref{multi_rho_example}) in the $(N-2,D-2)$ plane. The points represent both the various vertices of the Lagrangian as well as the various contributions to the scattering amplitudes.  For illustration of the general discussion in Section \ref{section_multi_rho}, we have depicted one particular point $P_\Gamma$ corresponding to $A^{(1/2)}_{10}$ and two vertices which can contribute to it via one-propagator graph $\Gamma$. }
\label{fig:1}
\end{figure}

The theory is an example of a multi-$\rho $ theory with $0\leq \rho \leq 1$ (see Fig. \ref{fig:1}).
We easily identify the subtheory with $\rho =\rho _{\min }=0$, which is
nothing else but the nonlinear sigma model and corresponds to the $\Lambda
\rightarrow \infty $ limit of (\ref{multi_rho_example}) 
\begin{equation}
\mathcal{L}_{0}=\mathcal{L}_{\mathrm{NLSM}}=\frac{1}{8\lambda ^{2}}\bigl\langle
 \partial _{\mu }U^{\dagger}\partial ^{\mu }U \bigr\rangle \,.
 \label{eq:NLSM_Lagrangian}
\end{equation}%
The latter shares the symmetries of the original Lagrangian (\ref{multi_rho_example}) and the corresponding amplitudes $A^{\left( 0\right) }$
have the soft exponent $\sigma =\sigma _{\min }=1$.

The subtheory with $\rho =\rho _{\max }=1$ can be identified with (multi)
DBI theory with the Lagrangian, which can be obtained as the $\lambda
\rightarrow 0$ limit of (\ref{multi_rho_example}) 
\begin{equation}
\mathcal{L}_{1}=\mathcal{L}_{\mathrm{DBI}}=-\Lambda ^{4}\sqrt{-\det \left( \eta _{\mu
\nu }-\frac{\partial _{\mu }\phi ^{a}\partial _{\nu }\phi ^{a}}{\Lambda ^{4}}%
\right) }+\Lambda ^{4} \,.
\end{equation}
This subtheory shares the $U(N) _{V}$ symmetry of the original
theory and a linearized form of the shift symmetry (\ref{chiral_shift_symmetry}) obtained formally as its limit for $\lambda
\rightarrow 0$
\begin{equation}
\phi ^{\prime }=\phi +\alpha . 
\end{equation}
On top of these two it possesses also\ the higher polynomial shift symmetry\footnote{This symmetry is a non-linearly realized  $(N^2+4)$-dimensional Lorentz symmetry which mixes extra dimensions $b=1,\dots,N^2$ with $4D$ coordinates $x^\beta$.}
\begin{equation}
\phi ^{a\prime }=\phi ^{a}+\omega _{\beta b}\left( x^{\beta }\delta
^{ab}\Lambda ^{2}-\eta ^{\beta \mu }\phi ^{b}\partial _{\mu }\phi
^{a}\Lambda ^{-2}\right) +O\left( \omega ^{2}\right) . 
\end{equation}
This symmetry is responsible for the enhanced Adler zero of the amplitudes $A^{(1)}$ with soft exponent $\sigma =\sigma _{\max }=2$.
Therefore the multi-$\rho $ theory (\ref{multi_rho_example}) satisfies (\ref{hierarchy_reconstructible}) and can be then reconstructed by means of
recursion based on the graded soft theorem (\ref{graded_soft_theorem}).

%%%%%%%%%%%%%%%%%%%%%%%%%%%%%%%%%%%%%%%%%%%%%%%%%%%%%%%%%%%%
\section{Extended DBI theory}\label{sec:eDBI}
%%%%%%%%%%%%%%%%%%%%%%%%%%%%%%%%%%%%%%%%%%%%%%%%%%%%%%%%%%%%

In this section we introduce the main subject of our further studies, which represents a nontrivial generalization of the simple example from the previous section and which we will use as a  theoretical laboratory illustrating the above general considerations.
We will start with the Lagrangian of the theory and rewrite it in a more formal mathematical language, which will be well suited for the investigation of its symmetries and their geometrical origin.
We will also discuss various limiting cases.

\subsection{Lagrangian formulation}

In~\cite{Cachazo:2014xea}, Cachazo \textit{et al.} derived a theory that
interpolates between Dirac--Born--Infeld (DBI) theory (see~\cite{Tseytlin:1999dj} for a review) and the non-linear sigma
model~\cite{Cronin:1967jq,Weinberg:1966fm,Weinberg:1968de} with $\mathrm{U}(N)$ target space. They constructed the model via
the scattering equations approach to the S-matrix (the CHY formalism). The latter allows
to express on-shell $n$-point amplitudes as integrals of CHY integrands over
moduli spaces of $n$-punctured Riemann surfaces (genus zero at tree level).
The authors started with the known integrand corresponding to tree-level amplitudes for Einstein gravity and applied on it  a set of non-trivial operations (dimensional reduction and squeezing)
that resulted in a new CHY integrand that defined a new theory.
This construction allowed the authors to compute on-shell
tree-level amplitudes up to 10-points and consequently partially reconstruct the
action of the newly proposed effective field theory. Quite remarkably,
they were able to extrapolate and conjectured a complete action (to all orders in
fields). It takes the form\footnote{Note that the
  sign in front of the $\tfrac{1}{\Lambda^2}$ term is irrelevant since
  $\det(S+A)=\det((S+A)^T)=\det(S-A)$ for $S$ symmetric and $A$ anti-symmetric
  matrices. The other signs are on the other hand very important to have canonically normalized
  kinetic terms.}
\begin{align}\label{eq:eDBI}
S_{\mathrm{eDBI}}=\int_{\mathbb{R}^{1,3}}{\rm{d}}^4x\Bigg\{\Lambda^4\bigg[1-\sqrt{-\det\bigg( \eta_{\mu\nu}-\frac{1}{\Lambda^4}g_{\mu\nu}-\frac{1}{\Lambda^2}\Big( cW_{\mu\nu}+F_{\mu\nu} \Big)\bigg)} \bigg]\Bigg\}
\end{align}   
and depends on three real couplings $\Lambda,\;\lambda$ and $c$ of mass dimensions
\begin{align}
&[\Lambda]=1,&&[\lambda]=-1,&[c]=0.  
\end{align}
Note that compared
 to~\cite{Cachazo:2014xea} we introduced one extra coupling $c$. The original
 action corresponds to $c=(\Lambda \lambda)^{-2}$.
 
Now let us define the various building blocks appearing in the above expression.
As in the previous section, the term
\begin{equation}
g_{\mu\nu}=\tfrac{1}{4\lambda^2}\langle\partial_\mu U^{\dagger} \partial_\nu U\rangle
\end{equation}
represents the building block for the NLSM Lagrangian (at leading order in
derivative expansion,   $\mathcal{L}_{\mathrm{NLSM}}=\tfrac{1}{2}\eta^{\mu\nu}g_{\mu\nu}$, cf. (\ref{eq:NLSM_Lagrangian})), where $U\in\mathrm{U}(N)$ is
defined as
\begin{align}\label{eq:coset_rep}
  U=\frac{\mathbb{1}+\lambda \phi}{\mathbb{1}-\lambda \phi}
  \end{align}
with $\phi=\phi^aT^a\in \mathfrak{u}(N)$ anti-hermitian (and
additionally traceless for $\mathrm{SU}(N)$). Normalization of the generators is
chosen according to (\ref{eq:generators}). 
The next piece $F_{\mu\nu}$=$\partial_{\mu}A_\nu-\partial_\nu A_\mu$ is the field strength of an abelian gauge field
$A_{\mu}$.
Finally, $W_{\mu\nu}$ is an anti-symmetric tensor field constructed
from the scalars $\phi$ in the form of a weak field expansion as
\begin{align}\label{eq:W}
  W_{\mu\nu}=  \sum_{m=1}^\infty \sum_{k=0}^{m-1} \frac{2(m-k)}{2m+1}\lambda^{2m+1}\left\langle \partial_{[\mu}\phi \phi^{2k} \partial_{\nu]}\phi \phi^{2(m-k)-1} \right\rangle. 
\end{align}
At first sight it looks rather daunting, but we will argue later on that it is a
unique and very elegant expression originating from topology of (special)
unitary groups. 

 At least
classically, the theory can be written down in a space-time of arbitrary
dimension (as the nice feature of CHY formalism is that it holds in a general dimension). However, then the mass dimension of $\lambda$ would be
$[\lambda]=1-\tfrac{d}{2}$ and powers of the mass scale $\Lambda$ in front of various terms would be
different, which effects the important interpolating properties of the action.
In any case, in this paper we will study this model in four dimensional
Minkowski space-time.  

Following the terminology of~\cite{Cachazo:2014xea}, we will call this model the
\emph{extended DBI theory}. The action in~\eqref{eq:eDBI} is the starting point
for our analysis of the extended DBI model. In~\eqref{eq:2eDBI} we will give
a further generalization of this theory. 

\subsection{Note on inclusion of the extra coupling $c$}
When we introduced the extended DBI theory in \eqref{eq:eDBI}, we included one extra dimensionless coupling $c$ and stated that the original theory in~\cite{Cachazo:2014xea} corresponds to the choice $c=(\Lambda \lambda)^{-2}$. Let us explain our motivation for enlarging the theory in this way and comment on the (im)possibility of generalizing the CHY formalism to encompass this additional coupling. The approach of the authors of~\cite{Cachazo:2014xea} was to invent a CHY integrand that interpolates between NLSM and DBI models. 
On the  level of the original Lagrangian it was achieved by the limiting procedure according to the scheme\footnote{Strictly speaking, the NLSM comes with a decoupled free photon.}
\begin{equation}
\mathcal{L}_{\mathrm{eDBI}}%
\begin{array}{c}
\overset{\Lambda \rightarrow \infty }{\nearrow } \\ 
\underset{\lambda \rightarrow 0}{\searrow }%
\end{array}%
\begin{array}{l}
\;\mathcal{L}_{\mathrm{NLSM}} \\ 
\\ 
\\ 
\;\mathcal{L}_{\mathrm{DBI}}%
\end{array}%
\label{eq:interpolation}
\end{equation}
 In other words, this interpolation property was encoded at the heart of their construction. Our point of view is slightly different. 
 We relax the particular interpolation assumption (\ref{eq:interpolation}) and instead search for the most general theory of this type (i.e. a multi-$\rho$ theory including NLSM and DBI as boundary subtheories corresponding to $\rho_{\min}=0$ and $\rho_{\max}=1$ respectively) that has sufficiently constraining soft theorems. By this we mean that the soft theorems are powerful enough to yield the tree-level S-matrix on-shell constructible by soft BCFW recursion. This reasoning led us to enlarge the extended DBI theory of~\cite{Cachazo:2014xea} by the extra coupling $c$ in~\eqref{eq:eDBI}. Analysis of soft theorems associated with this action and study of their implications for on-shell reconstructibility of the tree-level S-matrix will be the main subject for the rest of this paper.

Imposing the NLSM/DBI interpolation property of~\cite{Cachazo:2014xea}  in~\eqref{eq:eDBI} leads to $c=c_{\mathrm{num}}(\Lambda\lambda)^{-a}$ with $a\in(0,2)$ and $c_{\mathrm{num}}$ purely numerical, i.e. independent of the dimensionless combination $\Lambda\lambda$. The authors of~\cite{Cachazo:2014xea} presented the theory with the choice $c_{\mathrm{num}}=1$ and $a=2$. From a simple Feynman diagram analysis, we will show that $a$ can indeed be fixed to the particular value $a=2$. However, fixing $c_{\mathrm{num}}$ is more difficult. It is not known a priori which property makes the choice $c_{\mathrm{num}}=1$ special.

Now, let us take an opposite approach and investigate the possibility of generalizing the CHY formalism in order to include a general $c$. This appears to be much harder, in fact we do not know if such a modification can be consistently made.  Let us remind, that the CHY integrands are classified according to the number of external legs, their division to scalars and photons and additionally according to the algebraic structure of traces (over the group associated with NLSM). So each scattering amplitude specified by these data has a uniquely associated CHY integrand. In order to include a generic $c$, these unique integrands would have to be further subdivided based on additional criteria that are not obvious. 

\begin{figure}[ht!]
\centering
\includegraphics[width=10cm]{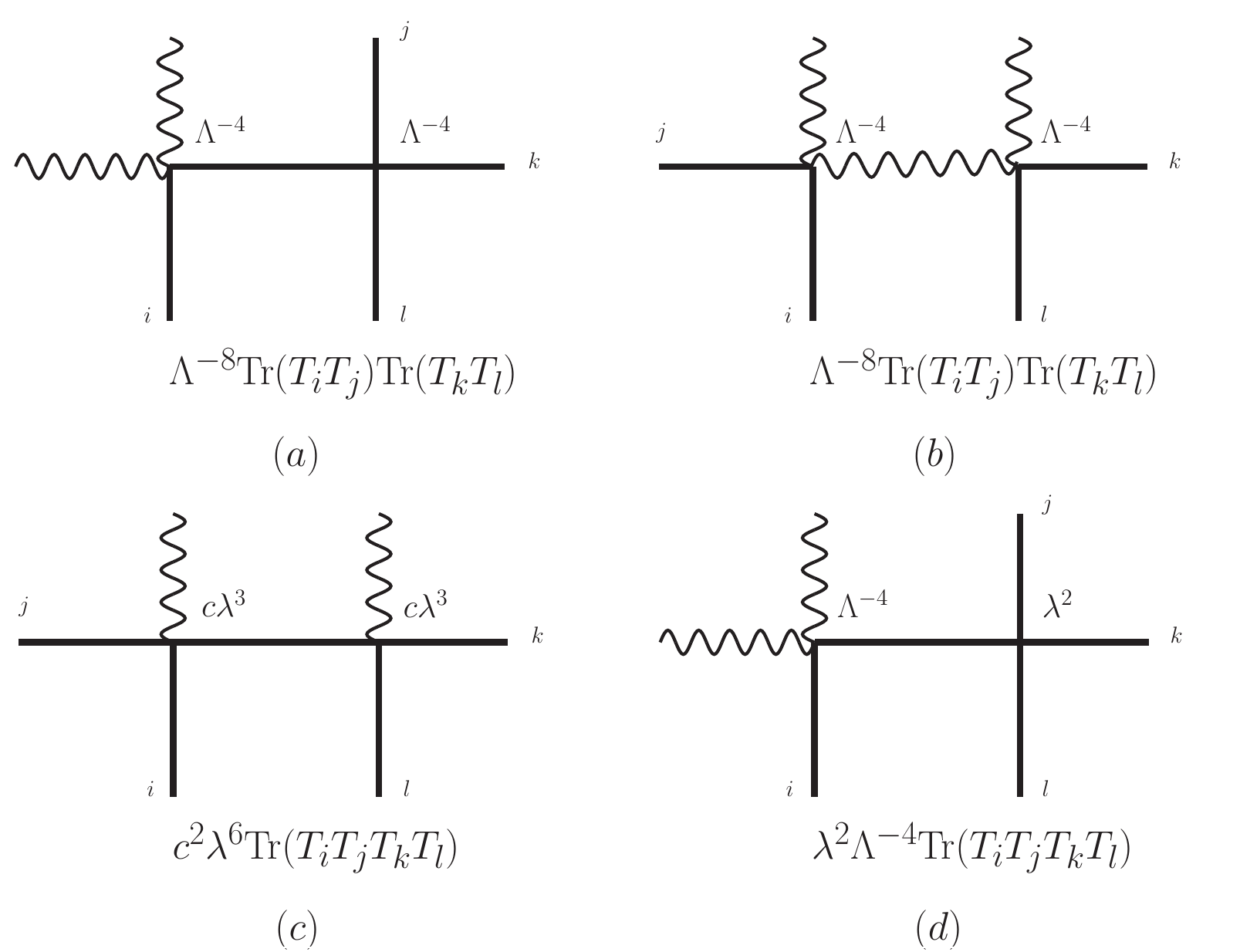}
\caption{Four Feynman diagrams contributing to the scattering of four scalars and two photons in the extended DBI theory with generic coupling $c$.}
\label{fig:4s2g}
\end{figure}

We now turn to a simple example illustrating the above issue. Let us consider a 6-pt amplitude with four scalars and two photons (see Fig.~\ref{fig:4s2g}). 
There are four graphs contributing to this process. Those in the first line, Fig.~\ref{fig:4s2g}(a,b), are pure DBI and are irrelevant for our discussion (they have a different structure of traces). So let us concentrate on the second row of graphs (c) and (d) in Fig.~\ref{fig:4s2g}. The important property to observe is that they both have the same structure of traces $\mathrm{Tr}(TTTT)$, but different dependence on couplings (the amplitude in (c) is proportional to $c^2\lambda^6$ while the one in (d) to $\lambda^2\Lambda^{-4}$). However, both of them must arise from a unique CHY integrand as they have the same trace structure (and clearly the same number of scalars/ photons in external legs). Since this unique CHY integrand comes with a given normalization (dependence on couplings), it implies that normalizations of these amplitudes have to be proportional $c^2\lambda^6\sim \lambda^2\Lambda^{-4}$. This condition results in $c=c_{\mathrm{num}}(\Lambda \lambda)^{-2}$ and therefore fixes $a=2$ as we anticipated. It does not impose any condition on $c_{\mathrm{num}}$ though. Thus to keep $c$ generic, away from this special value, we would have to split the unique CHY integrand into two independent ones. Of course this subdivision would have to be performed on all CHY integrands containing the $c$-vertex in a consistent way. It is not clear to us whether this is possible.

\subsection{Geometrical formulation of the action}\label{sec:geom_form}

The coset manifold $M$ of the $\mathrm{U}(N)$ NLSM is
\begin{align}\label{eq:coset_mfd}
 M:= \frac{\mathrm{U}(N)\times \mathrm{U}(N)}{\mathrm{U}(N)_{\mathrm{diag}}}\simeq \mathrm{U}(N)
\end{align}
and in parallel we can consider also the $\mathrm{SU}(N)$ version. Let $U\in
\mathrm{U}(N)$ be a coset representative. In what follows, we will use the local coordinates on $U(N)$ or $SU(N)$  which correspond to the  Cayley
parameterization~\eqref{eq:coset_rep}. 
The left(right) invariant
Maurer--Cartan form $\sigma_\mathrm{L,R}$ on $\mathrm{U}(N)$ defines the left(right) invariant
 vielbein $e_{\mathrm{L,R}}^a$ 
\begin{align} 
  &\sigma_{\mathrm{L}}=U^{-1}{\rm{d}}U=U^\dagger {\rm{d}}U=e_{\mathrm{L}}^a T_a; \;\; T_a \in \mathfrak{u}(N) \label{eq:LMC_form}\\
  &\sigma_{\mathrm{R}}={\rm{d}}UU^{-1}={\rm{d}}UU^\dagger =e_{\mathrm{R}}^a T_a\label{eq:RMC_form}
\end{align}
and the bi-invariant metric on the coset manifold is then given as
\begin{align}\label{eq:metric}
  h=\frac{1}{4\lambda^2}\mathrm{Tr}({\rm{d}}U^\dagger\otimes {\rm{d}}U)&=\frac{1}{4\lambda^2}\mathrm{Tr}(\sigma_{\mathrm{L}}\otimes\sigma_{\mathrm{L}})=\delta_{ab}e_{\mathrm{L}}^a e_{\mathrm{L}}^b\notag \\ &=\frac{1}{4\lambda^2}\mathrm{Tr}(\sigma_{\mathrm{R}}\otimes\sigma_{\mathrm{R}})=\delta_{ab}e_{\mathrm{R}}^a e_\mathrm{R}^b.
\end{align}
Next, let us define a local 2-form $B~$ on the coset manifold, which is expressed in coordinates corresponding to the  Cayley parametrization \eqref{eq:coset_rep} as
\begin{align}\label{eq:2form}
  B= \sum_{m=1}^\infty \sum_{k=0}^{m-1} \frac{2(m-k)}{2m+1}\lambda^{2m+1}\mathrm{Tr}\left( {\rm{d}}\phi\wedge \phi^{2k}{\rm{d}}\phi \phi^{2(m-k)-1} \right).
\end{align}
A map from Minkowski spacetime to the coset manifold
\begin{align}\label{eq:map_Mink_coset}
  X:\mathbb{R}^{1,3}\to \mathrm{U}(N),
\end{align}
induces a pull-back $X^*$ of the metric $h$ and the 2-form $B$
\begin{align}
&g=X^*h=g_{\mu\nu}{\rm{d}}x^{\mu}\otimes {\rm{d}}x^{\nu}, &&[g]=4 \\
&W=X^*B=W_{\mu\nu}{\rm{d}}x^{\mu}\wedge {\rm{d}}x^{\nu}, &&[W]=2
\end{align}
to spacetime. 

Finally, we denote by $\eta$ the Minkowski metric and by $F$ a 2-form field
strength of an abelian $\mathrm{U}(1)$ gauge field on spacetime, which
conveniently combines with the pull-back of the 2-form $B$ into a unified 2-form $\mathcal{F}$
on $\mathbb{R}^{1,3}$
\begin{align}
  \mathcal{F}=cW+F=cW+{\rm{d}}A,&&[\mathcal{F}]=2
\end{align}
With these definitions at our disposal the action of the extended DBI
theory~\eqref{eq:eDBI} can be written in a more compact form
\begin{align}\label{eq:action}
  S=\int_{\mathbb{R}^{1,3}}{\rm{d}}^4x\Bigg\{\Lambda^4\Big[1-\sqrt{-\det{\left(\eta-\Lambda^{-4} g-\Lambda^{-2} \mathcal{F}  \right)_{\mu\nu}}}\Big]\Bigg\}
\end{align}
with all the matrices entering the determinant manifestly dimensionless.

\subsection{Symmetries of the action }
The isometry group of the coset space $M=G\big/H$ in~\eqref{eq:coset_mfd} is
$G=\mathrm{U}(N)\times \mathrm{U}(N)$. By definition, the metric $h$ is invariant
with respect to these isometries. The first $\mathrm{U}(N)$ corresponds to the left
action on a point $U\in M$: $U\mapsto g_{\mathrm{L}}U$, while the second $\mathrm{U}(N)$
corresponds to the right action $U\mapsto Ug_{\mathrm{R}}$. It is useful to take
linear combinations of these two groups of generators in
$\mathfrak{u}_{\mathrm{L,R}}$, such that they form the Lie algebra of
$H=\mathrm{U}_{\mathrm{diag}}(N)$ and the rest will form the tangent space to $M$
\begin{align}
\mathfrak{g}=\mathfrak{u}_{\mathrm{L}}\oplus\mathfrak{u}_{\mathrm{R}}\simeq\mathfrak{h}\oplus T_{e}M  \,.
\end{align}  
The new generators on the right hand side are formed in terms of the old ones
$X_{\mathrm{L,R}}\in\mathfrak{u}_{\mathrm{L,R}}$ as
\begin{align}\label{eq:Kill_vec}
&\mathbf{v}:= X_{\mathrm{L}}-X_{\mathrm{R}} \;\in \mathfrak{h}\simeq\mathfrak{u}(N)_{\mathrm{diag}} \\  
&\mathbf{a}:= X_{\mathrm{L}}+X_{\mathrm{R}} \;\in T_{e}M.  
\end{align}
In the sigma model community it is customary to denote the generators
$\mathbf{v}$ as vector symmetries while $\mathbf{a}$ as axial symmetries,
hence the notation. In this language, the isometry group becomes
\begin{align}
  G=\mathrm{SU}(N)_{\mathrm{V}}\times\mathrm{SU}(N)_{\mathrm{A}}\times\mathrm{U}(1)_{\mathrm{V}}\times\mathrm{U}(1)_{\mathrm{A}}.
\end{align}
We already stated that the metric is bi-invariant, i.e. its Lie derivative
vanishes for both $\mathbf{v}$ and $\mathbf{a}$
\begin{align}
\mathcal{L}_{\mathbf{v}}h=\mathcal{L}_{\mathbf{a}}h=0.  
\end{align} 
On the other hand the 2-form $B$ is only invariant with respect to $H$
\begin{align}
&\mathcal{L}_{\mathbf{v}}B=0,  &&\mathcal{L}_{\mathbf{a}}B\neq0.
\end{align}
However, we could restore a full $\mathrm{U}(N)_{\mathrm{L}}\times
\mathrm{U}(N)_{\mathrm{R}}$ invariance of the action~\eqref{eq:action} if
$\mathcal{L}_{\mathbf{a}}B$ would be a closed and hence exact 2-form (since
$H^2_{\mathrm{dR}}(\mathrm{U}(N))=0$ and the same holds for $\mathrm{SU}(N)$).
For the 2-form $\mathcal{L}_{\mathbf{a}}B$ we would have $\mathcal{L}_{\mathbf{a}}B= {\rm{d}}{\beta}$ for some 1-form ${\beta}$ (we denote its pull-back to
spacetime $X^*{\beta}$ as $b$). In that case, under an infinitesimal
transformation in the direction of the Killing vector field $\mathbf{a}$, the terms of the Lagrangian
density~\eqref{eq:action}
 inside the determinant  would transform as
\begin{align}
\eta+\Lambda^{-4} g-\Lambda^{-2}(cW+{\rm{d}}A)\xmapsto{\mathcal{L}_{\mathbf{a}}}&\eta+\Lambda^{-4} g-\Lambda^{-2}(cW+c{\rm{d}}b+{\rm{d}}A) \notag \\
=&\eta+\Lambda^{-4} g-\Lambda^{-2}(cW+{\rm{d}}[cb+A]). 
\label{axialL}
\end{align}
It is clear that we can arrange a full $\mathrm{U}(N)_{\mathrm{L}}\times
\mathrm{U}(N)_{\mathrm{R}}$ invariance if the term in square brackets stays
invariant, which we can easily achieve by imposing a shift symmetry for the
gauge field $A$ under an infinitesimal transformation in the direction $\mathbf{a}$
\begin{align}\label{eq:shift_symm}
  A\mapsto A-cb.
\end{align}
The proof that $\mathcal{L}_{\mathbf{a}}B={\rm{d}}{\beta}$ is carried out by an
explicit computation for $\mathrm{U}(N)$ in appendix~\ref{sec:appA}. In section~\ref{sec:SU2_theory} it is shown for the special case of
$\mathrm{SU}(2)\simeq S^3$ using just standard differential geometry on the 3-sphere.
Here we wish to give an argument based on group cohomology. 

Suppose that we are looking for a local two-form $B$ defined on the coordinate chart \eqref{eq:coset_rep} which is invariant under $\mathbf{v}$-transformations and such, that its Lie derivative in direction of $\mathbf{a}$ is closed, i.e.
\begin{align}
  {\rm{d}}\mathcal{L}_{\mathbf{a}}B=0\,.
\end{align}
Using  the fact that the external
differential and Lie derivative commute, we get immediately 
\begin{equation}
    \mathcal{L}_{\mathbf{a}}{\rm{d}}B=0.
\end{equation}
Therefore ${\rm{d}}B$ has to be invariant under $\mathbf{a}$-transformations. However, we
required that $B$ and hence ${\rm{d}}B$  is invariant under
$\mathbf{v}$-transformations. This implies that ${\rm{d}}B$ is a local bi-invariant 3-form defined on the coordinate chart \eqref{eq:coset_rep} of a compact
group $\mathrm{U}(N)$ or $\mathrm{SU}(N)$, i.e. 
\begin{align}
  &\mathcal{L}_{\mathbf{v}}{\rm{d}}B=0, &&\mathcal{L}_{\mathbf{a}}{\rm{d}}B=0.
\end{align} 
On the other hand, it is a mathematical fact that a bi-invariant form on a compact connected group is
harmonic (in a bi-invariant metric which always exists under these assumptions). The bi-invariant 3-form $\mathrm{d}B$ is thus 
a representative of third cohomology of either $\mathrm{U}(N)$ or
$\mathrm{SU}(N)$. On each compact connected group there is at least one 3-form
satisfying these properties, which is the Cartan 3-form
\begin{align}
  \Omega=\mathrm{Tr}\left( {\sigma_{\mathrm{L}}\wedge \sigma_{\mathrm{L}}\wedge \sigma_{\mathrm{L}}}\right),
\end{align}
where $\sigma_{\mathrm{L}}$ is the left invariant Maurer--Cartan form defined
in~\eqref{eq:LMC_form}. But in our case the third Betti number for either of the
relevant groups is equal to one
\begin{align}
 & b_3(\mathrm{U}(N))=\mathrm{dim}H^3_{\mathrm{dR}}(\mathrm{U}(N))=1 \\
 &b_3(\mathrm{SU}(N))=\mathrm{dim}H^3_{\mathrm{dR}}(\mathrm{SU}(N))=1,
\end{align}
therefore there is precisely one bi-invariant 3-form -- the Cartan 3-form $\Omega$. 
Note that this form coincides for $\mathrm{U}(N)$ and $\mathrm{SU}(N)$. Every
unitary matrix $U$ can be written as $U=e^{i\alpha}\hat{U}$, $\hat{U}\in \mathrm{SU}(N)$. The left
invariant Maurer--Cartan form of $\mathrm{U}(N)$ then splits as
\begin{align}
  \sigma_{\mathrm{L}}=\hat{U}^{-1}{\rm{d}}\hat{U}+i{\rm{d}}\alpha,
\end{align}
however the $d\alpha$ piece completely drops out in the wedge product, leaving us
with
\begin{align}\label{eq:Cartan_3form}
 & \Omega(\mathrm{U}(N))=\mathrm{Tr}\left( U^{-1}{\rm{d}}U\wedge U^{-1}{\rm{d}}U\wedge U^{-1}{\rm{d}}U\right) \notag \\ 
& =\mathrm{Tr}\left( \hat{U}^{-1}{\rm{d}}\hat{U}\wedge \hat{U}^{-1}{\rm{d}}\hat{U}\wedge \hat{U}^{-1}{\rm{d}}\hat{U}\right)=\Omega(\mathrm{SU}(N)).
\end{align}
 Since the unique bi-invariant 3-form $\Omega$ is closed, locally in every
coordinate patch $\{V_\alpha\}$ of $\mathrm{U}(N)$ or $\mathrm{SU}(N)$ there exists a
unique (up to an exact form)   2-form $C_\alpha$  such that 
\begin{align}\label{eq:Cartan_dB}
  \Omega={\rm{d}}C_\alpha
\end{align}
in $V_\alpha$. In the coordinate chart $V_{\rm{Cayley}}$ corresponding to the Cayley parametrization \eqref{eq:coset_rep} we can then set\footnote{We could add to $B$ an arbitrary exact 2-form ${\rm{d}}\gamma$, which could
be however absorbed in a redefinition of the gauge field $A\mapsto A+\gamma$.}
\begin{equation}
    B=\kappa\, C_{\rm{Cayley}},
\end{equation}
where $\kappa$ is an appropriate normalization.
Then ${\rm{d}}B$ is bi-invariant and as a consequence, $\mathcal{L}_{\mathbf{a}}B$ is closed as desired.
This finishes the proof of existence of $B$. In the coordinates (\ref{eq:coset_rep}) it coincides with (\ref{eq:2form}). For an explicit
confirmation of this fact and fixing of the normalization $\kappa$, see appendix~\ref{sec:appA}, in particular~\eqref{eq:dB_norm}.

To summarize, we have shown that there is a unique way how to ensure the shift
symmetry of the gauge field $A$ in~\eqref{eq:shift_symm} and thus a full
$\mathrm{U}(N)_{\mathrm{L}}\times \mathrm{U}(N)_\mathrm{R}$ symmetry of the
action~\eqref{eq:action}
\footnote{Let us remark that the 3-form $\Omega$ might have on top of its topological meaning also a more physical interpretation. The NLSM Lagrangian (enriched by four derivative terms) encompasses a special subsector known as the Skyrme model. The Skyrme model admits solitons which can be identified with baryons of its parent theory at high energies -- the $\mathrm{SU}(N)$ QCD theory. Their integral topological charge arises from a topological current $J_{\mathrm{top}}=\tfrac{1}{24\pi^2}\star\Omega$. Thus for the $W$ piece of the Lagrangian~\eqref{eq:eDBI}, one arrives at the relation $\star J_{\mathrm{top}}\sim {\rm{d}}W$.}.
In particular we have proved, that the basic building blocks $g_{\mu\nu}$ and ${\cal{F}}_{\mu\nu}$ are separately invariant. This means that any sensible Lagrangian built of these basic building blocks will be invariant too. In the next subsection we will give an explicit example of such Lagrangian.
The full
$\mathrm{U}(N)_{\mathrm{L}}\times \mathrm{U}(N)_\mathrm{R}$ invariance of the action
implies soft theorems for scalar particles described by the field $\phi(x)$ with
values in $\mathfrak{u}(N)$. Those will be discussed and exploited in section~\ref{sec:reconstr}
to prove on-shell constructibility of the tree-level S-matrix of the extended
DBI theory.

\subsection{Significant limits in coupling constants}\label{sec:limits}

The extended DBI theory was engineered by the authors of~\cite{Cachazo:2014xea} to
interpolate between NLSM and DBI theories. However, modification of the action
that we introduced in~\eqref{eq:eDBI} consisting in the introduction of one extra
coupling $c$ for $W_{\mu\nu}$ and altering the power of the mass scale $\Lambda$ of this
term slightly modifies these interpolating properties. Now we have more freedom
and thus the web of Lagrangians emerging from various limits in the three
couplings $\Lambda$, $\lambda$ and $c$ will be richer. In particular, we will obtain the
NLSM Lagrangian in a two step procedure with the intermediate Lagrangian
of interest in its own right. It is of the form
$\mathcal{L}[g_{\mu\nu},\mathcal{F}_{\mu\nu}]$ and as discussed in the previous subsection and explicitly shown in appendix~\ref{sec:appA}
all such theories enjoy the full chiral symmetry
$\mathrm{U}(N)_{\mathrm{L}}\times\mathrm{U}(N)_{\mathrm{R}}$ which subsequently
implies soft theorems for the scalars. It will be discussed in
section~\ref{sec:reconstr} that such theories have an on-shell constructible
tree-level S-matrix. 
 
In order to present the flow in the space of couplings $(\Lambda,\lambda,c)$ in
a more elegant way, it is beneficial to assign to the coupling $\lambda$ a geometrical
meaning as it is proportional to the square root of the scalar curvature $R_S$ of the compact coset
manifold
\begin{align}\label{eq:lambda_scal_curv}
 \lambda \sim \sqrt{R_S}.
\end{align}
Then it becomes clear that for $\lambda\to 0$ the coset manifold expands to a flat space
$\mathbb{R}^{N^2}$ and thus the theory flows to a DBI model.
Let us remark that for the original theory
in~\cite{Cachazo:2014xea} corresponding to the choice $c=(\Lambda\lambda)^{-2}$ there is one further limit which appears beyond the weak field expansion and which
reduces~\eqref{eq:action} to the BI theory. 
Note that  according to the above geometric interpretation of the parameter $\lambda$, for $\lambda\to \infty$
the coset manifold shrinks to a point. For the $\mathrm{SU}(2)$ extended DBI theory, we can show that both the metric $h$ and the 2-form $B$ vanish in this latter point-like limit and thus the theory flows to a BI theory when $\lambda\to\infty$. We however defer the discussion of this BI limit to the next section, since it requires a closed form expression for the 2-form $B$, which will be given in~\eqref{eq:B_explicit}.

The various limits of the Lagrangian $\mathcal{L}_{\mathrm{eDBI}}$ in~\eqref{eq:eDBI} are presented in Fig.~\ref{fig:lim_eDBI} below:
\begin{figure}[ht!]
\centering
\tikzstyle{block} = [rectangle, draw, text width=5em, text centered, rounded corners, minimum height=4em]
\tikzstyle{arrow} = [thick,->,>=stealth]
\resizebox{\columnwidth}{!}{
\begin{tikzpicture}[node distance = 3cm, auto]
    % Place nodes
    \node [block, fill=blue!20] (M1) {U$(N)$ DBI \\ \eqref{eq:UN_DBI}};
    \node [block, fill=teal!20, above of=M1] (start) {eDBI \\ \eqref{eq:eDBI}};
    \node [block, fill=green!20, left of=M1, xshift=-2cm] (L1) {min \\ \eqref{eq:min}};
    \node [block, fill=red!20, right of=M1, xshift=2cm] (R1) {reDBI \\ \eqref{eq:reDBI}};
    \node [block, fill=violet!20, below of=M1] (M2) {DBI \\ \eqref{eq:DBI}};
    \node [block, fill=yellow!20, left of=M2, xshift=-2cm] (L2) {$\mathrm{NLSM}\;\oplus\;[\mathrm{free}\;\gamma]$  \\ \eqref{eq:NLSM_gamma}};
    \node [block, fill=brown!20, right of=M2, xshift=2cm] (R2) {rmin \\ \eqref{eq:rmin}};
    \node [block, below of=M2, xshift=-2cm] (L3) {$[\mathrm{free}\;\phi]\;\oplus\;[\mathrm{free}\;\gamma]$ \\ \eqref{eq:free_sg}};
    \node [block, below of=M2, xshift=2cm] (R3) {[$\mathrm{free}\;\phi]$ \\ \eqref{eq:free_s}};
    % Draw arrows
    \draw [arrow] (start) -| node [near end, anchor=east] {$\Lambda\to\infty$} (L1);
    \draw [arrow] (start) -- node [anchor=east] {$c\to 0$} (M1);
    \draw [arrow] (start) -| node [near end, anchor=west] {$\begin{array}{ll}c\to c^*=(\lambda\mu)^{-3}\\ \lambda\to 0\end{array}$} (R1);
    \draw [arrow,green!60!black] (L1) -- node [anchor=east] {$c\to 0$} (L2);
    \draw [arrow,green!60!black] (L1.east) -- node [anchor=west,pos=0.8] {$\lambda\to 0$} (L3);
    \draw [arrow,blue!60!black] (M1) -- node [anchor=west,pos=0.75] {$\Lambda\to \infty$} (L2);
    \draw [arrow,blue!60!black] (M1) -- node [anchor=west,pos=0.5] {$\lambda\to 0$} (M2);
    \draw [arrow,red!60!black] (R1) -- node [anchor=west,pos=0.6] {$\mu\to \infty$} (M2);
    \draw [arrow,red!60!black] (R1) -- node [anchor=west,pos=0.5] {$\Lambda\to\infty$} (R2);
    \draw [arrow,yellow!60!black] (L2) -- node [anchor=east,pos=0.5] {$\lambda\to 0$} (L3);
    \draw [arrow,violet!60!black] (M2) -- node [anchor=west,pos=0.8] {$\Lambda\to \infty$} (L3);
    \draw [arrow,brown!60!black] (R2) -- node [anchor=west,pos=0.8] {$\mu\to\infty$} (R3);
\end{tikzpicture}}
\caption{Web of limits for the extended DBI theory.}
\label{fig:lim_eDBI}
\end{figure}
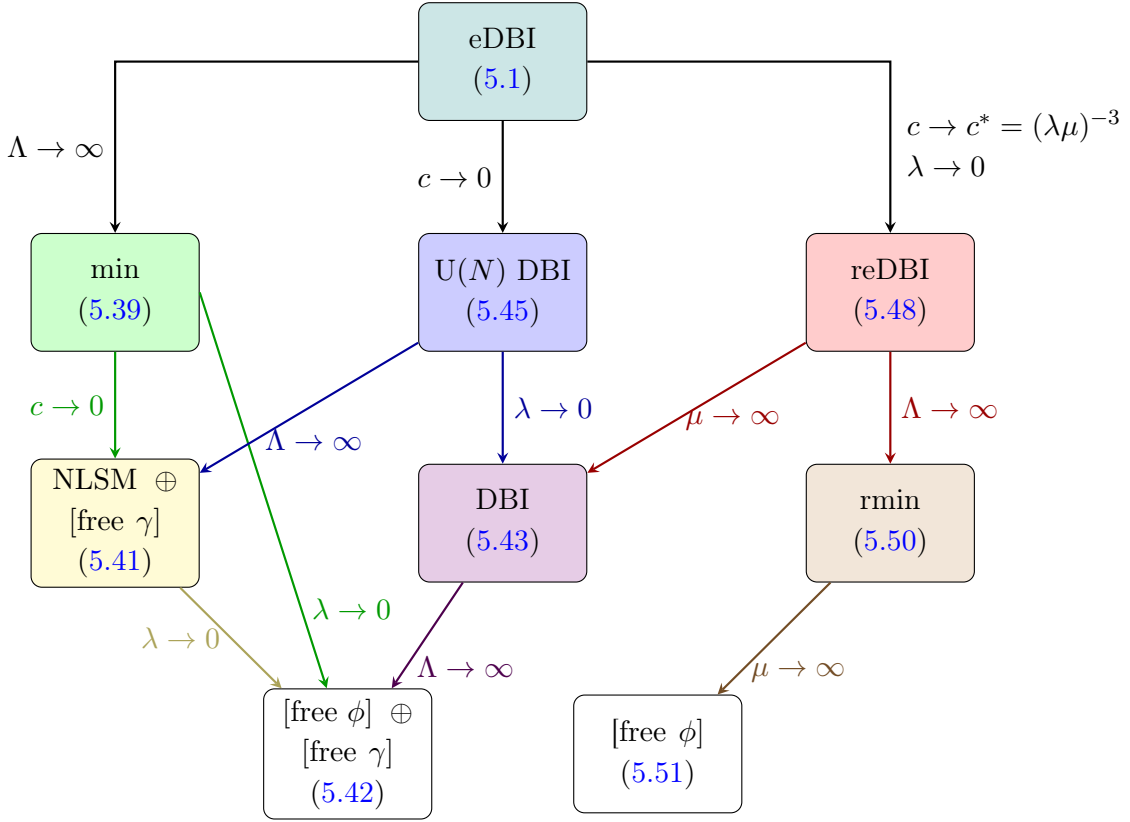

Not to get caught in the web of limits presented in Fig.~\ref{fig:lim_eDBI}, it is useful to have the following geometrical picture in mind. The extended DBI can be alternatively interpreted as a theory living  on a worldvolume of a flat 3-brane $\mathbb{R}^{1,3}$ embedded in an ambient space $\mathbb{R}^{1,3}\times\mathrm{U}(N)$ equipped with the metric
\begin{align}
ds^2&=\eta_{\mu\nu}dx^\mu dx^\nu-\Lambda^{-4}h_{ab}d\phi^ad\phi^b \notag \\  &=\Big(\eta_{\mu\nu}-\underbrace{\Lambda^{-4}h_{ab}\partial_\mu\phi^a\partial_\nu\phi^b}_{g_{\mu\nu}}\Big)dx^\mu dx^\nu \notag \\
&=H_{\mu\nu}dx^\mu dx^\nu,
\end{align}
where $\eta_{\mu\nu}$ and $h_{ab}$ are metrics on $\mathbb{R}^{1,3}$ and $\mathrm{U}(N)$, respectively. The pull-back of the $\mathrm{U}(N)$ metric to the 3-brane is denoted $g_{\mu\nu}$ as above, and finally $H_{\mu\nu}$ stands for the induced metric on the 3-brane worldvolume $\mathbb{R}^{1,3}$. These are natural objects from which actions of individual theories in the diagram are constructed. The last needed building block is provided by the generalized field strength
\begin{align}
\mathcal{F}&=F_{\mu\nu}dx^\mu\wedge dx^\nu+cB_{ab}d\phi^a\wedge d\phi^b \notag \\ 
&=\Big(F_{\mu\nu}+\underbrace{cB_{ab}\partial_\mu\phi^a\partial_\nu\phi^b}_{cW_{\mu\nu}}\Big)dx^\mu\wedge dx^\nu \notag \\
&=\mathcal{F}_{\mu\nu}dx^\mu\wedge dx^\nu.
\end{align}
Based on this view of a 3-brane living in a curved ambient space, the action of the  extended DBI theory~\eqref{eq:eDBI} can be rewritten as
\begin{align}
\mathcal{L}_{\mathrm{eDBI}}=\Lambda^4-\Lambda^4\sqrt{-\det(H_{\mu\nu}-\Lambda^{-2}\mathcal{F}_{\mu\nu})}    
\end{align}
and all limits following from this action have a natural interpretation:
\begin{itemize}
    \item $\mathbf{\lambda\to 0:}$ the curvature of the $\mathrm{U}(N)$ manifold goes to zero and thus the ambient space becomes flat $\mathbb{R}^{1,3}\times\mathbb{R}^{N^2}$, and at the same time the 2-form $B$ vanishes, so to obtain an action resulting from this limit, it is enough to replace the curved metric $h_{ab}$ by a flat one $\delta_{ab}$ (for the induced metric $H_{\mu\nu}\to\Delta_{\mu\nu}=\eta_{\mu\nu}-\Lambda^{-4}\delta_{ab}\partial_{\mu}\phi^a\partial_{\nu}\phi^b$) and the generalized filed strength ${\cal{F}}_{\mu\nu}$ with $F_{\mu\nu}$
    \item $\mathbf{c\to 0:}$ this limit turns off the 2-form $B$ on $\mathrm{U}(N)$ and so the generalized field strength simply reduces to an ordinary one, thus we just make a replacement $\mathcal{F}\to F$ in the starting Lagrangian
    \item $\mathbf{\Lambda\to\infty:}$ this is the ``decoupling limit'', one is instructed to extract the linear and quadratic leading order terms from the square root DBI-like Lagrangians in order to obtain the action resulting from these limits 
     \item ${\mathbf{c=c^*=(\lambda\mu)^{-3},\,\,\,\lambda\to 0}}$, where we introduced a new scale $\mu$ with dimension $[\mu]=1$:  this is the ``reduced'' $\lambda\to 0$ limit. As before, the ambient space becomes flat which results in the replacement $H_{\mu\nu}\to\Delta_{\mu\nu}$, but the form $B$ survives in this limit being replaced with $B\to B^*=\frac{2}{3}\mu^{-3}\langle\phi {\rm d}\phi\wedge{\rm {d}}\phi \rangle $
\end{itemize}
Let us now describe various branches of Fig. \ref{fig:lim_eDBI} in more detail.

\subsubsection*{$\Lambda\to\infty$: path towards NLSM via a minimal model with shift symmetry}
To perform this limit one needs to expand the square root to order $\mathcal{O}(\Lambda^{-4})$
(order $\mathcal{O}(\Lambda^{-2})$ vanishes due to anti-symmetry of $\mathcal{F}$ while order
$\mathcal{O}(1)$ is cancelled by the $\Lambda^4$ term in the action). This
immediately yields the result of the limit $\Lambda\to\infty$
\begin{equation}\label{eq:min}
  \mathcal{L}_{\mathrm{eDBI}}\xrightarrow{\Lambda\to\infty} \mathcal{L}_{\mathrm{min}}=\frac{1}{2}\eta^{\mu\nu}g_{\mu\nu}-\frac{1}{4}\mathcal{F}_{\mu\nu}\mathcal{F}^{\mu\nu}=
  \frac{1}{2}h_{ab}\partial_\mu\phi^a\partial_\nu\phi^b\eta^{\mu\nu}-\frac{1}{4}\mathcal{F}_{\mu\nu}\mathcal{F}^{\mu\nu}, 
\end{equation}
which is the minimal invariant Lagrangian of the form ${\cal{L}}[g_{\mu\nu},{\cal{F}}]$. It non-trivially couples  scalars to massless vectors. The first term is the Lagrangian of NLSM and the second term is a minimal
Lagrangian for the generalized field strength $\mathcal{F}$. Thanks to the shift
symmetry~\eqref{eq:shift_symm}, this theory is invariant with respect to the
full chiral symmetry. As discussed in section~\ref{sec:reconstr}, this property is sufficient for showing that its tree-level S-matrix is on-shell constructible.
One can further reduce it by taking either $c\to0$ or $\lambda\to0$. This results in the following chain
of theories
\begin{align}
  \mathcal{L}_{\mathrm{eDBI}}\xrightarrow{\Lambda\to\infty}\mathcal{L}_{\mathrm{min}}\begin{cases}\xrightarrow{c\to0}\mathcal{L}_{\mathrm{NLSM}\oplus\mathrm{free}(\gamma)} \\ \xrightarrow{\lambda\to0}\mathcal{L}_{\mathrm{free}(\phi)\oplus\mathrm{free}(\gamma)}\end{cases},  
\end{align}
where the last Lagrangians
\begin{align}
&\mathcal{L}_{\mathrm{NLSM}\oplus\mathrm{free}(\gamma)}=\frac{1}{2}h_{ab}\partial_\mu\phi^a\partial_\nu\phi^b\eta^{\mu\nu}-\frac{1}{4}F_{\mu\nu}F^{\mu\nu} \label{eq:NLSM_gamma}
\\  &\mathcal{L}_{\mathrm{free}(\phi)\oplus\mathrm{free}(\gamma)}=\frac{1}{2}\delta_{ab}\partial_\mu\phi^a\partial_\nu\phi^b\eta^{\mu\nu}-\frac{1}{4}F_{\mu\nu}F^{\mu\nu} \label{eq:free_sg}
\end{align}
are decoupled theories of either NLSM or free scalars together with a free $\mathrm{U}(1)$ photon.

\subsubsection*{$\lambda\to0$: DBI theory}
As discussed above, in this limit the metric $H$ goes to a flat metric $\Delta$ on flat ambient space 
$\mathbb{R}^{1,3}\times\mathbb{R}^{N^2}$ and the 2-form $W$ vanishes, thus $\mathcal{F}\to F$.
  Therefore we get the ordinary (multi-)DBI theory
\begin{align}\label{eq:DBI}
\mathcal{L}_{\mathrm{eDBI}}\xrightarrow{\lambda\to0} \mathcal{L}_{\mathrm{DBI}}=\Lambda^4\Big[1-\sqrt{-\det{\left(\eta_{\mu\nu}-\Lambda^{-4} \delta_{ab}\partial_\mu\phi^a\partial_{\nu}\phi^b-\Lambda^{-2} F_{\mu\nu}  \right)}}\Big]
\end{align}
and a further limit $\Lambda\to\infty$ results in
\begin{align}
 \mathcal{L}_{\mathrm{eDBI}}\xrightarrow{\lambda\to0}\mathcal{L}_{\mathrm{DBI}}\xrightarrow{\Lambda\to\infty}\mathcal{L}_{\mathrm{free}(\phi)\oplus\mathrm{free}(\gamma)} \,.
\end{align}

\subsubsection*{$c\to0$: decoupling of $W_{\mu\nu}$ and DBI on the $\mathrm{U}(N)$ group}
Sending the strength of the $W_{\mu\nu}$ interaction $c\to0$ trivially leads to
DBI theory, describing a 3-brane in $\mathbb{R}^{1,3}\times\mathrm{U}(N)$ 
\begin{align}\label{eq:UN_DBI}
  \mathcal{L}_{\mathrm{eDBI}}\xrightarrow{c\to0}\mathcal{L}_{\mathrm{U}(N)\mathrm{DBI}}=\Lambda^4\Big[1- \sqrt{-\det\Big(\eta_{\mu\nu}-\Lambda^{-4}h_{ab}\partial_{\mu}\phi^a\partial_{\nu}\phi^b-\Lambda^{-2}F_{\mu\nu}\Big)}\Big].
\end{align}
The above theory still depends on two couplings $\Lambda$, $\lambda$ that can be sent to
infinity and zero, respectively. The theories one obtains are either NLSM with a
decoupled free photon or DBI, which we have already seen
\begin{align}
  \mathcal{L}_{\mathrm{eDBI}}\xrightarrow{c\to0}  \mathcal{L}_{\mathrm{U}(N)\mathrm{DBI}}\begin{cases}\xrightarrow{\Lambda\to\infty}&\mathcal{L}_{\mathrm{NLSM}\oplus\mathrm{free}(\gamma)}
\\
\xrightarrow{\lambda\to0}&\mathcal{L}_{\mathrm{DBI}}\end{cases}
\end{align}

\subsubsection*{$c\to c^*=(\lambda \mu)^{-3},\,\lambda\to 0$: reduced extended DBI theory}
\begin{figure}[t]
\centering
\includegraphics[width=12cm]{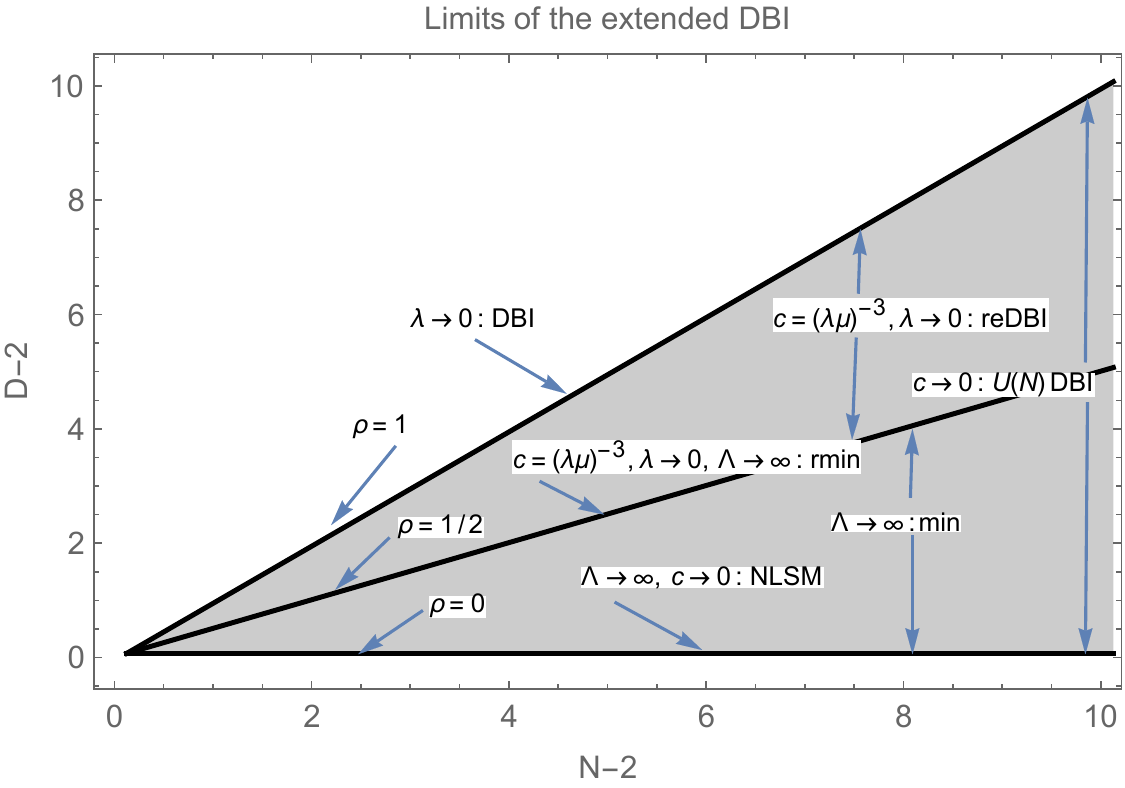}
\caption{Graphical representation of the various limits of the extended DBI in the $(N-2,D-2)$ plane. }
\label{fig:3}
\end{figure}

In order to define this limit we introduce an additional mass scale $\mu$ and then
send $c$ to the special value $c^*=(\lambda \mu)^{-3}$. The effect of this
operation is that only the first term in the expansion of $W_{\mu\nu}$ survives,
while all higher order terms vanish for $\lambda\to 0$
\begin{align}
  c^*W_{\mu\nu}\xrightarrow {\lambda\to 0}W^*_{\mu\nu}=\frac{2}{3}\mu^{-3}\mathrm{Tr}\left( \partial_{[\mu}\phi\partial_{\nu]}\phi\phi \right).
\end{align}
Defining $\mathcal{F}^*_{\mu\nu}=F_{\mu\nu}+W^*_{\mu\nu}$, we can write the resulting
reduced extended DBI theory as
\begin{align}\label{eq:reDBI}
  \mathcal{L}_{\mathrm{eDBI}}\xrightarrow{c\to c^*,\,\lambda\to 0}\mathcal{L}_{\mathrm{reDBI}}= \Lambda^4\Big[1- \sqrt{-\det\Big(\eta_{\mu\nu}-\Lambda^{-4}\delta_{ab}\partial_{\mu}\phi^a\partial_{\nu}\phi^b-\Lambda^{-2}\mathcal{F}^*_{\mu\nu}\Big)}\Big].
\end{align}
As before, by sending $\mu\to\infty$ or $\Lambda\to\infty$, we complete the full chain of
flows
\begin{align}
\mathcal{L}_{\mathrm{eDBI}}\xrightarrow{c\to c^*,\,\lambda\to 0}\mathcal{L}_{\mathrm{reDBI}}\begin{cases}\xrightarrow{\mu\to \infty}&\mathcal{L}_{\mathrm{DBI}}\xrightarrow{\Lambda\to \infty}\mathcal{L}_{\mathrm{free}(\phi)\oplus\mathrm{free}(\gamma)} \\
\xrightarrow{\Lambda\to\infty}&\mathcal{L}_{\mathrm{rmin}}\xrightarrow{\mu\to \infty}\mathcal{L}_{\mathrm{free}(\phi)},  
\end{cases}  
\end{align}
where the Lagrangians of the reduced minimal theory and of free scalars take the form
\begin{align}
  &\mathcal{L}_{\mathrm{rmin}}=\frac{1}{2}\delta_{ab}\partial_{\mu}\phi^a \partial_{\nu}\phi^b\eta^{\mu\nu}-\frac{1}{4}\mathcal{F}^*_{\mu\nu}\mathcal{F}^{*\mu\nu} \label{eq:rmin} \\
&\mathcal{L}_{\mathrm{free}(\phi)}=\frac{1}{2}\delta_{ab}\partial_{\mu}\phi^a \partial_{\nu}\phi^b\eta^{\mu\nu}\,.  \label{eq:free_s}
\end{align}
\begin{center}\rule{5cm}{0.4pt}\end{center}
Note that by the limiting procedure we got single-$\rho$ theories as
well as multi-$\rho$ theories. The list of $\rho$ values for the former ones is:
$\rho_{\mathrm{NLSM}\oplus\mathrm{free}(\gamma)}=0$, $\rho_{\mathrm{rmin}}=\tfrac{1}{2}$ and
$\rho_{\mathrm{DBI}}=1$. The range of $\rho$ values for the second class of multi-$\rho$
theories is: $0\le\rho_{\mathrm{eDBI}}\le1$, $0\le\rho_{\mathrm{U}(N)\mathrm{DBI}}\le1$, $0\le\rho_{\min}\le 1/2$ and
$\tfrac{1}{2}\le\rho_{\mathrm{reDBI}}\le1$.
Their graphical representation in the $(N-2,D-2)$ plane is depicted in Fig. \ref{fig:3}.

%%%%%%%%%%%%%%%%%%%%%%%%%%%%%%%%%%%%%%%%%%%%%%%%%%%%%%%%%%%%
\section{Geometry of $\mathrm{SU}(2)$ extended DBI theory}\label{sec:SU2_theory}
%%%%%%%%%%%%%%%%%%%%%%%%%%%%%%%%%%%%%%%%%%%%%%%%%%%%%%%%%%%%

In this section we present a detailed example of the
constructions discussed so far for the case of $\mathrm{SU}(2)$ theory. We specialize therefore to the coset space
\begin{align}
  \frac{\mathrm{SU}(2)\times \mathrm{SU}(2)}{\mathrm{SU}(2)_{\mathrm{diag}}}\simeq \mathrm{SU}(2)\simeq S^3 \simeq \mathrm{SO}(4) \big / \mathrm{SO}(3)
\end{align}
of the NLSM. The advantage of doing so is that we will be able to carry out all
computations explicitly. In particular we will derive a closed form expression
for the 2-form $B$ on $S^3 \simeq \mathrm{SU}(2)$.

So let us start with basic differential geometry of $S^3$. We will work
in the Cartan formalism and express all objects in terms of the vielbein (triad
in this case) and its dual. Computing the metric defined in~\eqref{eq:metric},
we get
\begin{align}
  h=\frac{1}{\Big[1+\lambda^2\left( \frac{\|\phi\|}{\sqrt{2}}\right)^2\Big]^2}\mathrm{d}s^2_{\mathbb{R}^3}.
\end{align}
We see that the $\phi$-coordinates are those that yield $S^3$ as conformally flat,
i.e. they are the coordinates of a stereographic projection of $S^3$ to
$\mathbb{R}^3$ from the north, respectively south pole (we need two coordinate
patches $V_N$ and $V_S$ to cover $S^3$). We will be interested almost exclusively in local
considerations, so we will work in the north patch $V_N$.

It will turn out to be very useful to introduce
spherical coordinates on the $\mathbb{R}^3$ image of the stereographic
projection
\begin{align}
\phi^1&=\sqrt{2}R\sin(\theta)\sin(\varphi) \notag \\ 
\phi^2&=\sqrt{2}R\sin(\theta)\cos(\varphi) \notag \\
\phi^3&=\sqrt{2}R\cos(\theta).
\end{align}
The reason is that these coordinates are well adapted to the 2-form
$B$ as we will see in a moment. A standard computation then gives the
fundamental objects of differential geometry -- the triad $e^a$, its dual $E_a$,
the spin connection 1-form $\omega^a_{\phantom{a}b}$ and the curvature 2-form $R^a_{\phantom{a}b}$. The triad
defined in~\eqref{eq:LMC_form} reads (we work with the left invariant version from now
on, unless otherwise stated)
\begin{align}
e^1&=\frac{\sqrt{2}}{1+\lambda^2R^2}\mathrm{d}R \notag \\
e^2&=\frac{\sqrt{2}}{1+\lambda^2R^2}R\mathrm{d}\theta \notag \\
e^3&=\frac{\sqrt{2}}{1+\lambda^2R^2}R\sin(\theta)\mathrm{d}\varphi
\end{align}
and provides an explicit expression for the metric
\begin{align}\label{eq:metric_expl}
h=\sum_{a=1}^3e^a\otimes e^a=\frac{2}{\left( 1+\lambda^2R^2\right)^2}\Big[\mathrm{d}R^2+R^2\left( \mathrm{d}\theta^2+\sin^2{\theta}\mathrm{d}\varphi^2 \right)\Big].
\end{align}
The dual triad takes the form
\begin{align}
E_1&=\frac{1+\lambda^2R^2}{\sqrt{2}}\partial_R \notag \\  
  E_2&=\frac{1+\lambda^2R^2}{\sqrt{2}R}\partial_{\theta} \notag \\  
  E_3&=\frac{1+\lambda^2R^2}{\sqrt{2}R\sin(\theta)}\partial_{\varphi}. 
\end{align}
Cartan's first structure equation
\begin{align}
 \mathrm{d}e^a+\omega^a_{\phantom{a}b}\wedge e^b=0 
\end{align}
yields the unique spin connection (remind that it is torsionless and hence anti-symmetric)
\begin{align}
\omega^1_{\phantom{1}2}&=-\frac{1-\lambda^2R^2}{2R}e^2 \notag \\  
  \omega^1_{\phantom{1}3}&=-\frac{1-\lambda^2R^2}{2R}e^3 \notag \\  
  \omega^2_{\phantom{2}3}&=-\frac{1+\lambda^2R^2}{2R}\cot(\theta)e^3.
\end{align}
Cartan's second structure equation gives the curvature 2-form
\begin{align}
R^a_{\phantom{a}b}=\mathrm{d}\omega^a_{\phantom{a}b}+\omega^a_{\phantom{a}c}\wedge \omega^c_{\phantom{c}b}  
\end{align}
with the result
\begin{align}
R^1_{\phantom{1}2}&=\lambda^2 e^1\wedge e^2 \notag \\  
  R^1_{\phantom{1}3}&=\lambda^2 e^1\wedge e^3 \notag \\  
  R^2_{\phantom{2}3}&=\lambda^2 e^2\wedge e^3.
\end{align}
From the above one obtains the Ricci tensor
\begin{align}
 R_{11}=R_{22}=R_{33}=2\lambda^2, 
\end{align}
which leads to a scalar curvature $R_S=6\lambda^2$. Thus we see that in this case the
coupling $\lambda$ is equal to the inverse radius of the 3-sphere
\begin{align}
  \lambda=r^{-1}_{S^3},
\end{align}
a concrete incarnation of the general formula~\eqref{eq:lambda_scal_curv}.

Next, in order to perform explicit calculations, we need the Killing vectors
$\mathbf{v}$ and $\mathbf{a}$ given in~\eqref{eq:Kill_vec} in terms of
left(right) invariant vector fields $X_{\mathrm{L,R}}$ on $\mathrm{SU}(2)$. They have the following
form
\begin{align}
 \mathbf{v}^1&=-2\frac{R}{1+\lambda^2R^2}\left( \sin{\varphi}E_2+\cos{\theta}\cos{\varphi}E_3 \right) \notag \\
  \mathbf{v}^2&=2\frac{R}{1+\lambda^2R^2}\left( \cos{\varphi}E_2-\cos{\theta}\sin{\varphi}E_3 \right) \notag \\
  \mathbf{v}^3&=2\frac{R}{1+\lambda^2R^2}\sin{\theta}E_3  \\
\notag \\
  \mathbf{a}^1&=\frac{1}{\lambda}\bigg\{\sin{\theta}\cos{\varphi}E_1+\frac{1-\lambda^2R^2}{1+\lambda^2R^2}\cos{\theta}\cos{\varphi}E_2-\frac{1-\lambda^2R^2}{1+\lambda^2R^2}\sin{\varphi}E_3\bigg\} \notag \\
  \mathbf{a}^2&=\frac{1}{\lambda}\bigg\{\sin{\theta}\sin{\varphi}E_1+\frac{1-\lambda^2R^2}{1+\lambda^2R^2}\cos{\theta}\sin{\varphi}E_2+\frac{1-\lambda^2R^2}{1+\lambda^2R^2}\cos{\varphi}E_3\bigg\} \notag \\
  \mathbf{a}^3&=\frac{1}{\lambda}\bigg\{\cos{\theta}E_1-\frac{1-\lambda^2R^2}{1+\lambda^2R^2}\sin{\theta}E_2\bigg\}.
\end{align}

Finally, we derive a closed form expression for the 2-form $B$. Then we will have all the
ingredients to check the $\mathrm{U}(N)_{\mathrm{L}}\times \mathrm{U}(N)_\mathrm{R}$
invariance of the action.

We start with the definition of $B$ in~\eqref{eq:2form} and as a first step
realize that for $\phi\in\mathfrak{su}(2)$
\begin{align}
  \left\{\begin{array}{l}\phi^{2n}=\left( -\frac{1}{2} \right)^n \|\phi\|^{2n}\mathbb{1}=\left( -R^2 \right)^n\mathbb{1}  \\ \phi^{2n+1}=\left( -\frac{1}{2} \right)^n \|\phi\|^{2n}\phi=\left( -R^2 \right)^n\phi\,.\end{array}  \right.  
\end{align}

\begin{figure}[ht!]
\centering
\includegraphics[width=6cm]{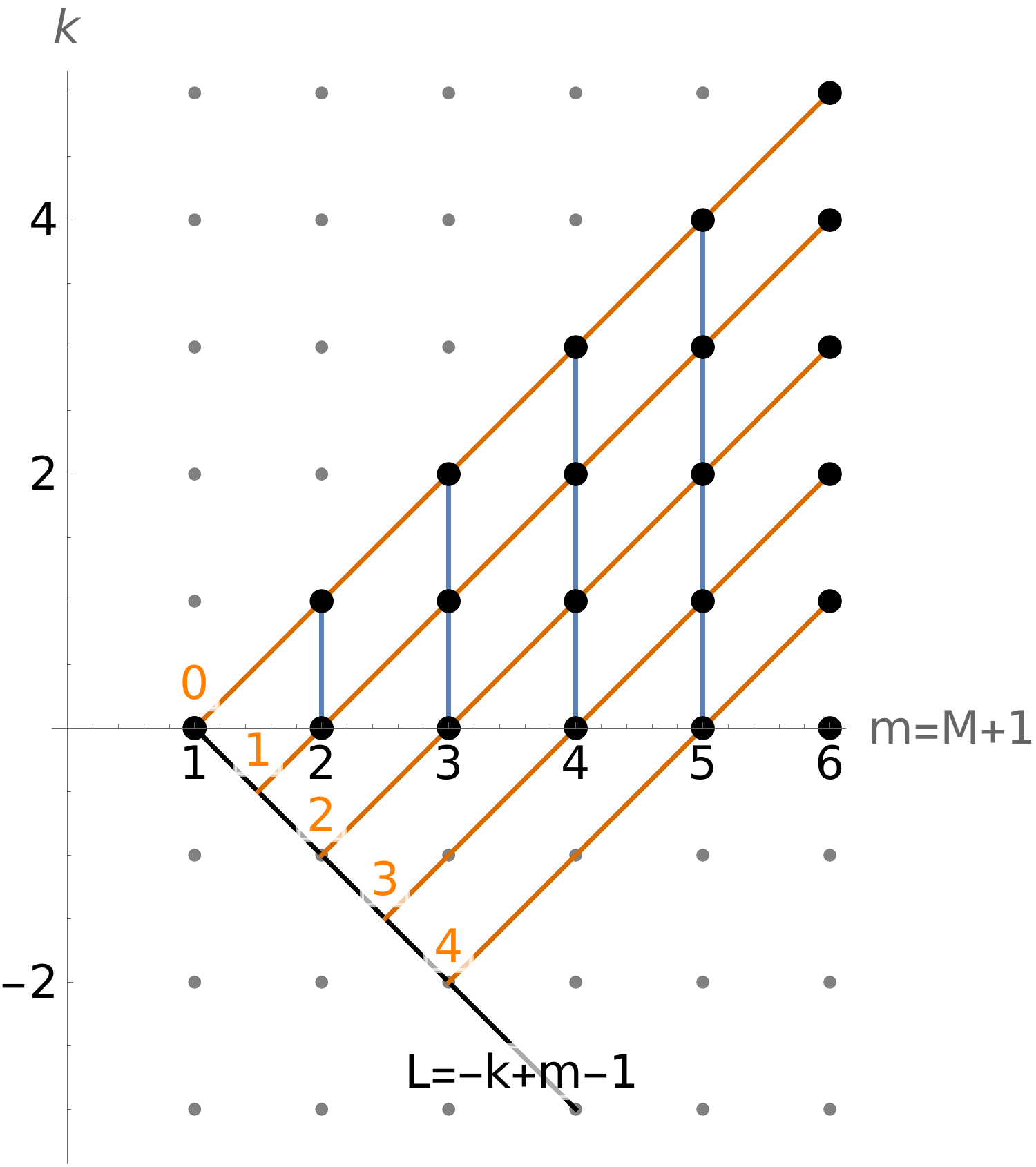}
\caption{The sum that defines the 2-form $B$ runs over the black marked points in the lattice (lower wedge of first quadrant). In the original definition, we first sum over $k$ (vertical blue lines) and then add up those. It is however convenient to change coordinates to $(L=-k+m-1,\;M=m-1)$, $L\geq0,\;M\geq L$, such that the summation runs first along the diagonals (orange lines with fixed $L=0,1,2,\ldots$) and in the final step we add up their contributions.}
\label{fig:B_sum}
\end{figure}
The double summation (see Fig.~\ref{fig:B_sum})  ranges over the lower wedge which divides the first quadrant of the
$\mathbb{Z}^2$ lattice in two equal parts (with $m$ on the $x$-axis and $k$ on
the $y$-axes). The sum is originally arranged in such a way that we first sum over points in the vertical
direction $k$ (blue lines) and then add up the results in the horizontal direction $m$. We wish to
change variables such that we first sum across diagonals (orange lines) and then add up those.
In other words, we define
\begin{align}
&L=-k+m-1,\;\;L\in\mathbb{Z}_{\ge0} &&M=m-1,\;\;M\in\mathbb{Z}_{\ge L} \,,
\end{align}
which transforms the expression for $B$ into
\begin{align}
  B=\lambda^2\mathrm{Tr}\bigg\{\mathrm{d}\phi\wedge \sum_{L\ge 0}(2L+2)(-R^2)^{-L}\bigg[\sum_{M\ge L}\frac{\lambda^{2M+1}}{2M+3}(-R^2)^M\bigg]\mathrm{d}\phi(-R^2)^L\phi\bigg\}.
\end{align}
The inner sum $[\ldots]$ evaluates to
\begin{align}
  [\ldots]=\frac{\lambda}{2L+3}\left( -\lambda^2R^2 \right)^L{}_2F_1\left(1,L+\frac{3}{2};L+\frac{5}{2}\bigg\vert-(\lambda R)^2 \right)
\end{align} 
and luckily the hypergeometric function is of special type and can be further
simplified. If we express it in a standard way as a sum of Pochhammer symbols,
there are telescopic cancellations among them, leaving us with
\begin{align}
  {}_2F_1\left(1,L+\frac{3}{2};L+\frac{5}{2}\bigg\vert-(\lambda R)^2 \right)=\sum_{n\ge 0}\frac{2L+3}{2L+2n+3}(-\lambda^2R^2)^n.
\end{align}
Collecting partial results, they fortunately combine into a very simple
expression that admits a closed form presentation
\begin{align}
B=\lambda^3\bigg\{\sum_{L\ge 0}\sum_{n\ge 0}\frac{2L+2}{2L+2n+3}\left( -\lambda^2R^2 \right)^{L+n}\bigg\}\mathrm{Tr}\left( \mathrm{d}\phi\wedge \mathrm{d}\phi\phi \right)  \,.
\end{align}
Now we express the trace in spherical coordinates and recognize the double sum
$\{\ldots\}$ as a Taylor expansion of an elementary function. All together, one
obtains a remarkably simple formula for the 2-form $B$ 
\begin{align}\label{eq:B_explicit}
B&=\lambda^2f(R)e^2 \wedge e^3 \notag \\
  &=\frac{2\lambda^2}{\left( 1+\lambda^2R^2\right)^2}f(R)\big[R^2\sin{\theta}\mathrm{d}\theta\wedge \mathrm{d}\varphi\big]=\frac{2\lambda^2}{\left( 1+\lambda^2R^2\right)^2}f(R)\mathrm{\mathbf{vol}}_{S^2_R}
\end{align}
with $f(R)$ given as
\begin{align}
f(R)=\frac{1}{2\lambda R}\bigg\{-\left( 1-\lambda^2 R^2\right)+\left( 1+\lambda^2R^2 \right)^2\frac{\arctan{(\lambda R)}}{\lambda R}\bigg\}.
\end{align}
The expression for the 2-form $B$ justifies the choice of spherical coordinates for
the stereographic projection as was anticipated. We see that $B$ is supported
and constant on 2-spheres corresponding to a fixed radius $R$ in the
$\mathbb{R}^3$ plane of the stereographic projection.

Hodge decomposition for an arbitrary 2-form, in particular $B$ takes the form
\begin{align}
B=\mathrm{d}\gamma+\delta\omega+h,  
\end{align}
where the codifferential is expressed in terms of the Hodge star as $\delta=\star \mathrm{d}\star$.
Note, however, that second cohomology of $S^3$ is trivial and thus the harmonic
2-form $h$ is missing. We already commented that the shift by an exact form $\mathrm{d}\gamma$
is trivial as it can be absorbed by a redefinition of the gauge field $A$. Thus
we get (now already for our particular form $B$)
\begin{align}
B=\star \mathrm{d}\star F(R)\mathrm{\mathbf{vol}}_{S^3}=\star \mathrm{d}F(R)  
\end{align}
for a yet to be determined function $F(R)$. Taking the Hodge star
of~\eqref{eq:B_explicit} results in a differential equation for $F(R)$
\begin{align}
\frac{\mathrm{d}F(R)}{\mathrm{d}R}=\frac{\sqrt{2}\lambda^2}{1+\lambda^2R^2}f(R),  
\end{align}
which can be easily solved leading to (a trivial additive integration constant
has been dropped)
\begin{align}
&B=\star \mathrm{d}F(R),  &&F(R)=-\frac{\lambda}{2\sqrt{2}}(1-\lambda^2R^2)\frac{\arctan(\lambda R)}{\lambda R}.
\end{align}

It is now straightforward to compute both sides of~\eqref{eq:Cartan_dB} and fix
the normalization $\kappa$. It is derived for the more general $\mathrm{U}(N)$ or $\mathrm{SU}(N)$ case in~\eqref{eq:dB_norm} with the result $\kappa=1/12$. Here, we just give an expression for $\mathrm{d}B$, as it will be useful for further analysis 
\begin{align}
\mathrm{d}B&=2\sqrt{2}\lambda^3 e^1\wedge e^2 \wedge e^3=2\sqrt{2}\lambda^3\mathrm{\mathbf{vol}}_{S^3}. 
\end{align}
Since $\mathrm{d}B$ is essentially the volume form,
it is evident that $\mathcal{L}_{\mathbf{a}}\mathrm{d}B=\mathrm{d}\mathcal{L}_{\mathbf{a}}B=0$ as
the volume form and equally the metric are invariant under the full isometry
group. Thus $\mathcal{L}_{\mathbf{a}}B$ is a closed and hence exact 2-form, $\mathcal{L}_{\mathbf{a}}B=\mathrm{d}{\beta}$, and
we are all set to compute the 1-form ${\beta}$. We do not display the results
in full generality for all three Killing vectors
$\{\mathbf{a}^1,\mathbf{a}^2,\mathbf{a}^3\}$. Rather, for the sake of brevity,
we pick the simplest case of $\mathbf{a}^3$, and illustrate the shift in the
gauge field $A$ generated by the flow in this direction. So ${\beta}$
corresponding to $\mathbf{a}^3$ is
\begin{align}
{\beta}=\frac{\lambda}{2}\frac{1+\lambda^2R^2}{\lambda^2R^2}\Big[\lambda R-\left( 1-\lambda^2 R^2\right)\arctan{(\lambda R)}\Big]\sin{\theta}e^3.  
\end{align}
Using this expression, we conclude by writing down the shift symmetry in the
abelian gauge field $A$ defined in~\eqref{eq:shift_symm} under the flow
generated by the Killing vector $\mathbf{a}^3$
\begin{align}
A_\mu\xmapsto{\mathcal{L}_{\mathbf{a}^3}} A_\mu-\frac{c}{\sqrt{2}}\left(1-\frac{1-\lambda^2R^2}{\lambda R}\arctan{(\lambda R)}  \right)\sin^2{\theta}\partial_\mu\varphi.  
\end{align}
Similar, just more complicated, results hold also for the other two directions
$\mathbf{a}^1$ and $\mathbf{a}^2$. 

Provided this shift symmetry is postulated, the action is invariant under both
$\mathcal{L}_{\mathbf{v}}$ (trivially) and $\mathcal{L}_{\mathbf{a}}$, i.e.
under the full isometry group $\mathrm{U}(N)_{\mathrm{L}}\times
\mathrm{U}(N)_{\mathrm{R}}$. So we managed to prove what we set out to do, and
in due course computed all quantities explicitly as an illustration (at least in
the simplest case of $\mathrm{SU}(2)$).

We conclude this section by exploring one extra limit of the extended DBI Lagrangian that was briefly anticipated in Section~\ref{sec:limits}. We postponed its discussion until this point, since we have proven it just for the $\mathrm{SU}(2)$ extended DBI theory. In fact, it is different in nature compared to the web of limits presented in Section~\ref{sec:limits}. Those were ``weakly coupled limits'', valid without resummation of the $W_{\mu\nu}$ interactions. On the other hand, this limit demands a closed form expression for the 2-form $B$ (or equivalently its pull-back to spacetime $W_{\mu\nu}$), in order to have its $\lambda\to\infty$ behavior fully under control.
\subsubsection*{$\lambda\to\infty$: BI theory}
For the original theory~\cite{Cachazo:2014xea} corresponding to $c=(\Lambda\lambda)^{-2}$ it
can be verified that the metric $h$ and the 2-form $B$ on the coset
go smoothly to zero as the coset manifold shrinks to a point in the $\lambda\to\infty$ limit.
Indeed, looking at the explicit formulae~\eqref{eq:metric_expl}
and~\eqref{eq:B_explicit}, we readily compute
\begin{align}
\lim_{\lambda\to\infty}h=\lim_{\lambda\to\infty}cB=0.  
\end{align}
This fact then trivially implies that the Lagrangian reduces to a BI theory
\begin{align}
\mathcal{L}_{\mathrm{eDBI}}\xrightarrow{\lambda\to\infty} \mathcal{L}_{\mathrm{BI}}=\Lambda^4\Big[1-\sqrt{-\det{\left(\eta_{\mu\nu}-\Lambda^{-2}F_{\mu\nu}  \right)}}\Big]\,.
\end{align}

%%%%%%%%%%%%%%%%%%%%%%%%%%%%%%%%%%%%%%%%%%%%%%%%%%%%%%%%%%%%
\section{Soft theorems and reconstructibility}\label{sec:reconstr}

According to our classification discussed in Section \ref{section_multi_rho}, the theory with
Lagrangian (\ref{eq:eDBI}) is a multi-$\rho $ theory with $\rho _{\min }=$ $0$ and $\rho
_{\max }=1$. The subtheories which correspond to these boundary values of
$\rho $ are the nonlinear sigma model
\begin{equation}
\mathcal{L}_{0}=\frac{1}{8\lambda ^{2}}\langle \partial _{\mu }U^{\dagger}\partial
^{\mu }U\rangle
\end{equation}
and the multi DBI theory coupled to a $U(1) $ gauge field
\begin{equation}
\mathcal{L}_{1}=-\Lambda ^{4}\sqrt{-\det \left( \eta _{\mu \nu }-\frac{%
\partial _{\mu }\phi ^{a}\partial _{\nu }\phi ^{a}}{\Lambda ^{4}}-\frac{%
F_{\mu \nu }}{\Lambda ^{2}}\right) }+\Lambda ^{4}.
\label{multiDBILagrangian}
\end{equation}
In the previous sections we have demonstrated that the Lagrangian (\ref{eq:eDBI}) is
invariant with respect to the generalized shifts defined as (cf. (\ref{chiral_shift_symmetry}) and (\ref{axialL}), (\ref{eq:shift_symm}))
\begin{equation}\label{eq:shifts}
\delta _{A}\phi =\alpha -\lambda ^{2}\phi \alpha \phi ,~~~~\delta _{A}A_{\mu
}=-cb_{\mu }\,,
\end{equation}
where $\alpha $ is a generator of $U(N)$ and $\lambda$ an infinitesimal parameter. Here $b_{\mu }$ is at least
quadratic in $\phi $ and its derivatives and it is determined by the
condition
\begin{equation}
\delta _{A}W_{\mu \nu }=\partial _{\mu }b_{\nu }-\partial _{\nu }b_{\mu }.
\end{equation}
According to the general theorem \cite{Cheung:2016drk}, this symmetry of the action is
responsible for the Adler zero of the scattering amplitudes when one of the
scalar particles becomes soft
\begin{equation}
A\left( p^{\phi },1^{\phi },\ldots ,k^{\phi },\left( k+1\right) ^{h},\ldots
,n^{h}\right) \overset{p\rightarrow 0}{=}O(p) .
\label{all_rho_soft_theorem}
\end{equation}
Here we use condensed notation for the momenta $p_{i}\equiv i$ and the
superscript denotes either the type of the particle or its helicity.

The above symmetry is valid also for the lowest $\rho $ subtheory $\mathcal{L%
}_{0}$. On the other hand, the highest $\rho $ subtheory $\mathcal{L}_{1}$
obeys instead the following linearized shift symmetry 
\begin{equation}
\delta _{\alpha }\phi =\alpha 
\end{equation}
and also a higher polynomial shift symmetry which corresponds to the nonlinear realization of the higher dimensional Lorentz symmetry  extended to the $U(1)$ gauge field living on the brane \cite{Gliozzi:2011hj}
\begin{eqnarray}
\delta _{\omega }\phi ^{a} &=&\omega _{\beta b}\left( x^{\beta }\delta
^{ab}\Lambda ^{2}-\eta ^{\beta \mu }\phi ^{b}\partial _{\mu }\phi
^{a}\Lambda ^{-2}\right)   \label{DBI_symmetry}\notag \\
\delta _{\omega }A_{\mu } &=&-\Lambda ^{-2}\eta ^{\nu \beta }\omega _{\beta
b}\partial _{\mu }\phi ^{b}A_{\nu }-\Lambda ^{-2}\eta ^{\nu \beta }\omega
_{\beta b}\phi ^{b}\partial _{\nu }A_{\mu }.  \label{vector_DBI_symmetry}
\end{eqnarray}
The latter implies enhanced Adler zero for soft scalars, namely%
\begin{equation}
A^{\left( \rho _{\max }\right) }\left( p^{\phi },1^{\phi },\ldots ,k^{\phi
},\left( k+1\right) ^{h},\ldots ,n^{h}\right) \overset{p\rightarrow 0}{=}%
O\left( p^{2}\right) .  \label{max_rho_soft_theorem}
\end{equation}
Note that the tree amplitudes $A^{\left( \rho _{\max }\right) }$ without
scalars are determined by the Lagrangian of the Born-Infeld electrodynamics
\begin{equation}
\mathcal{L}_{BI}=-\Lambda ^{4}\sqrt{-\det \left( \eta _{\mu \nu }-\frac{%
F_{\mu \nu }}{\Lambda ^{2}}\right) }+\Lambda ^{4}  \label{DBI_branch}
\end{equation}
and therefore they vanish under the multichiral soft limit \cite{Cheung:2018oki} when all the
particles with the same helicity are simultaneously soft. This type of soft
limit is formulated within the spinor-helicity formalism. It is taken in
such a way that for helicity plus particles, the holomorphic spinors are
sent to zero (in the case of helicity minus particles we use instead the
antiholomorphic spinors). This choice ensures that no artificial soft
suppression stemming from the polarizations of the soft particles appears. We
get then
\begin{equation}
\lim_{|1\rangle ,\ldots |n\rangle \rightarrow 0}A^{\left( \rho _{\max
}\right) }\left( 1^{+},\ldots ,n^{+},\left( n+1\right) ^{-},\ldots
,2n^{-}\right) =0  \label{multichiral_soft_theorem}
\end{equation}
and similarly for the helicity minus multichiral soft limit.

The extended DBI theory with Lagrangian (\ref{eq:eDBI}) obeys therefore graded soft
theorems (\ref{all_rho_soft_theorem}) and (\ref{max_rho_soft_theorem}) and
for the pure vector highest $\rho $ amplitudes we have also the multichiral
soft limit (\ref{multichiral_soft_theorem}).

Let us write the scattering amplitudes with $n_{\phi }$ scalars and $%
n_{\gamma }$ vectors in the form (\ref{eq:graded_A})%
\begin{equation}
A_{n_{\phi }n_{\gamma }}=\sum_{0\leq \rho \leq 1}A_{n_{\phi }n_{\gamma
}}^{(\rho )},
\end{equation}%
and define the all-line shift (which is guaranteed to exist for $n_{\phi
}+n_{\gamma }\geq 6$) 
\begin{eqnarray}
\widehat{p}_{i}(z) &=&\left( 1-a_{i}z\right) p_{i}\notag\\
|\widehat{j}(z) \rangle &=&\left( 1-b_{j}z\right) |j\rangle \notag\\
|\widehat{k}(z) ] &=&\left( 1-b_{k}z\right) |k]
\label{eq:soft_shift}
\end{eqnarray}%
for scalars, helicity plus and minus vectors respectively. Then the various
components $A^{\left( \rho \right) }$ with $\rho $ fixed behave under such a
shift as%
\begin{equation}
\widehat{A}_{n_{\phi }n_{\gamma }}^{\left( \rho \right) }(z)
=O\left( z^{\rho \left( n_{\phi }+n_{\gamma }-2\right) +2-n_{\gamma
}}\right) .
\end{equation}
Note that the improvement with respect to the naive scaling $O\left( z^{\rho
\left( n_{\phi }+n_{\gamma }-2\right) +2}\right) $ (based on the
power-counting parameter $\rho $) is due to our choice of the shift for the
vector particles, since to each external vector, an undeformed pair of
spinors is attached according to the little group scaling. Therefore
provided $\rho <1$, the function (cf. (\ref{eq:f_n}))
\begin{equation}
f_{n_{\phi }n_{\gamma }}^{\left( \rho \right) }(z) =\frac{%
\widehat{A}_{n_{\phi }n_{\gamma }}^{\left( \rho \right) }(z) }{%
\prod\limits_{i=1}^{n_{\phi }}\left( 1-a_{i}z\right) }=O\left( z^{\left(
\rho -1\right) \left( n_{\phi }+n_{\gamma }-2\right) }\right)
\end{equation}
vanishes for $z\rightarrow \infty $ and has only the unitarity poles due to
the validity of the soft theorem (\ref{all_rho_soft_theorem}). Similarly,
for $\rho =1$ and $n_{\phi }\neq 0$, the function
\begin{equation*}
f_{n_{\phi }n_{\gamma }}^{(1) }(z) =\frac{\widehat{A}%
_{n_{\phi }n_{\gamma }}^{(1) }(z) }{%
\prod\limits_{i=1}^{n_{\phi }}\left( 1-a_{i}z\right) ^{2}}=O\left(
z^{-n_{\phi }}\right)
\end{equation*}
has the same properties as a consequence of the soft theorem (\ref%
{max_rho_soft_theorem}). Therefore, we can reconstruct all the amplitudes $%
A_{n_{\phi }n_{\gamma }}$ with $n_{\phi }\neq 0$ from their residues at the
unitarity poles, i.e. from the amplitudes $A_{m_{\phi }m_{\gamma }}$ with $%
m_{\phi }+m_{\gamma }<n_{\phi }+n_{\gamma }$, using the soft BCFW recursion
based on the graded soft scalar theorem
\begin{eqnarray}
A_{n_{\phi }n_{\gamma }}(p) -A_{n_{\phi }n_{\gamma }}^{(1) } &=&O(p)  \notag \\
A_{n_{\phi }n_{\gamma }}^{(1) } &=&O\left( p^{2}\right)
\end{eqnarray}
precisely as in Section \ref{sec:Graded_soft}. What remains are the amplitudes $A_{0,n_{\gamma
}}^{(1) }$. These coincide with the amplitudes of the Born-Infeld
electrodynamics and according to \cite{Cheung:2018oki} they can be recursively reconstructed
using the soft theorem (\ref{multichiral_soft_theorem}).
 Here we can use e.g.
the all but two shift: for all helicity plus particles\footnote{%
Note that the amplitudes of Born-Infeld electrodynamics conserve helicity,
i.e. only the amplitudes with the same number of helicity plus and helicity
minus particles are nonzero.} $i=1^{+},\ldots ,\left( n_{\gamma }/2\right)
^{+}$ we take
\begin{equation}
|\widehat{i}(z) \rangle =\left( 1-bz\right) |i\rangle ,
\end{equation}
and for two helicity minus particles $j^{-}$ and $k^{-}$ we compensate the
violation of the momentum conservation as 
\begin{eqnarray}
|\widehat{j}(z)] &=&|j]+\frac{bz}{\langle j|k\rangle }\sum_{i=1}^{\left(
n_{\gamma }/2\right) }|i]\langle i|k\rangle ,  \notag \\
|\widehat{k}(z)] &=&|k]-\frac{bz}{\langle j|k\rangle }\sum_{i=1}^{\left(
n_{\gamma }/2\right) }|i]\langle i|j\rangle .
\end{eqnarray}
Then the function
\begin{equation}
f_{0,n_{\gamma }}^{(1) }(z) =\frac{\widehat{A}%
_{0,n_{\gamma }}^{(1) }(z) }{\left( 1-bz\right) }%
=O\left( z^{-1}\right)
\end{equation}
vanishes for $z\rightarrow \infty $ and has only the unitarity poles and
thus the amplitude $A_{0,n_{\gamma }}^{(1) }$ can be
reconstructed recursively.

To summarize, the scattering amplitudes of the theory with the Lagrangian (\ref{eq:eDBI})
are fully reconstructible either using the graded soft theorem (in the case
when $n_{\phi }\neq 0$) or using the multichiral soft limit (for amplitudes
with $n_{\phi }=0$).
The corresponding seed amplitudes are the 4pt ones. They can be easily calculated using the Feynman rules derived from the Lagrangian (\ref{eq:eDBI}) (see Appendix \ref{app:seed4pt} for details). The explicit form for the only nonzero seed amplitude reads (momenta are implicit, they are labelled by numbers corresponding to external particles)\footnote{Here we suppose invariance with respect to parity, i.e. the amplitudes with opposite helicity configuration can be obtained from these by the exchange $[i,j]\leftrightarrow\langle j,i\rangle$.}
\begin{eqnarray}
A^{\left( 0\right) }\left( 1_{a}^{\phi },2_{b}^{\phi },3_{c}^{\phi
},4_{d}^{\phi }\right)  &=&\lambda ^{2}\sum_{\sigma\in S_4 }\langle T^{\sigma
\left( a\right) }T^{\sigma \left( b\right) }T^{\sigma \left( c\right)
}T^{\sigma \left( d\right) }\rangle \left( \sigma (1) \cdot
\sigma \left( 3\right) \right) \notag \\
A^{(1) }\left( 1_{a}^{\phi },2_{b}^{\phi },3_{c}^{\phi
},4_{d}^{\phi }\right)  &=&-\frac{1}{8\Lambda ^{4}}\sum_{\sigma\in S_4 }\langle
T^{\sigma \left( a\right) }T^{\sigma \left( b\right) }\rangle \langle
T^{\sigma \left( c\right) }T^{\sigma \left( d\right) }\rangle \left( \sigma
(1) \cdot \sigma \left( 2\right) \right) ^{2} \notag\\
&&+\frac{1}{4\Lambda ^{4}}\sum_{\sigma\in S_4 }\langle T^{\sigma \left( a\right)
}T^{\sigma \left( b\right) }\rangle \langle T^{\sigma \left( c\right)
}T^{\sigma \left( d\right) }\rangle \left( \sigma (1) \cdot
\sigma \left( 3\right) \right) ^{2}\notag \\
A^{\left( 1/2\right) }\left( 1^{+},2_{a}^{\phi },3_{b}^{\phi },4_{c}^{\phi
}\right)  &=&\frac{1}{3\sqrt{2}}c\lambda ^{3}\sum_{\sigma\in S_3 }\langle T^{\sigma
\left( a\right) }T^{\sigma \left( b\right) }T^{\sigma \left( c\right)
}\rangle \lbrack 1|\sigma \left( 2\right) \sigma \left( 3\right) |1]\notag \\
A^{(1) }\left( 1^{+},2^{-},3_{a}^{\phi },4_{b}^{\phi }\right) 
&=&\frac{1}{2\Lambda ^{4}}\langle T^{a}T^{b}\rangle \lbrack 1|3|2\rangle
\lbrack 1|4|2\rangle \notag \\
A^{(1) }\left( 1^{+},2^{+},3^{-},4^{-}\right)  &=&\frac{1}{%
8\Lambda ^{4}}\left[ 1,2\right] ^{2}\langle 3,4\rangle ^{2}\,.
\label{eq:seeds_eDBI}
\end{eqnarray}
%

%%%%%%%%%%%%%%%%%%%%%%%%%%%%%%%%%%%%%%%%%%%%%%%%%%%%%%%%%%%%
\section{Bottom-up reconstruction of the three-scalar and photon case}
\label{sec:bottomup}

The results of the previous section suggest that we have several
complementary possibilities how to define the extended DBI theory. The
original one is based on the explicit form of the CHY integrand ~\cite{Cachazo:2014xea} and this
definition was conjectured to be equivalent to the theory  described by the
Lagrangian (\ref{eq:eDBI}) with particular value of the coupling $c$. The second possible
definition is just to start with the Lagrangian (\ref{eq:eDBI}). The latter was found to
posses strong symmetries leading to a remarkable set of soft theorems which
determine uniquely all the tree-level amplitudes of the theory provided the
seed amplitudes are given. This therefore suggests the third possibility how
to fix the theory, namely postulating just the soft theorems and
power-counting as its defining property. This possibility naturally raises a
question whether such an amplitude-based definition leads uniquely to the
theory with Lagrangian (\ref{eq:eDBI}) or whether there is some space for possible
generalizations. 

Since the soft BCFW recursion described in the previous section is based
solely on the power-counting and the soft theorems, the only possibility for
generalization is connected with the freedom to choose more general seed
amplitudes which serve as the initial conditions for the recursion. As we
have seen, the five nonzero seeds (\ref{eq:seeds_eDBI}) are parametrized by three parameters,
i.e. there exist nontrivial constraints on the seed provided we wish to
reproduce the amplitudes generated by the Lagrangian (\ref{eq:eDBI}). On the other hand,
completely arbitrary choice of the seeds might be inconsistent, since the
soft BCFW recursion could lead to objects which cannot be identified with
amplitudes of any sensible theory. 

Note that the recursive construction depends on a set of free parameters 
$a_{i}$ and $b_{i}$ which parametrize the soft shift (cf. (\ref{eq:soft_shift})), while the
resulting amplitudes have to be independent of them. The consistency check
based on this $a_{i}$ and $b_{i}$ independence typically requires some
constraints on the seeds. We can also proceed differently and try to
construct the higher-point amplitudes using a generic ansatz with
appropriate pole structure and including a full set of generic contact
terms. Then we demand both the right factorization on the poles and the
appropriate soft behavior corresponding to the soft theorems. This procedure
can be treated as a recursion starting with the seed amplitudes. At each $n-$th step, we construct full basis of $n$-point contact terms with free coefficients which
parametrize the ansatz and the soft theorems are used to fix them. In
general we obtain a set of linear equations for these free coefficients and
the conditions of existence of a solution for this set can put constraints
on the seeds. Both these strategies were used for classification of theories based on soft theorems and are known as the {\emph{soft bootstrap}}  \cite{Cheung:2016drk,Cheung:2018oki,Elvang:2018dco,Kampf:2019mcd}. In
this section we apply this approach to the problem of possible generalizations
of the extended DBI.

Instead of discussing the most general $U(N)$ case, we will assume a theory with only three scalars $\phi^{\pm}$, $\phi^0$ in the spectrum. Here we are  anticipating the $SU(2)$ extended DBI, but we relax the full $SU(2)$ symmetry and suppose only global $U(1)$, with respect to which  the particles $\phi^{\pm}$ are charged. 

As a warm up, let us start with a {\it pure scalar\/} theory corresponding to the above particle content and let us define its amplitudes by the following requirements:
\begin{itemize}
 \item multi-$\rho$ power counting with $0\le \rho\le1$  
 \item all odd-point amplitudes vanish identically, $A_{2n+1}=0$
 \item global $U(1)$ symmetry, i.e. the corresponding charge conservation
       \item graded soft theorem of the form  \begin{eqnarray}
A_{2n}(p) -A_{2n}^{\left(
1\right) } &=&O(p)  \notag \\
A_{2n}^{(1) } &=&O\left( p^{2}\right)
\end{eqnarray}
i.e. we suppose Adler zero for any $\rho$
and enhanced soft limit for $\rho=1$.
\end{itemize}
Note that some of the assumptions might be superfluous, however, their inclusion from the beginning simplifies considerably the study. 
 
According to the above discussion, we will suppose the existence of a theory with amplitudes satisfying the above requirements and try to reconstruct it recursively. We expect that consistency of this procedure will put nontrivial constraints on the free parameters of the seed amplitudes. 
As there are amplitudes with only even number of legs, the seed amplitudes are the contact 4pt vertices. 
We can create a basis of all allowed monomials; the number of individual terms is summarized in Table~\ref{tab:4pt}. 
\begin{table}[tb]
\centering
 \begin{tabular}{|c | c | c|} \hline
 4pt & $\rho=0$ (2der) & $\rho=1$ (4der) \\
 \hline
 $0000$ & 0 & 1 \\ 
 ${+}{-}00$ & 1 & 2 \\
 ${+}{+}{-}{-}$ & 1 & 2 \\
 \hline
 \end{tabular}
 \caption{The number of independent monomials for pure scalar 4pt vertices with 2 derivatives ($\rho=0$) and 4 derivatives ($\rho=1$).}
 \label{tab:4pt}
\end{table}
There are all together seven independent constants $c_i^{(\rho)}$ and $\tilde{c}_i^{(\rho)}$ which parametrize the seed amplitudes, explicitly 
\begin{align}
&A(1^\phi_0,2^\phi_0,3^\phi_0,4^\phi_0) = c^{(1)}_1 (s_{12}^2+s_{13}^2+s_{14}^2)\notag\\
&A(1^\phi_+,2^\phi_-,3^\phi_0,4^\phi_0) = c^{(0)}_2 s_{12} + c^{(1)}_2 s_{13}s_{14} + {\tilde c}^{(1)}_2 s_{12}^2\notag\\
&A(1^\phi_+,2^\phi_+,3^\phi_-,4^\phi_-) = c_3^{(0)}(s_{13}+s_{14}) + c_3^{(1)}(s_{13} + s_{14})s_{12} + \tilde{c}_3^{(1)} (s_{13}^2+s_{14}^2)\,,
\end{align}
where $s_{ij}=(p_i+p_j)^2$ are the Mandelstam variables.
\begin{table}[tb]
\centering
 \begin{tabular}{|c | c | c| c|} 
 \hline
 6pt & $\rho=0$ (2der) & $\rho=1/2$ (4der) & $\rho=1$ (6der) \\
 \hline
 $000000$ & 0 & 1 & 2\\ 
 ${+}{-}0000$ & 1 & 4 & 9 \\
 ${+}{+}{-}{-}00$ & 2 & 8 & 19 \\
 ${+}{+}{+}{-}{-}{-}$ & 1 & 3 & 7 \\
 \hline
 \end{tabular}
 \caption{The number of independent monomials for pure scalar 6pt vertices.}
 \label{tab:6pt}
\end{table}

 One possible strategy of the soft bootstrap continues then with a general ansatz for the 6pt amplitudes. To get it, we glue the on-shell seed amplitudes together with corresponding propagators as if they were vertices corresponding to Feynman rules and then add a linear combination of a complete basis of the 6pt contact terms with free coefficients.   
 The number of independent terms for various values of $\rho$ is given in  Table \ref{tab:6pt}.
 Such a construction guarantees the right factorization properties of the resulting amplitude. The free parameters of the ansatz should be then fixed applying the graded soft theorem.
 As discussed above, this procedure might create constraints on the parameters $c_i^{(\rho)}$ and $\tilde{c}_i^{(\rho)}$ of the seed amplitudes as the necessary condition for the existence of the solution for the 6pt couplings.
  
 However, we can proceed more economically, without the necessity of classifying the 6pt contact terms.    
 Using the all-line shift for the 6pt amplitude  we can construct the meromorphic function $f_6(z)$ of the shift parameter $z$ according to the prescription (\ref{eq:f_n}) and (\ref{eq:soft_shift_general}). 
 Following the discussions in Sections~\ref{sec:Graded_soft} and~\ref{sec:reconstr} it is easy to verify that 
\begin{equation}
f_6(z) \sim 1/z^2\,, \qquad \text{for } z\to\infty \,.   
\end{equation}
Employing the residue theorem on  $f_6(z)/z$ we can try to reconstruct  the amplitude $A_6(0)$ using (\ref{recursion}) and check the consistency of such reconstruction investigating the independence on the parameters $a_i$ of the shift.
However, we can use the residue theorem directly on $f_6(z)$. Provided the 6pt amplitude exists and is consistent, the sum of the residues still vanishes:
\begin{equation}
    \sum_{\mathcal{F},I=\pm }\mathrm{res~}
    (f_{6},z_{\mathcal{F}}^{I})=0\,.
\end{equation}
These conditions, which are sometimes called {\emph{ bonus relations}},  are not automatically satisfied for the most general seeds and can give us relations among the inputs, i.e. the parameters $c_i^{(\rho)}$ and $\tilde{c}_i^{(\rho)}$ of the 4pt vertices. 

Going through all 6pt amplitudes  and using all the above strategies in order to check the result we got
\begin{equation}\label{eq:cond1}
  \tilde c_2^{(1)} = \tilde c_3^{(1)} = 0\,,
  \end{equation}
and 
\begin{equation}\label{eq:cond2}
  c_2^{(1)} = -2 c_1^{(1)}\,,\quad  
  c_3^{(0)} = c_2^{(0)}\,,\quad
  c_3^{(1)} = - 2 c_1^{(1)}\,,
\end{equation}
reducing originally seven 4pt constants down to two, one standing for $\rho=0$ and one for $\rho=1$ vertices.
Of course, the next recursion step can in principle bring further constraints on the inputs, but we can prove that this is not the case.
Note that we know one particular example of a theory which satisfies the above requirements. It is nothing else but the $SU(2)$ variant of the theory discussed in Section \ref{section: simple example} and described  by Lagrangian (\ref{multi_rho_example}). 
Its seed amplitudes correspond to the choice
\begin{equation}
    c_1^{(1)}=\frac{1}{4\Lambda^4},\,\,\,c_2^{(0)}=-4\lambda^2
\end{equation}
and the very existence of this theory with two independent constants 
proves that no new constraints can occur in the higher recursion steps.  
We can interpret this result also the other way round: the $SU(2)$ variant of the theory with Lagrangian (\ref{multi_rho_example}) is uniquely defined by the requirements of power counting, the graded soft theorem and global $U(1)$
symmetry as formulated above. Note that these requirements are strong enough to ensure even stronger $SU(2)$ global symmetry, which is in this case emergent.

\begin{table}[tb]
\centering
 \begin{tabular}{|c | c | c|} 
 \hline
 4pt & vertex degree & constant \\
 \hline
 ${+}{-}0\gamma^+$      & $\rho=1/2\;$ (3der)   & $d_\gamma^{(1/2)}$ \\ 
 $00\gamma^+\gamma^-$   & $\rho=1\;$ (4der)     & $d_{\gamma\gamma}^{(1)}$ \\
 ${+}{-}\gamma^+\gamma^-$ & $\rho=1\;$ (4der)   & $\tilde{d}_{\gamma\gamma}^{(1)}$\\
 $\gamma^+\gamma^+\gamma^-\gamma^-$ &   $\rho=1\;$ (4der)     &$d_{4\gamma}^{(1)}$\\
 $\gamma^+\gamma^+\gamma^+\gamma^+$ &   $\rho=1\;$ (4der)     &$\tilde{d}_{4\gamma}^{(2)}$\\
 \hline
 \end{tabular}
 \caption{List of possible 4pt vertices of a scalar triplet and a photon.}
 \label{tab:4ptg}
\end{table}

Now we will put also a {\it massless vector particle\/} (photon) into the game under the same assumptions on the amplitudes as before supplemented by the multichiral soft limit (\ref{multichiral_soft_theorem}) for the pure vector amplitudes. Note this additional condition is equivalent to demanding that the $\rho=1$ theory in the pure photon sector is Born-Infeld.
Allowing interaction of the massless vector with the scalar particles $\phi^\pm$ and $\phi^0$ and limiting the power counting by $0\leq\rho\leq1$, we have five additional 4pt seed amplitudes parametrized by the constants $d_A^{(\rho)}$, namely (cf. also  Tab.~\ref{tab:4ptg}).
\begin{align}
    &A(1^\phi_+,2^\phi_-,3^\phi_0,4^+) =  d_\gamma^{(1/2)} \langle12\rangle [14][24]\notag\\
    &A(1^\phi_0,2^\phi_0,3^+,4^-) = d_{\gamma\gamma}^{(1)}\langle14\rangle\langle24\rangle [13][23]\notag\\
    &A(1^\phi_+,2^\phi_-,3^+,4^-) = \tilde{d}_{\gamma\gamma}^{(1)}\langle14\rangle\langle24\rangle [13][23]\notag\\
    &A(1^+,2^+,3^-,4^-) = d_{4\gamma}^{(1)}\langle34\rangle^2 [12]^2\notag\\
    &A(1^+,2^+,3^+,4^+) = \tilde{d}_{4\gamma}^{(1)}( [12]^2[34]^2+ [13]^2[24]^2+ [14]^2[23]^2)
    \,.
\end{align}
Note that there is always only one possible monomial for the given combination of photon helicities and/or scalar flavours. It is not possible to construct e.g. the vertices $ A(1^\phi_0,2^\phi_0,3^\phi_0,4^\pm)$ and $ A(1^\pm,2^\pm,3^\pm,4^\mp)$. Though it is possible to write down a candidate for the $ A(1^\phi_0,2^\phi_0,3^+,4^+)$ or $A(1^+,2^-,3^+,4^+)$ vertex, such  terms violate the anticipated enhanced soft limit already by itself, i.e. already at the 4pt order. 
As in the previous pure scalar case, we can summarize also the types of independent 6pt vertices (see Appendix \ref{sec:basis6pt}).
However, this is again only for reference purposes because as before we can use soft BCFW recursion to reconstruct the six- and higher- point amplitudes without explicit knowledge of the corresponding basis.
Repeating the same reasonings as in the pure scalar case, we can use the bonus relations connected with 6pt amplitudes to obtain the constraints on the 4pt seeds.
The bonus relations will provide us with the following conditions on top of (\ref{eq:cond1}) and (\ref{eq:cond2})
\begin{equation}\label{eq:cond3}
    d_{\gamma\gamma}^{(1)} = \tilde{d}_{\gamma\gamma}^{(1)} =2 c_1^{(1)}\,.
\end{equation}
The above mentioned  multichiral soft photon limit applied directly to the seed amplitude $A(1^+,2^+,3^-,4^-)$ leads to an additional constraint
\begin{equation}\label{eq:cond4}
     \tilde{d}_{4\gamma}^{(1)} = 0\,.
\end{equation}
We end up therefore with four independent constants which parametrize the sought theory,  two in the scalar sector, $c^{(1)}_1$, $c^{(0)}_2$, one in the photon sector, namely $d_{4\gamma}^{(1)}$ and the mixed one $d_{\gamma}^{(1/2)}$.
As in the pure scalar case we can ask whether the higher recursion steps will give rise to some additional constraints. 
Let us note that a particular solution of our amplitude reconstruction problem, namely the $SU(2)$ variant of the extended DBI with Lagrangian (\ref{eq:eDBI}) corresponds to the choice (cf. (\ref{eq:seeds_eDBI}))
\begin{equation}
   c_1^{(1)} = \frac{1}{4 \Lambda^4}\,,\qquad 
    c_2^{(0)} = -4 \lambda^2\,,\qquad
    d_{\gamma}^{(1/2)} = c\lambda^3\,,\qquad
    d_{4\gamma}^{(1)}=\frac{1}{8\Lambda^4}   
\end{equation}
i.e. in this theory there is an additional constraint
\begin{equation}
    d_{4 \gamma}^{(1)} = 2 c_1^{(1)}\,.
    \label{eq:cond5}
\end{equation}
The original CHY~\cite{Cachazo:2014xea} variant of the same theory is even more restrictive, since it demands on top of it $c=(\Lambda\lambda)^{-2}$, i.e.
\begin{equation}
     (d_\gamma^{(1/2)})^2 =  | c_1^{(1)} c_2^{(0)}|\,,
\end{equation}
but this theory is a special case of the previous one.
Provided the $SU(2)$ variant of the extended DBI (\ref{eq:eDBI}) is the unique solution, the answer to our question should be positive, i.e. the higher recursion steps would demand the constraint (\ref{eq:cond5}) as a consistency condition of the amplitude bootstrap.
However, similarly to the pure scalar case, the final decision whether the above four parameters are really free  would be given provided we found a theory which fulfills all the above requirements on its amplitudes and which is parametrized just by four unconstrained parameters $c^{(1)}_1$, $c^{(0)}_2$, $d_{4\gamma}^{(1)}$ and  $d_\gamma^{(1/2)}$.
Search for such a theory is described in the next section.

%To summarize, in order to get a non-trivial reconstructible theory, we have to add the condition of multichiral soft limit for pure photon amplitudes on top of the conditions listed at the beginning of this section. This can be done setting the 4pt inputs to BI (we have then altogether 4 parameters: $c_1^{\rho=1}$, $c_2^{\rho=0}$, $c^\gamma$ and $c_1^{4\gamma}$) or to DBI (we have then only 3 parameters as $c_1^{4\gamma}$ is fixed). In order to obtain the theory with only two parameters, corresponding to the CHY construction~\cite{Cachazo:2014xea}, one has to fix also $c^\gamma$. However, we have found no special amplitude property justifying this last requirement.

\section{Generalization of extended DBI}\label{sec:generalization}

From the consistency of  on-shell recursion discussed in the previous section, we learned that there might be space for generalizing the  $SU(2)$ variant of extended DBI theory. Let us concentrate on its $\rho=1$ subtheory. Within the bottom-up approach, it can be obtained from the general case by sending all the input constants $c_i^{(\rho)}$ and $d_A^{(\rho)}$  for $\rho<1$ to zero and by fixing the remaining two as
\begin{align}
&c_1^{(1)} = \frac{1}{4 \Lambda^4},    &&d_{4\gamma}^{(1)}=\frac{1}{8\Lambda^4}. 
\end{align}
However, as was noted, the first recursive step does not require the 4pt coupling $d_{4\gamma}^{(1)}$  to be fixed to this value, which indicates that it can be kept free. Even in that case all the consistency conditions dictated by the first step of the recursive amplitude construction are still satisfied. This indicates that a 2-parametric $\rho=1$ theory satisfying all the requirements dictated by the soft theorems might exist. We will show that it indeed exists and its Lagrangian which will be given in~\eqref{eq:2DBI} corresponds to the following choice of 4pt seed couplings
\begin{align}
&c_1^{(1)} = \frac{1}{4 \Lambda^4},    &&d_{4\gamma}^{(1)}=\frac{1}{8M^4}
\end{align}
parametrized by two mass scales $\Lambda$ and $M$.

As discussed in Section~\ref{sec:reconstr}, the $\rho_{\mathrm{max}}=1$ theory can be regarded as independent from the point of view of the soft bootstrap. Reconstruction of its tree-level amplitudes relies on its own  soft theorems that are valid just for the special $\rho_{\mathrm{max}}=1$ subtheory of the given multi-$\rho$ theory (extended DBI in this case). Therefore, once the existence of the 2-scale $\rho=1$ theory is established, we can just replace the $\rho=1$ DBI subtheory of extended DBI by this newly constructed theory. Since soft reconstructibility of $\rho<1$ branches of the extended DBI theory is unaffected by modification of its $\rho_{\mathrm{max}}=1$ subtheory, this procedure is expected to output a consistent multi-$\rho$ theory -- a generalization of extended DBI parametrized by four 4pt seed couplings
\begin{equation}
   c_1^{(1)} = \frac{1}{4 \Lambda^4}\,,\qquad 
    c_2^{(0)} = -4 \lambda^2\,,\qquad
    d_\gamma^{(1/2)} = c\lambda^3\,,\qquad
    d_{4\gamma}^{(1)}=\frac{1}{8M^4}.   
\end{equation}
Lagrangian of this generalized extended DBI theory will be presented in~\eqref{eq:2eDBI}.

\subsection{2-scale single-scalar DBI theory}

In the previous section we have seen that amplitudes $A_{n_{\phi }n_{\gamma
}}^{(1) }$ with $n_{\phi }\geq 1$ of the single-$\rho =1$
multi-DBI theory (\ref{multiDBILagrangian}) can be reconstructed by the soft
bootstrap based on the enhanced $O\left( p^{2}\right) $ Adler zero. In the
same theory, the pure photon amplitudes $A_{0,n_{\gamma }}^{(1) }
$can be reconstructed using the multichiral soft limit. It is then a natural
question whether these two soft theorems define a unique theory or whether
there is a wider class of theories obeying the same soft behavior. Of
course, once the 4pt seed amplitudes are fixed, the soft BCFW recursion
determines the whole tree-level $S-$matrix. Apparently then, such a class of
theories could be uniquely parametrized by the set of all possible seed
amplitudes. However, as we have seen, the consistency of the recursion might in general
require some additional constraints on the input and the couplings of the
4pt seeds might be correlated. Of course, this is the case of the theory
with Lagrangian (\ref{multiDBILagrangian}), where all the 4pt couplings are
expressed in terms of just one dimensionful coupling $\Lambda $. But, as we
will demonstrate now, this is not the most general solution. 

Let us try to construct a $\rho =1$ theory of massless scalars and a $U(1) $ gauge field (photon in what follows), whose amplitudes $A_{n_{\phi
}n_{\gamma }}^{(1) }$ satisfy the following requirements

\begin{itemize}
\item for $n_{\phi }\geq 1$ the amplitudes $A_{n_{\phi }n_{\gamma }}^{\left(
1\right) }$obey the enhanced $O\left( p^{2}\right) $ Adler zero for soft
scalars 

\item the amplitudes $A_{0,n_{\gamma }}^{(1) }$ obey the
multichiral soft limit for soft vectors
\end{itemize}

Assume just a single scalar case for simplicity. It is natural to suppose that
the action we are looking for is invariant with respect to the polynomial
shift symmetry (\ref{DBI_symmetry}), i.e.
that the scalar is the DBI one, since such symmetry automatically guarantees
the required enhanced Adler zero. The most general Lagrangian, which couples
a photon to a single DBI scalar in a way invariant with respect to this symmetry
reads\footnote{Here the trace is taken over the Lorentz indices of the corresponding matrices.}
\begin{equation}
\mathcal{L}=-\Lambda ^{4}+\Lambda ^{4}\sqrt{g}+\sqrt{g}\frac{1}{4}{\rm{Tr}}
\left( g^{-1}F\right) ^{2} +\sqrt{g}\sum\limits_{n}\sum\limits_{%
\alpha ,~|\alpha |>2}c_{\alpha }\prod\limits_{i=1}^{n}{\rm{Tr}}\left( \frac{%
g^{-1}F}{M^{2}}\right) ^{\alpha _{i}} \,,
\label{eq:general_DBI}
\end{equation}
where $\alpha =\left\{ \alpha _{j}\right\} _{j=1}^{n}$ is a multi-index and $%
|\alpha |=\sum\limits_{j=1}^{n}\alpha _{j}$ . In the above Lagrangian, $%
c_{\alpha }$ are free dimensionless couplings while $\Lambda $ and $M$ are
two scales with mass dimension $\left[ \Lambda \right] =\left[ M\right] =1$.
The induced metric is defined as
\begin{equation}
g_{\mu \nu }=\eta _{\mu \nu }-\frac{\partial _{\mu }\phi \partial _{\nu
}\phi }{\Lambda ^{4}}\,,
\end{equation}
where $g^{\mu \nu }$ is its inverse, $g=|\det g_{\mu \nu }|=-\det g_{\mu \nu
}$ and we have denoted for simplicity $\left( g^{-1}F\right) _{~~\nu }^{\mu
}=g^{\mu \alpha }F_{\alpha \nu }$. In this theory, the tree-level pure photon amplitudes are
generated by the Lagrangian of the BI type
\begin{equation}
\mathcal{L}_{\gamma }=\frac{1}{4}{\rm{Tr}}\left( \eta ^{-1}F\right)
^{2} +\sum\limits_{n}\sum\limits_{\alpha ,~|\alpha |>2}c_{\alpha
}\prod\limits_{i=1}^{n}{\rm{Tr}} \left( \frac{\eta ^{-1}F}{M^{2}}\right)
^{\alpha _{i}} ,  \label{general_DBI_vector}
\end{equation}
which can be obtained form (\ref{eq:general_DBI}) setting $\phi\to 0$.
As it is known \cite{Cheung:2018oki}, the multichiral soft limit together with helicity
conservation applied to this pure photon theory fixes uniquely the constants 
$c_{\alpha }$ in such a way that $\mathcal{L}_{\gamma }$ appears to be
the BI Lagrangian
\begin{eqnarray}
\mathcal{L}_{\gamma } &=&\mathcal{L}_{\mathrm{BI}}\left( M\right) =M^{4}\sqrt{-\det
\left( \eta _{\mu \nu }-\frac{F_{\mu \nu }}{M^{2}}\right) }-M^{4}  \notag \\
&=&M^{4}\sqrt{\det \left( \delta _{\nu }^{\mu }-\frac{\eta ^{\mu \alpha
}F_{\alpha \nu }}{M^{2}}\right) }-M^{4}
\end{eqnarray}
with some scale $M$. With these $c_{\alpha }$ we can therefore sum up the
above Lagrangian (\ref{eq:general_DBI}) to the form
\begin{eqnarray}
\mathcal{L} &=&-\Lambda ^{4}+\Lambda ^{4}\sqrt{g}+\sqrt{g}M^{4}\sqrt{\det
\left( \delta _{\nu }^{\mu }-\frac{g^{\mu \alpha }F_{\alpha \nu }}{M^{2}}%
\right) }-\sqrt{g}M^{4}  \notag \\
&=&-\Lambda ^{4}+\left( \Lambda ^{4}-M^{4}\right) \sqrt{g}+M^{4}\sqrt{-\det
\left( g_{\mu \nu }-\frac{F_{\mu \nu }}{M^{2}}\right) }\,,
\end{eqnarray}
which can be finally rewritten as
\begin{eqnarray}\label{eq:2DBI}
\mathcal{L}_{\mathrm{2DBI}} &\mathcal{=}&-\Lambda ^{4}+\left( \Lambda ^{4}-M^{4}\right) 
\sqrt{-\det \left( \eta _{\mu \nu }-\frac{\partial_{\mu }\phi \partial
_{\nu }\phi }{\Lambda ^{4}}\right) }  \notag \\
&&+M^{4}\sqrt{-\det \left( \eta _{\mu \nu }-\frac{\partial _{\mu }\phi
\partial _{\nu }\phi }{\Lambda ^{4}}-\frac{F_{\mu \nu }}{M^{2}}\right) }\,.
\end{eqnarray}
The resulting Lagrangian is two-scale. 
The pure scalar amplitudes are
governed by the Lagrangian with the scale $\Lambda $ which sets the strength
of the pure scalar interaction%
\begin{equation}
\mathcal{L}_{\phi }=\mathcal{L}_{\mathrm{DBI}}\left( \Lambda \right) =-\Lambda
^{4}+\Lambda ^{4}\sqrt{-\det \left( \eta _{\mu \nu }-\frac{\partial _{\mu
}\phi \partial _{\nu }\phi }{\Lambda ^{4}}\right) },
\end{equation}%
while the pure photon amplitudes by the Lagrangian 
\begin{equation}
\mathcal{L}_{\gamma }=\mathcal{L}_{\mathrm{BI}}\left( M\right) =M^{4}\sqrt{-\det
\left( \eta _{\mu \nu }-\frac{F_{\mu \nu }}{M^{2}}\right) }-M^{4},
\end{equation}%
with the parameter $M$ which sets the scale of the nonlinearities in the
photon sector. 

Since the complete Lagrangian satisfies the same soft theorems as the one
scale $M=\Lambda $ classical DBI Lagrangian, the amplitudes of this
two-parametric theory are reconstructible in the same way as in the $%
M=\Lambda $ case. The seed amplitudes are the 4pt ones, but now depending on
two independent parameters $\Lambda $ and $M$. 

On the other hand, the most general helicity conserving 4pt amplitudes in
the single scalar case, which are compatible with the enhanced $O\left(
p^{2}\right) $ Adler zero, read (here $s_{ij}=\left( p_{i}+p_{j}\right) ^{2}$)%
\begin{eqnarray}
A_{4,0}\left( 1^\phi,2^\phi,3^\phi,4^\phi\right)  &=&c_{40}\left(
s_{12}^{2}+s_{13}^{2}+s_{23}^{2}\right)   \notag \\
A_{2,2}\left( 1^{+},2^{-},3^\phi,4^\phi\right)  &=&c_{22}[1|3|2\rangle \lbrack
1|4|2\rangle   \notag \\
A_{0,4}\left( 1^{+},2^{+},3^{-},4^{-}\right)  &=&c_{04}[1,2]^{2}\langle
3,4\rangle ^{2}.
\end{eqnarray}%
Note that they apparently depend on three free parameters. However, in the same way as in the previous  section we can prove that the consistency of the soft BCFW recursion in
fact reduces this freedom and only two free parameters $c_{40}$ and $c_{04}$ remain\footnote{Note also that  identifying $\phi$ with the neutral particle $\phi^0$ of the $SU(2)$ case discussed in the previous section, and taking into account the $U(1)$ symmetry within the latter theory, we can identify also the tree amplitudes with only neutral scalars and photons with those of the single scalar theory. The reason is that the recursive reconstruction of these amplitudes does not depend on the amplitudes with the external charged scalars as a result of charge conservation.}
. This means,
that the two-scale DBI Lagrangian is the most general solution in the single
scalar case.
Generalization to the $U(N)$ case is then straightforwardly obtained by replacing $\partial_{\mu}\phi\partial_{\nu}\phi\to \delta_{ab}\partial_{\mu}\phi^a\partial_{\nu}\phi^b$ in the Lagrangian (\ref{eq:2DBI}).

%%%%%%%%%%%%%%%%%%%%%%%%%%%%%%%%%%%%%%%%%%%%%%%%%%%%%%%%%%%%
\subsection{2-scale extended DBI theory}
%%%%%%%%%%%%%%%%%%%%%%%%%%%%%%%%%%%%%%%%%%%%%%%%%%%%%%%%%%%%
In previous sections we learned that:
\begin{enumerate}[(i)]
\item{the branch $0\le\rho<1$ of extended DBI theory has an on-shell reconstructible tree-level S-matrix from
    soft theorems for the scalars $\phi$~\eqref{all_rho_soft_theorem}, which follow from shift invariance~\eqref{eq:shifts}
    and a resulting full chiral symmetry $\mathrm{U}(N)_{\mathrm{L}}\times
    \mathrm{U}(N)_{\mathrm{R}}$ of the action}
\item{the most general $\rho=1$ theory (that we know of) with an on-shell
    reconstructible tree-level S-matrix is the two-scale DBI theory introduced
    in~\eqref{eq:2DBI}, whose constructibility is based on the $\mathcal{O}(p^2)$ enhanced Adler zero for soft scalars as well as the multichiral
    soft limit for amplitudes with photons exclusively }  
\end{enumerate}
These two branches of the S-matrix can be merged into a multi-$\rho$ theory with
$0\le\rho\le1$. It will be called \emph{2-scale extended DBI theory}. Its relation to
extended DBI is the same as that of 2-scale DBI to DBI. The fusion is simple, it
merely consists of replacing the flat metric $\delta$ on the scalar coset manifold with
the metric $h$ on the group manifold $\mathrm{U}(N)$ together with the replacement of
the photon field strength with the generalized field strength
$F_{\mu\nu}\rightarrow\mathcal{F}_{\mu\nu}$ in the action of the 2-scale DBI theory~\eqref{eq:2DBI}. This readily yields the Lagrangian of the 2-scale $(\Lambda,M)$ extended DBI theory
\begin{align}\label{eq:2eDBI}
\mathcal{L}_{2\mathrm{eDBI}}=&\Lambda^4-(\Lambda^4-M^4)\sqrt{-\det\Big[\eta_{\mu\nu}-\Lambda^{-4}h_{ab}\partial_{\mu}\phi^a\partial_{\nu}\phi^b\Big]} \notag \\ 
&-M^4 \sqrt{-\det\Big[\eta_{\mu\nu}-\Lambda^{-4}h_{ab}\partial_{\mu}\phi^a\partial_{\nu}\phi^b-M^{-2}\mathcal{F}_{\mu\nu}\Big]},
\end{align}
which interpolates between NLSM ($\rho=0$) and two-scale DBI ($\rho=1$).
By construction this theory clearly enjoys all the symmetries (namely the generalized shift symmetry (\ref{eq:shifts}) and  the polynomial shift symmetry (\ref{vector_DBI_symmetry}) for its $\rho=1$ branch) and soft theorems
that guarantee on-shell constructibility of its tree-level S-matrix. An explicit expansion of this Lagrangian and calculation of the corresponding 4pt vertices is done in Appendix~\ref{app:seed4pt}. Here we only quote the resulting 4pt seed amplitudes
\begin{eqnarray}
A\left( 1_{a}^{\phi },2_{b}^{\phi },3_{c}^{\phi },4_{d}^{\phi }\right) 
&=&\lambda ^{2}\sum_{\sigma\in S_4 }\langle T^{\sigma \left( a\right) }T^{\sigma
\left( b\right) }T^{\sigma \left( c\right) }T^{\sigma \left( d\right)
}\rangle \left( \sigma (1) \cdot \sigma \left( 3\right) \right) 
\notag \\
&&-\frac{1}{8\Lambda ^{4}}\sum_{\sigma\in S_4 }\langle T^{\sigma \left( a\right)
}T^{\sigma \left( b\right) }\rangle \langle T^{\sigma \left( c\right)
}T^{\sigma \left( d\right) }\rangle \left( \sigma (1) \cdot
\sigma \left( 2\right) \right) ^{2}  \notag \\
&&+\frac{1}{4\Lambda ^{4}}\sum_{\sigma\in S_4 }\langle T^{\sigma \left( a\right)
}T^{\sigma \left( b\right) }\rangle \langle T^{\sigma \left( c\right)
}T^{\sigma \left( d\right) }\rangle \left( \sigma (1) \cdot
\sigma \left( 3\right) \right) ^{2}\notag\\
A\left( 1^{+},2_{a}^{\phi },3_{b}^{\phi },4_{c}^{\phi }\right)  &=&\frac{1}{3%
\sqrt{2}}c\lambda ^{3}\sum_{\sigma \in S_3 }\langle T^{\sigma \left( a\right)
}T^{\sigma \left( b\right) }T^{\sigma \left( c\right) }\rangle \lbrack
1|\sigma \left( 2\right) \sigma \left( 3\right) |1]\notag \\
A\left( 1^{+},2^{+},3_{a}^{\phi },4_{b}^{\phi }\right)  &=&0 \notag\\
A\left( 1^{+},2^{-},3_{a}^{\phi },4_{b}^{\phi }\right)  &=&\frac{1}{2\Lambda
^{4}}\langle T^{a}T^{b}\rangle \lbrack 1|3|2\rangle \lbrack 1|4|2\rangle \notag \\
A\left( 1^{+},2^{+},3^{-},4^{-}\right)  &=&\frac{1}{8M^{4}}\left[ 1,2\right]
^{2}\langle 3,4\rangle ^{2}\,.
\label{eq:2eDBIseeds}
\end{eqnarray}
Comparing with the outcome of the bottom-up approach in Sec.~\ref{sec:bottomup} we can conclude that it is a unique Lagrangian (up to a reparametrization) fulfilling the mentioned symmetries and soft theorems.

%%%%%%%%%%%%%%%%%%%%%%%%%%%%%%%%%%%%%%%%%%%%%%%%%%%%%%%%%%%%
\subsection{Significant limits of 2-scale extended DBI theory}
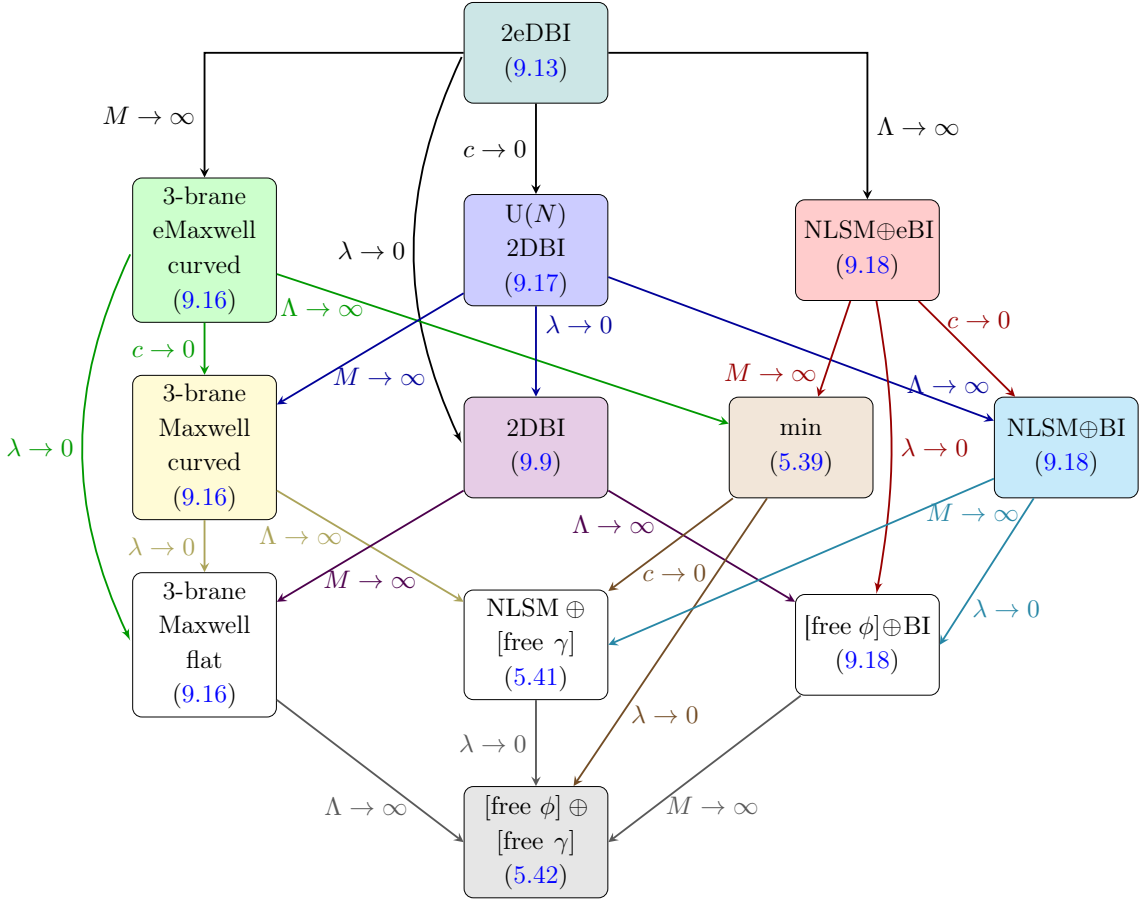
\begin{figure}[ht!]
\centering
\tikzstyle{block} = [rectangle, draw, text width=5em, text centered, rounded corners, minimum height=4em]
\tikzstyle{arrow} = [thick,->,>=stealth]
\resizebox{\columnwidth}{!}{
\begin{tikzpicture}[node distance = 3cm, auto]
    % Place nodes
    \node [block, fill=blue!20] (M1) {U$(N)$ 2DBI \\ \eqref{eq:U_N2DBI}};
    \node [block, fill=teal!20, above of=M1] (start) {2eDBI \\ \eqref{eq:2eDBI}};
    \node [block, fill=green!20, left of=M1, xshift=-2cm] (L1) {3-brane eMaxwell curved \\ \eqref{eq:3b_Maxwell}};
    \node [block, fill=red!20, right of=M1, xshift=2cm] (R1) {NLSM$\oplus$eBI \\ \eqref{eq:eBI}};
    \node [block, fill=violet!20, below of=M1] (M2) {2DBI \\ \eqref{eq:2DBI}};
    \node [block, fill=yellow!20, left of=M2, xshift=-2cm] (L2) {3-brane Maxwell curved \\ \eqref{eq:3b_Maxwell}};
    \node [block, fill=brown!20, right of=M2, xshift=1cm] (R2) {min \\ \eqref{eq:min}};
    \node [block, fill=cyan!20, right of=R2, xshift=1cm] (RR2) {NLSM$\oplus$BI \\ \eqref{eq:eBI}};
    \node [block,  below of=M2] (M3) {$\mathrm{NLSM}\oplus[\mathrm{free}\;\gamma]$ \\ \eqref{eq:NLSM_gamma}};
    \node [block, left of=M3, xshift=-2cm] (L3) {3-brane Maxwell flat \\ \eqref{eq:3b_Maxwell}};
    \node [block, right of=M3, xshift=2cm] (R3) {[$\mathrm{free}\;\phi]\oplus\mathrm{BI}$ \\ \eqref{eq:eBI}};
    \node [block, fill=gray!20, below of=M3] (M4) {$[\mathrm{free}\;\phi]\oplus[\mathrm{free}\;\gamma]$ \\ \eqref{eq:free_sg}};
    % Draw arrows
    \draw [arrow] (start) -| node [near end, anchor=east] {$M\to\infty$} (L1);
    \draw [arrow] (start) -- node [anchor=east] {$c\to 0$} (M1);
    \draw [arrow] (start) -| node [near end, anchor=west] {$\Lambda\to\infty$} (R1);
    \draw[->, >=latex', shorten >=2pt, shorten <=2pt, bend right=25, thick] 
    (start.west) to node[auto, swap] {$\lambda\to 0$}(M2.west);
    \draw[->, >=latex', shorten >=2pt, shorten <=2pt, bend right=25, thick, green!60!black] 
    (L1.west) to node[auto, swap] {$\lambda\to 0$}(L3.west);
    \draw [arrow,green!60!black] (L1) -- node [anchor=east] {$c\to 0$} (L2);
    \draw [arrow,green!60!black] (L1) -- node [anchor=north,pos=0.1] {$\Lambda\to \infty$} (R2);
    \draw [arrow,blue!60!black] (M1) -- node [anchor=west,pos=0.75] {$M\to \infty$} (L2);
    \draw [arrow,blue!60!black] (M1) -- node [anchor=west,pos=0.2] {$\lambda\to 0$} (M2);
    \draw [arrow,blue!60!black] (M1) -- node [anchor=west,pos=0.75] {$\Lambda\to \infty$} (RR2);
    \draw [arrow,red!60!black] (R1) -- node [anchor=east,pos=0.75] {$M\to \infty$} (R2);
    \draw [arrow,red!60!black] (R1) -- node [anchor=west,pos=0.2] {$c\to 0$} (RR2);
     \draw[->, >=latex', bend left=10, thick, red!60!black] 
    (R1) to node[anchor=west, pos=0.5] {$\lambda\to 0$}(R3);
    \draw [arrow,yellow!60!black] (L2) -- node [anchor=east,pos=0.5] {$\lambda\to 0$} (L3);
    \draw [arrow,yellow!60!black] (L2) -- node [anchor=east,pos=0.4] {$\Lambda\to \infty$} (M3);
    \draw [arrow,violet!60!black] (M2) -- node [anchor=west,pos=0.8] {$M\to \infty$} (L3);
    \draw [arrow,violet!60!black] (M2) -- node [anchor=east,pos=0.3] {$\Lambda\to \infty$} (R3);
    \draw [arrow,brown!60!black] (R2) -- node [anchor=west,pos=0.8] {$c\to 0$} (M3);
    \draw [arrow,brown!60!black] (R2) -- node [anchor=west,pos=0.75] {$\lambda\to 0$} (M4);
    \draw [arrow,cyan!60!black] (RR2) -- node [anchor=west,pos=0.2] {$M\to \infty$} (M3.east);
    \draw [arrow,cyan!60!black] (RR2) -- node [anchor=west,pos=0.75] {$\lambda\to 0$} (R3.east);
    \draw [arrow,gray!70!black] (L3) -- node [anchor=east,pos=0.75] {$\Lambda\to \infty$} (M4.west);
    \draw [arrow,gray!70!black] (M3) -- node [anchor=east,pos=0.5] {$\lambda\to 0$} (M4);
    \draw [arrow,gray!70!black] (R3) -- node [anchor=west,pos=0.75] {$M\to \infty$} (M4.east);
\end{tikzpicture}}
\caption{Web of limits for the 2-scale extended DBI theory.}
\label{fig:lim_2eDBI}
\end{figure}

%%%%%%%%%%%%%%%%%%%%%%%%%%%%%%%%%%%%%%%%%%%%%%%%%%%%%%%%%%%%
In Fig.~\ref{fig:lim_2eDBI}, we present a web of theories (with on-shell constructible tree-level S-matrices by soft bootstrap) emerging from particular limits of the single 2-scale extended DBI theory. The Lagrangians as well as the seed amplitudes for these theories can be obtained taking the corresponding limits of (\ref{eq:2eDBI}) and (\ref{eq:2eDBIseeds}) respectively.

The list of various types of possible limits was introduced already in Section \ref{sec:limits}.
There is, however, one more decoupling limit on top of $\Lambda\to\infty$, namely $M\to\infty$, which has an analogous effect as the former and which has not been mentioned yet.
The limit $M\to\Lambda$ is not included in Fig.~\ref{fig:lim_2eDBI}, since it reduces the 2-scale extended DBI theory to the extended DBI theory, whose limits were already analyzed in Sec.~\ref{sec:limits} and visualized in Fig.~\ref{fig:lim_eDBI},\ref{fig:3}. Using the same notation as in Section \ref{sec:limits} we can rewrite the Lagrangian (\ref{eq:2eDBI}) in the form
\begin{align}
\mathcal{L}_{\mathrm{eDBI}}=\Lambda^4-
(\Lambda^4-M^4)\sqrt {H}-M^4\sqrt{-\det(H_{\mu\nu}-M^{-2}\mathcal{F}_{\mu\nu})}.    
\end{align}
As in Section \ref{sec:limits}, $H_{\mu\nu}$ and $\Delta_{\mu\nu}$ is the induced metric in curved and flat ambient space respectively and $H^{\mu\nu}$ and $\Delta^{\mu\nu}$ denote the matrix inverse to  $H_{\mu\nu}$ and $\Delta_{\mu\nu}$. We also define $H=-\det(H_{\mu\nu})$ and similarly $\Delta=-\det(\Delta_{\mu\nu})$.

Let us list all the non-trivial Lagrangians appearing in Fig.~\ref{fig:lim_2eDBI}. Our naming conventions are partially inspired by the brane interpretation introduced in Section \ref{sec:limits}, however it was not easy to make them completely coherent. For instance the letter ``e'' in front of a theory name means extended (i.e. the generalized field strength $\mathcal{F}_{\mu\nu}$ appears in the Lagrangian instead of the ordinary one), while $\mathrm{U}(N)$ in front of a theory name means that the ambient space is curved (i.e. the $\mathrm{U}(N)$ metric $h_{ab}$ enters the action instead of the flat $\delta_{ab}$). Let us first explore theories in the left column of Fig.~\ref{fig:lim_2eDBI}, which have not appeared so far
\begin{align}\label{eq:3b_Maxwell}
&\mathcal{L}_{\mathrm{2eDBI}}\xrightarrow{M\to\infty}\underbrace{\Lambda^4(1-\sqrt{H})-\frac{1}{4}\sqrt{H}H^{\mu\alpha}H^{\nu\beta}\mathcal{F}_{\mu\nu}\mathcal{F}_{\alpha\beta}}_{\text{3-brane extended Maxwell in curved ambient space}} \notag \\
&\xrightarrow{c\to 0}\underbrace{\Lambda^4(1-\sqrt{H})-\frac{1}{4}\sqrt{H}H^{\mu\alpha}H^{\nu\beta}F_{\mu\nu}F_{\alpha\beta}}_{\text{3-brane Maxwell in curved ambient space}}\xrightarrow{\lambda\to 0}\underbrace{\Lambda^4(1-\sqrt{\Delta})-\frac{1}{4}\sqrt{\Delta}\Delta^{\mu\alpha}\Delta^{\nu\beta}F_{\mu\nu}F_{\alpha\beta}}_{\text{3-brane Maxwell in flat ambient space}}.
\end{align}
We named these theories after the second hallmark term (supplemented by a cosmological term on the 3-brane), which is a Maxwell kinetic term. One just uses the induced metrics $H_{\mu\nu}$, $\Delta_{\mu\nu}$ on the 3-brane instead of the Minkowski metric (and in the first Lagrangian replaces the field strength by a generalized one). All theories in this sequence have $0\leq~\rho\leq~1$, except for the last one, which is a $\rho=1$ theory.

Next, we proceed with the middle column of Fig.~\ref{fig:lim_2eDBI}
\begin{align}\label{eq:U_N2DBI}
&\mathcal{L}_{\mathrm{2eDBI}}\xrightarrow{c\to 0}\mathcal{L}_{\mathrm{U}(N)\mathrm{2DBI}}=\Lambda^4-(\Lambda^4-M^4)\sqrt{H}-M^4\sqrt{-\det(H_{\mu\nu}-M^{-2}F_{\mu\nu})} \notag \\
&\xrightarrow{\lambda\to 0}\mathcal{L}_{\mathrm{2DBI}}=\Lambda^4-(\Lambda^4-M^4)\sqrt{\Delta}-M^4\sqrt{-\det(\Delta_{\mu\nu}-M^{-2}F_{\mu\nu})}.
\end{align}
The first two theories in this chain have $0\leq\rho\leq1$, while the last 2-scale extended DBI theory has $\rho=1$ by construction.

Finally, we list theories in the rightmost column of Fig.~\ref{fig:lim_2eDBI}
\begin{align}\label{eq:eBI}
 &\mathcal{L}_{\mathrm{2eDBI}}\xrightarrow{\Lambda\to \infty}\underbrace{\frac{1}{2}\eta^{\mu\nu}g_{\mu\nu}+M^4\big(1-\sqrt{-\det{\eta_{\mu\nu}-M^{-2}\mathcal{F}_{\mu\nu}}}\big)}_{\text{NLSM}\;\oplus\;\text{extended BI}} \notag \\
 &\xrightarrow{\lambda\to 0}\underbrace{\frac{1}{2}\eta^{\mu\nu}\delta_{ab}\partial_\mu\phi^a\partial_\nu\phi^b+M^4\big(1-\sqrt{-\det{\eta_{\mu\nu}-M^{-2}F_{\mu\nu}}}\big)}_{[\mathrm{free}\;\phi]\;\oplus\;\text{BI}}.
\end{align}
The first NLSM $\oplus$ BI theory has $0\leq\rho\leq1$, while the second one [free $\phi$] $\oplus$ BI is a $\rho=1$ theory.

The minimal theory $\mathcal{L}_{\mathrm{min}}=\tfrac{1}{2}\eta^{\mu\nu}g_{\mu\nu}-\tfrac{1}{4}\mathcal{F}_{\mu\nu}\mathcal{F}^{\mu\nu}$ as well as the remaining two theories $\mathrm{NLSM}\oplus[\mathrm{free}\;\gamma]$, $[\mathrm{free}\;\phi]\oplus[\mathrm{free}\;\gamma]$ already appeared while analyzing the limits of extended DBI in Sec.~\ref{sec:limits}.

There is one special series of limits of the 2eDBI theory which has not been depicted in Fig. \ref{fig:lim_2eDBI}, namely $c=(\lambda\mu)^{-3},\,\,\,\lambda\to 0$.
This limit can replace all the $\lambda\to 0$ arrows in Fig. \ref{fig:lim_2eDBI} and leads to nontrivial new theories provided it is not preceded by a $c\to 0$ limit. In this way we obtain the reduced 2eDBI theory with $1/2\le\rho\le 1$
\begin{equation}
{\cal L}_{\rm {2eDBI}}\to  {\cal L}_{\rm {r2eDBI}}  =\Lambda^4-(\Lambda^4-M^4)\sqrt{\Delta}-M^4 \sqrt{-\det \left(\Delta_{\mu\nu}-M^{-2}{\cal F}^*_{\mu\nu}\right )},
\end{equation}
where 
\begin{equation}
{\cal F}^*_{\mu\nu}=F_{\mu\nu}+\frac{2}{3}\mu^{-3}\langle \phi\partial_{[\mu}\phi\partial_{\nu ]}\phi\rangle,
\end{equation}
and its descendants which correspond either to the $M\to\infty$ limit, which can be identified with
the reduced 3-brane extended Maxwell theory ($1/2\le\rho\le 1$)
\begin{equation}
{\cal{L}}_{\rm{rU(N)DBI}}=
    \Lambda^4 (1-\sqrt{\Delta})-\frac{1}{4}\sqrt{\Delta}\Delta^{\mu\alpha}\Delta^{\nu\beta}{\cal{F}}^*_
    {\mu\nu}{\cal{F}}
    ^*_{\alpha\beta},
\end{equation}
or to the $\Lambda\to\infty$ limit, which gives reduced NLSM$\oplus$eBI  theory ($1/2\le\rho\le 1$)
\begin{equation}
  {\cal{L}}_{\rm{rNLSM\oplus eBI}}= \frac{1}{2}\eta^{\mu\nu}\delta_{ab}\partial_{\mu}\phi^{a}\partial_{\nu}\phi^{b}-M^4\sqrt{-\det\left(\eta_{\mu\nu}-M^{-2}\cal{F}^*_{\mu\nu}\right)} .
\end{equation}
Finally, taking both $M,\,\Lambda\to\infty$ we get the reduced minimal theory ($\rho=1/2$)
\begin{equation}
 {\cal{L}}_{\rm{rmin}}= \frac{1}{2}\eta^{\mu\nu}\delta_{ab}\partial_{\mu}\phi^{a}\partial_{\nu}\phi^{b}-\frac{1}{4}{\cal{F}}^*_{\mu\nu}{\cal{F}}^{*\mu\nu}.
\end{equation}
All these reduced theories are invariant with respect to the reduced shift symmetry
\begin{align}
  \delta \phi =\alpha,&&\delta A_{\mu }=-\frac{2}{3}\mu^{-3}\langle \alpha
  \phi \partial _{\mu }\phi \rangle \,,
\end{align}
which is the $c\to(\lambda\mu)^{-3}$, $\lambda\to 0$ limit of the symmetry (\ref{eq:shifts}).

%%%%%%%%%%%%%%%%%%%%%%%%%%%%%%%%%%%%%%%%%%%%%%%%%%%%%%%%%%%%
\section{Conclusions and outlook}\label{sec:conclusions}
%%%%%%%%%%%%%%%%%%%%%%%%%%%%%%%%%%%%%%%%%%%%%%%%%%%%%%%%%%%%
The purpose of this work was to explore \emph{soft bootstrap} methods for multi-$\rho$ effective field theories (i.e. theories whose Lagrangian consists of operators with mixed power counting). The condition deciding whether a given multi-$\rho$ EFT has an on-shell constructible tree-level S-matrix was formulated as the \emph{graded soft theorem} in~\eqref{graded_soft_theorem}. The takeaway message of this analysis is that multi-$\rho$ EFTs have two distinguished branches:
\begin{itemize}
    \item a branch consisting of terms with the minimal number of derivatives per field in the Lagrangian ($\rho_{\mathrm{min}}$ branch)
    \item a branch consisting of terms with the maximal number of derivatives per field in the Lagrangian ($\rho_{\mathrm{max}}$ branch)
\end{itemize}
Each of these two branches represents by itself an independent subtheory which can have its own soft behavior $\mathcal{O}(p^\sigma)$, with $\sigma=\sigma_{\mathrm{min}}$ or $\sigma=\sigma_{\mathrm{max}}$, respectively. We can assume the following hierarchy 
\begin{equation}
\rho_{\min}\le\sigma_{\min},\,\,\,\,\rho_{\max}\le\sigma_{\max}\,,
\end{equation}
which guarantees that both subtheories are reconstructible. 
Provided the amplitudes of the full theory behave as $O(p^{\sigma_{\min}})$, it follows that a given multi-$\rho$ EFT has an on-shell constructible tree-level S-matrix by soft BCFW recursion if there is no gap between $\sigma_{\mathrm{min}}$ and $\rho_{\mathrm{max}}$, i.e. $\sigma_{\mathrm{min}}\geq\rho_{\mathrm{max}}$ (cf.~\eqref{hierarchy_reconstructible}). The full theory can be then reconstructed by soft BCFW recursion based on the graded soft theorem \eqref{graded_soft_theorem}.

The goal was to apply this formalism to a particular theory of this class -- the \emph{extended DBI theory} -- proposed by the authors of~\cite{Cachazo:2014xea}. We discussed in detail the symmetries of this theory and their geometrical origin and managed to show that its tree-level $S$-matrix is indeed reconstructible from soft theorems. Along the route we introduced other interesting reconstructible models corresponding to particular limits of the original theory. The soft bootstrap method allowed us to construct generalizations of some theories belonging to the web of theories associated to extended DBI. In particular, we found a 2-parametric generalization of the DBI theory, presented as the \emph{2-scale DBI theory} in~\eqref{eq:2DBI} as well as the \emph{2-scale extended DBI theory} (see~\eqref{eq:2eDBI}) that generalizes the extended DBI theory.
Various limits of this theory provided us with additional interesting examples of on-shell reconstructible theories.

Let us mention some open questions connected with other approaches to construction of the (tree-level) S-matrix. We wish to highlight the following two:
\begin{itemize}
  \item the Cachazo--He--Yuan (CHY) formalism introduced
    in~\cite{Cachazo:2013hca,Cachazo:2013iea} and further developed in other
    works including~\cite{Cachazo:2014xea}
\item{the color-kinematics duality (BCJ duality) and the associated double-copy
    structure~\cite{Bern:2008qj}}
\end{itemize}

Since the tree-level S-matrix of the original extended DBI theory (i.e. with the extra coupling $c$ fixed to the special value $c=(\Lambda\lambda)^{-2}$) has a CHY formulation, it is
logical to ask whether also the generalized 2-scale extended DBI theory
in~\eqref{eq:2eDBI} might have such a formulation. The existence of a CHY representation for a given theory is closely connected with its double-copy structure. The latter was worked out for the (original) extended DBI theory from its CHY representation in~\cite{Low:2020ubn}. Thus a natural question arises, whether also the 2-scale extended DBI theory admits a double-copy structure. Its existence seems plausible, as many theories derived from the 2-scale extended DBI theory (e.g. (extended) DBI, NLSM) already have a known double-copy structure.

%%%%%%%%%%%%%%%%%%%%%%%%%%%%%%%%%%%%%%%%%%%%%%%%%%%%%%%%%%%%
\acknowledgments

We thank Congkao Wen for helpful discussions.
This work was supported by the Czech Government project GA \v{C}R 21-26574S. Additional supports by the Ministry of Education grant LTAUSA17069 and Charles University Research Center UNCE/SCI/013.

%%%%%%%%%%%%%%%%%%%%%%%%%%%%%%%%%%%%%%%%%%%%%%%%%%%%%%%%%%%%
%%%%%%%%%%%%%%%%%%%%%%%%%%%%%%%%%%%%%%%%%%%%%%%%%%%%%%%%%%%%
\appendix
%%%%%%%%%%%%%%%%%%%%%%%%%%%%%%%%%%%%%%%%%%%%%%%%%%%%%%%%%%%%
\section{Explicit derivation of $\mathrm{U}(N)_{\mathrm{L}}\times\mathrm{U}(N)_{\mathrm{R}}$ invariance of the action}\label{sec:appA}
%%%%%%%%%%%%%%%%%%%%%%%%%%%%%%%%%%%%%%%%%%%%%%%%%%%%%%%%%%%%

Let us assume the Lagrangian 
\begin{equation}\label{eq:Lag_gen}
  \mathcal{L}=\mathcal{L}\left[ g_{\mu \nu },\mathcal{F}_{\mu \nu }\right]\,,
\end{equation}
where
\begin{eqnarray}
  g_{\mu \nu } &=&-\frac{1}{4\lambda^2}\left\langle \left( U^{\dagger}\partial _{\mu }U\right) \left(
               U^{\dagger}\partial _{\nu }U\right) \right\rangle ,~~~\mathcal{F}_{\mu \nu
               }=F_{\mu \nu }+cW_{\mu \nu }.
\end{eqnarray}
The symbol $\langle \cdots \rangle$ is used for trace over $\mathrm{U}(N)$ in this appendix.
The group element $U\in\mathrm{U}(N)$ is given in~\eqref{eq:coset_rep},
$F_{\mu\nu}$ is the abelian field strength and finally $W_{\mu\nu}$ is defined in~\eqref{eq:W}.

The purpose of this appendix is to prove by explicit computation that the Lagran\-gian~\eqref{eq:Lag_gen} is invariant with respect to the full chiral
symmetry $\mathrm{U}(N)_{\mathrm{L}}\times
\mathrm{U}(N)_{\mathrm{R}}\simeq\mathrm{U}(N)_{\mathrm{V}}\times
\mathrm{U}(N)_{\mathrm{A}}$. As usual, the vector subgroup
$\mathrm{U}(N)_{\mathrm{V}}$ will be realized linearly, while the axial one
non-linearly.

We begin by computing the exterior derivative of the 2-form $B$ on
$\mathrm{U}(N)$ (the pull-back of $B$ to space-time is $W$)
\begin{equation}
  B=\sum_{m=1}^{\infty }\sum_{k=0}^{m-1}\lambda ^{2m+1}\frac{2\left(
      m-k\right) }{2m+1}\left\langle T^{a}\phi ^{2k}T^{b}\phi ^{2\left( m-k\right)
      -1}\right\rangle \mathrm{d}\phi ^{a}\wedge \mathrm{d}\phi ^{b},
\end{equation}%
which yields
\begin{eqnarray}
  \mathrm{d}B &=&2\mathrm{d}\phi ^{a}\wedge \mathrm{d}\phi ^{b}\wedge \mathrm{d%
                  }\phi ^{c}\sum_{m=1}^{\infty }\frac{\lambda ^{2m+1}}{2m+1}\left[
                  \sum_{l=0}^{2m-2}m\left\langle T^{a}T^{b}\phi ^{l}T^{c}\phi
                  ^{2m-l-2}\right\rangle \right. \notag \\
              &&\left. -\sum_{k=1}^{m-1}\sum_{l=0}^{2k-1}\left( m-k\right) \left\langle
                 T^{a}\phi ^{l}T^{b}\phi ^{2k-l-1}T^{c}\phi ^{2\left( m-k\right)
                 -1}\right\rangle \right. \notag \\
              &&\left. +\sum_{k=1}^{m-1}\sum_{l=0}^{2(m-k)-2}\left( m-k\right)
                 \left\langle T^{a}\phi ^{2k}T^{b}\phi ^{l}T^{c}\phi ^{2\left( m-k\right)
                 -l-2}\right\rangle \right].
\end{eqnarray}%
Using total anti-symmetry of $\mathrm{d}\phi ^{a}\wedge \mathrm{d}\phi
^{b}\wedge \mathrm{d}\phi ^{c}$ this can be further simplified as a sum over cyclic permutations $\sigma$ of the indices $(a,b,c)$
\begin{eqnarray}
  \mathrm{d}B &=&\frac{1}{3}\mathrm{d}\phi ^{a}\wedge \mathrm{d}\phi
                  ^{b}\wedge \mathrm{d}\phi ^{c}\sum_{m=1}^{\infty }\frac{\lambda
                  ^{2m+1}}{2m+1}\sum_{\sigma \in %
                  % TCIMACRO{\U{2124} }%
                  % BeginExpansion
                  \mathbb{Z}
                  % EndExpansion
                  _{3}} \notag\\
              &&\times \left[ \sum_{l=0}^{2m-2}m\left\langle T^{\sigma \left( a\right)
                 }T^{\sigma \left( b\right) }\phi ^{l}T^{\sigma \left( c\right) }\phi
                 ^{2m-l-2}\right\rangle \right. \notag\\
              &&\left. -\sum_{k=1}^{m-1}\sum_{l=0}^{2k-1}\left( m-k\right) \left\langle
                 T^{\sigma \left( a\right) }\phi ^{l}T^{\sigma \left( b\right) }\phi
                 ^{2k-l-1}T^{\sigma \left( c\right) }\phi ^{2\left( m-k\right)
                 -1}\right\rangle \right. \notag\\
              &&\left. +\sum_{k=1}^{m-1}\sum_{l=0}^{2(m-k)-2}\left( m-k\right)
                 \left\langle T^{\sigma \left( a\right) }\phi ^{2k}T^{\sigma \left( b\right)
                 }\phi ^{l}T^{\sigma \left( c\right) }\phi ^{2\left( m-k\right)
                 -l-2}\right\rangle -\left( a\longleftrightarrow b\right) \right].\;
\end{eqnarray}
Note that for a cyclic permutation $\sigma$ and matrices $A_{i},\,i=1,2,3$, the following identity holds%
\begin{equation}
  \langle T^{\sigma \left( a\right) }A_{1}T^{\sigma \left( b\right)
    }A_{2}T^{\sigma \left( c\right) }A_{3}\rangle =\langle
    T^{a}A_{\sigma ^{-1}(1) }T^{c}A_{\sigma ^{-1}\left( 2\right)
    }T^{c}A_{\sigma ^{-1}\left( 3\right) }\rangle.
    \label{eq:cyclic}
\end{equation}%
Since $\phi $ is anti-hermitian, we can find a unitary matrix $V$ such that
$
  \phi =V\varphi V^{\dagger}
$, %
where $\varphi $ is diagonal with eigenvalues $f_k$
\begin{equation}
  \varphi =\mathrm{diag}\left( f_{1},f_{2},\ldots ,f_{N}\right).
\end{equation}%
Let us denote%
$
  \mathcal{T}^{a}=V^{\dagger}T^{a}V
$, 
then e.g.
\begin{eqnarray}
  \left\langle T^{a}\phi ^{l}T^{b}\phi ^{2k-l-1}T^{c}\phi ^{2\left( m-k\right)
  -1}\right\rangle &=&\left\langle \mathcal{T}^{a}\varphi ^{l}\mathcal{T}%
                       ^{b}\varphi ^{2k-l-1}\mathcal{T}^{c}\varphi ^{2\left( m-k\right)
                       -1}\right\rangle \notag\\
                   &=&\sum_{i,j,k=1}^{N}\mathcal{T}_{ij}^{a}f_{j}^{l}\mathcal{T}%
                       _{jk}^{b}f_{k}^{2k-l-1}\mathcal{T}_{ki}^{c}f_{i}^{2\left( m-k\right) -1}
\end{eqnarray}
and thus, according to (\ref{eq:cyclic}) 
\begin{equation}
  \left\langle T^{\sigma \left( a\right) }\phi ^{l}T^{\sigma \left( b\right)
    }\phi ^{2k-l-1}T^{\sigma \left( c\right) }\phi ^{2\left( m-k\right)
      -1}\right\rangle =\sum_{i,j,k=1}^{N}\mathcal{T}_{ij}^{a}f_{\sigma
    ^{-1}(j) }^{l}\mathcal{T}_{jk}^{b}f_{\sigma ^{-1}(k)
  }^{2k-l-1}\mathcal{T}_{ki}^{c}f_{\sigma ^{-1}(i) }^{2\left(
      m-k\right) -1}.
\end{equation}
This allows us to sum up over $l$ and express the exterior derivative $dB$ as
\begin{eqnarray}
  \mathrm{d}B &=&\frac{1}{3}\mathrm{d}\phi ^{a}\wedge \mathrm{d}\phi
                  ^{b}\wedge \mathrm{d}\phi ^{c}\sum_{i,j,k=1}^{N}\left( \mathcal{T}_{ij}^{a}%
                  \mathcal{T}_{jk}^{b}\mathcal{T}_{ki}^{c}-\mathcal{T}_{ij}^{b}\mathcal{T}%
                  _{jk}^{a}\mathcal{T}_{ki}^{c}\right) \notag\\
              &&\times \sum_{\sigma \in 
                \mathbb{Z}      _{3}}\sum_{m=1}^{\infty }\frac{\lambda ^{2m+1}}{2m+1}\Biggl[ m\frac{%
                 f_{\sigma (i) }^{2m-1}-f_{\sigma (k) }^{2m-1}}{%
                 f_{\sigma (i) }-f_{\sigma (k) }} \\
              && \phantom{\times}-\sum_{k=1}^{m-1}( m-k) \frac{f_{\sigma (k)
                 }^{2k}-f_{\sigma (j) }^{2k}}{f_{\sigma (k)
                 }-f_{\sigma (j) }}f_{\sigma (i) }^{2 (m-k) -1} +\sum_{k=1}^{m-1}(m-k) f_{\sigma
                 (j) }^{2k}\frac{f_{\sigma (i) }^{2 (m-k)
                 -1}-f_{\sigma (k) }^{2 (m-k) -1}}{f_{\sigma (i) }-f_{\sigma (k) }}\Biggr].\notag
\end{eqnarray}
The sum can be calculated explicitly with the result
\begin{eqnarray}
  \mathrm{d}B &=&\frac{\lambda ^{3}}{3}\mathrm{d}\phi ^{a}\wedge \mathrm{d}%
                  \phi ^{b}\wedge \mathrm{d}\phi ^{c}\sum_{i,j,k=1}^{N}\left( \mathcal{T}%
                  _{ij}^{a}\mathcal{T}_{jk}^{b}\mathcal{T}_{ki}^{c}-\mathcal{T}_{ij}^{b}%
                  \mathcal{T}_{jk}^{a}\mathcal{T}_{ki}^{c}\right) \frac{1}{1-\lambda
                  ^{2}f_{j}^{2}}\frac{1}{1-\lambda ^{2}f_{k}^{2}}\frac{1}{1-\lambda
                  ^{2}f_{i}^{2}} \notag\\
              &=&\frac{\lambda ^{3}}{3}\mathrm{d}\phi ^{a}\wedge \mathrm{d}\phi ^{b}\wedge 
                  \mathrm{d}\phi ^{c}\left[ \left\langle T^{a}\frac{1}{1-\lambda \phi ^{2}}%
                  T^{b}\frac{1}{1-\lambda \phi ^{2}}T^{c}\frac{1}{1-\lambda \phi ^{2}}%
                  \right\rangle \right. \notag\\
              &&\left. -\left\langle T^{b}\frac{1}{1-\lambda \phi ^{2}}T^{a}\frac{1}{%
                 1-\lambda \phi ^{2}}T^{c}\frac{1}{1-\lambda \phi ^{2}}\right\rangle \right]
  \notag\\
              &=&\frac{2}{3}\lambda ^{3}\mathrm{d}\phi ^{a}\wedge \mathrm{d}\phi
                  ^{b}\wedge \mathrm{d}\phi ^{c}\left\langle T^{a}\frac{1}{1-\lambda \phi ^{2}}%
                  T^{b}\frac{1}{1-\lambda \phi ^{2}}T^{c}\frac{1}{1-\lambda \phi ^{2}}%
                  \right\rangle.
\end{eqnarray}
Computing the left invariant Maurer--Cartan form
\begin{equation}
  U^{\dagger}\mathrm{d}U=2\frac{1}{1+\lambda \phi }\lambda \mathrm{d}\phi \frac{1}{%
    1-\lambda \phi }
\end{equation}%
and plugging it into the expression for $\mathrm{d}B$ finally gives the final
elegant formula
\begin{equation}\label{eq:dB_norm}
  \mathrm{d}B=\frac{1}{12}\left\langle U^{\dagger}\mathrm{d}U\wedge U^{\dagger}\mathrm{d}%
    U\wedge U^{\dagger}\mathrm{d}U\right\rangle.
\end{equation}%
The right hand side is nothing else than the Cartan
3-form~\eqref{eq:Cartan_3form} and thus the normalization introduced
in~\eqref{eq:Cartan_dB} is fixed as $\kappa=1/12$.

From the formula above also easily follows that the 3-form $\mathrm{d}B$ is manifestly invariant with respect to
the chiral $\left( V_{R},V_{L}\right) \in \mathrm{U}(N) _{\mathrm{R}}\times
\mathrm{U}(N) _{\mathrm{L}}$ transformation 
\begin{equation}
  U\rightarrow V_{L}^{\dagger}UV_{R}.
\end{equation}%
For an infinitesimal transformation of $\phi $ with respect to a vector
transformations $V_{L}=V_{R}=\exp \left( \lambda \alpha \right) $, we get
\begin{equation}
  \delta _{V}\phi =-\lambda \left[ \alpha ,\phi \right],
\end{equation}
while for an axial transformation $V_{L}=V_{R}^{\dagger}=\exp \left( \lambda \alpha
\right) $ one obtains
\begin{equation}
  U[\phi +\delta _{A}\phi ]-U\left[ \phi \right] =\lambda \left\{ \alpha ,U%
    \left[ \phi \right] \right\} =\lambda \alpha \frac{1+\lambda \phi }{%
    1-\lambda \phi }+\lambda \frac{1+\lambda \phi }{1-\lambda \phi }\alpha.
    \label{eq:A16}
\end{equation}
On the other hand, one can express the variation as
\begin{equation}
  U[\phi +\delta _{A}\phi ]-U\left[ \phi \right] =\delta _{A}U\left[ \phi %
  \right] =\delta _{A}\frac{1+\lambda \phi }{1-\lambda \phi }=2\frac{1}{%
    1-\lambda \phi }\lambda \delta _{A}\phi \frac{1}{1-\lambda \phi }
    \label{eq:A17}
\end{equation}
and  therefore
\begin{equation}
  2\delta _{A}\phi =( 1-\lambda \phi ) \alpha ( 1+\lambda
                       \phi ) +( 1+\lambda \phi ) \alpha ( 1-\lambda \phi) 
                   =2 ( \alpha -\lambda ^{2}\phi \alpha \phi) .
                   \label{eq:A18}
\end{equation}
Now we have%
\begin{equation}
  \mathrm{d}\delta _{A}B=\delta _{A}\mathrm{d}B=0
\end{equation}%
and thus $\delta _{A}B$ is closed. When pulled back to Minkowski space ($W=X^*B$),
it must be exact due to the Poincare lemma, i.e. 
\begin{equation}
  \delta _{A}W=\mathrm{d}b
\end{equation}%
for some 1-form $b$, which we are going to compute now. To do so let us express the
variation of $W$ on the left hand side in terms of a Lie derivative 
\begin{equation}
  \delta _{A}W=\mathcal{L}_{\delta _{A}\phi }W,
\end{equation}%
where $\mathcal{L}_{\delta _{A}\phi }$ is the Lie derivative with respect to
the vector $\delta _{A}\phi $. Due to the Cartan formula%
\begin{equation}
  \mathcal{L}_{\delta _{A}\phi }=\mathrm{d}\circ i_{\delta _{A}\phi
  }+i_{\delta _{A}\phi }\circ \mathrm{d},
\end{equation}%
where $i_{\delta _{A}\phi }$ is the inner product we get 
\begin{equation}
  \mathcal{L}_{\delta _{A}\phi }W=\mathrm{d}\circ i_{\delta _{A}\phi
  }W+i_{\delta _{A}\phi }\circ \mathrm{d}W.
\end{equation}%
The first term is already exact, let us calculate the second one. As a first
step it is useful to compute
\begin{eqnarray}
  i_{\delta _{A}\phi }\mathrm{d}U &=&2\lambda \frac{1}{1-\lambda \phi }%
                                      i_{\delta _{A}\phi }\mathrm{d}\phi \frac{1}{1-\lambda \phi } = 2\lambda \frac{1}{1-\lambda \phi }\left( \alpha -\lambda ^{2}\phi \alpha
                                      \phi \right) \frac{1}{1-\lambda \phi } \notag\\
                                  &=&\lambda \frac{1}{1-\lambda \phi }\left[ \left( 1-\lambda \phi \right)
                                      \alpha \left( 1+\lambda \phi \right) +\left( 1+\lambda \phi \right) \alpha
                                      \left( 1-\lambda \phi \right) \right] \frac{1}{1-\lambda \phi } \notag\\
                                  &=&\lambda \alpha \frac{1+\lambda \phi }{1-\lambda \phi }+\lambda \frac{%
                                      1+\lambda \phi }{1-\lambda \phi }\alpha =\lambda \left\{ \alpha ,U\right\}
\end{eqnarray}%
and thus
\begin{equation}
  i_{\delta _{A}\phi }U^{\dagger}\mathrm{d}U=\lambda \alpha +\lambda U^{\dagger}\alpha U.
\end{equation}%
Using $\mathrm{d}\left( U^{\dagger}U\right) =0$, the second not manifestly
exact term in the Cartan formula for the Lie derivative can be expressed as
\begin{eqnarray}
  i_{\delta _{A}\phi }\mathrm{d}W &=&\frac{1}{12}i_{\delta _{A}\phi}\left\langle U^{\dagger}\mathrm{d}%
                                      U\wedge U^{\dagger}\mathrm{d}U\wedge U^{\dagger}\mathrm{d}U\right\rangle \notag\\
                                  &=&\frac{1}{12}\left\langle \left( i_{\delta _{A}\phi }U^{\dagger}\mathrm{d}%
                                      U\right) U^{\dagger}\mathrm{d}U\wedge U^{\dagger}\mathrm{d}U\right\rangle \notag\\
                                  &&-\left\langle U^{\dagger}\mathrm{d}U\left( i_{\delta _{A}\phi }U^{\dagger}\mathrm{d}%
                                     U\right) \wedge U^{\dagger}\mathrm{d}U\right\rangle +\left\langle U^{\dagger}\mathrm{d}%
                                     U\wedge U^{\dagger}\mathrm{d}U\left( i_{\delta _{A}\phi }U^{\dagger}\mathrm{d}U\right)
                                     \right\rangle \notag\\
                                  &=&\frac{1}{4}\left\langle \left( i_{\delta _{A}\phi }U^{\dagger}\mathrm{d}%
                                      U\right) U^{\dagger}\mathrm{d}U\wedge U^{\dagger}\mathrm{d}U\right\rangle \notag\\
                                  &=&\frac{1}{4}\lambda \left\langle \alpha U^{\dagger}\mathrm{d}U\wedge U^{\dagger}%
                                      \mathrm{d}U\right\rangle +\frac{1}{4}\lambda \left\langle U^{\dagger}\alpha UU^{\dagger}%
                                      \mathrm{d}U\wedge U^{\dagger}\mathrm{d}U\right\rangle \notag\\
                                  &=&-\frac{1}{4}\lambda \left\langle \alpha \mathrm{d}U^{\dagger}\wedge \mathrm{d}%
                                      U\right\rangle -\frac{1}{4}\lambda \left\langle \alpha \mathrm{d}U\wedge 
                                      \mathrm{d}U^{\dagger}\right\rangle \notag\\
                                  &=&-\frac{1}{4}\lambda \mathrm{d}\left\langle \alpha U^{\dagger}\mathrm{d}%
                                      U\right\rangle +\frac{1}{4}\lambda \mathrm{d}\left\langle \alpha \mathrm{d}%
                                      UU^{\dagger}\right\rangle =-\frac{1}{4}\lambda \mathrm{d}\left\langle \alpha \left[ U^{\dagger},\mathrm{d}U%
                                      \right] \right\rangle.
\end{eqnarray}%
Finally, this leads to an equation for the variation of $W$
\begin{equation}
  \delta _{A}W=\mathrm{d}b
\end{equation}%
with $b$ given by
\begin{equation}
  b=i_{\delta _{A}\phi }W-\frac{1}{4}\lambda \langle \alpha [ U^{\dagger},%
      \mathrm{d}U] \rangle.
\end{equation}%
The first term on the right hand side does not have a particularly nice form
\begin{eqnarray}
  i_{\delta _{A}\phi }W &=&\sum_{m=1}^{\infty }\sum_{k=0}^{m-1}\lambda
                    ^{2m+1}\frac{2\left( m-k\right) }{2m+1}\left( \left\langle \delta _{A}\phi
                    \phi ^{2k}\mathrm{d}\phi \phi ^{2\left( m-k\right) -1}\right\rangle
                    -\left\langle \mathrm{d}\phi \phi ^{2k}\delta _{A}\phi \phi ^{2\left(
                    m-k\right) -1}\right\rangle \right) \notag\\
                &=&\sum_{m=1}^{\infty }\sum_{k=0}^{m-1}\lambda ^{2m+1}\frac{2\left(
                    m-k\right) }{2m+1}\left( \left\langle \alpha \phi ^{2k}\mathrm{d}\phi \phi
                    ^{2\left( m-k\right) -1}\right\rangle -\left\langle \alpha \phi ^{2\left(
                    m-k\right) -1}\mathrm{d}\phi \phi ^{2k}\right\rangle \right) \notag\\
                &&-\sum_{m=1}^{\infty }\sum_{k=0}^{m-1}\lambda ^{2m+3}\frac{2\left(
                   m-k\right) }{2m+1}\left( \left\langle \alpha \phi ^{2k+1}\mathrm{d}\phi \phi
                   ^{2\left( m-k\right) }\right\rangle -\left\langle \alpha \phi ^{2\left(
                   m-k\right) }\mathrm{d}\phi \phi ^{2k+1}\right\rangle \right). \notag
\end{eqnarray}%
However, the main point is that if we define the transformation properties of
the gauge field $A_{\mu }$ as
\begin{align}
  {\delta _{V}A =0 },&&
  {\delta _{A}A =-cb},
\end{align}
we get
\begin{align}
  \delta _{V}\mathcal{F} =0,&& 
  \delta _{A}\mathcal{F} =\delta _{A}\mathrm{d}A+c\delta _{A}W=-c                       \mathrm{d}b+c\mathrm{d}b=0
\end{align}
and any Lagrangian of the form
\begin{equation}
  \mathcal{L}=\mathcal{L}\left[ g_{\mu \nu },\mathcal{F}_{\mu \nu }\right]
\end{equation}
is invariant with respect to the chiral symmetry $\mathrm{U}(N) _{\mathrm{R}}\times \mathrm{U}\left(
  N\right) _{\mathrm{L}}$ as we wanted to prove in this appendix. The minimal one can be written as
\begin{equation}
  \mathcal{L}_{\min }=\frac{1}{2}\eta ^{\mu \nu }g_{\mu \nu }-%
  \frac{1}{4}\mathcal{F}_{\mu \nu }\mathcal{F}^{\mu \nu },
\end{equation}
which is a multi-$\rho $ theory with $0\leq \rho \leq 1/2$.

Finally, let us comment that the reduced extended DBI Lagrangian $\mathcal{L}_{\mathrm{reDBI}}$, as well as
the reduced minimal Lagrangian $\mathcal{L}_{\mathrm{rmin}}$ (see
section~\ref{sec:limits} for definitions and notation) are invariant
with respect to the reduced shift symmetry%
\begin{align}
  \delta \phi =\alpha,&&\delta A_{\mu }=-\frac{2}{3}\mu^{-3}\langle \alpha
  \phi \partial _{\mu }\phi \rangle .
\end{align}

\section{Seed amplitudes for two-scale extended DBI}
\label{app:seed4pt}

The Lagrangian of the for two-scale extended DBI reads%
\begin{eqnarray}
\mathcal{L}_{\mathrm{2eDBI}} &=&\Lambda ^{4}-\left( \Lambda ^{4}-M^{4}\right) \sqrt{%
-\det \left( \eta -\frac{1}{4\Lambda ^{4}\lambda ^{2}}\langle \partial
U^{\dagger}\partial U\rangle \right) }  \notag \\
&&-M^{4}\sqrt{-\det \left( \eta -\frac{1}{4\Lambda ^{4}\lambda ^{2}}\langle
\partial U^{\dagger}\partial U\rangle -\frac{1}{M^{2}}\mathcal{F}\right) ,}
\end{eqnarray}%
where $\mathcal{F}_{\mu \nu }=F_{\mu \nu }+cW_{\mu \nu }$ and where%
\begin{eqnarray}
W_{\mu \nu } &=&\sum\limits_{m=1}^{\infty }\sum_{k=0}^{m-1}\lambda ^{2m+1}%
\frac{2\left( m-k\right) }{2m+1}\left\langle \partial _{\mu }\phi \phi
^{2k}\partial _{\nu }\phi \phi ^{2\left( m-k\right) -1}\right\rangle -\left(
\mu \leftrightarrow \nu \right)   \notag \\
&=&\frac{2}{3}\lambda ^{3}\left[ \left\langle \partial _{\mu }\phi \partial
_{\nu }\phi \phi \right\rangle -\left\langle \partial _{\nu }\phi \partial
_{\mu }\phi \phi \right\rangle \right] +\ldots 
\end{eqnarray}%
Expanding $\mathcal{L}_{\mathrm{2eDBI}}$ up to the terms quartic in the fields we get%
\begin{eqnarray}
\mathcal{L}_{\mathrm{2eDBI}} &=&-\frac{1}{2}\langle \partial \phi \cdot \partial \phi
\rangle -\frac{1}{4}F_{\mu \nu }F^{\mu \nu }-\lambda ^{2}\langle \phi
^{2}\partial \phi \cdot \partial \phi \rangle   \notag \\
&&-\frac{1}{8\Lambda ^{4}}\langle \partial \phi \cdot \partial \phi \rangle
^{2}+\frac{1}{4\Lambda ^{4}}\langle \partial _{\mu }\phi \partial _{\nu
}\phi \rangle \langle \partial ^{\mu }\phi \partial ^{\nu }\phi \rangle  
\notag \\
&&-\frac{2}{3}c\lambda ^{3}\langle \phi \partial _{\mu }\phi \partial \phi
_{\nu }\rangle F^{\mu \nu }-\frac{1}{8\Lambda ^{4}}\langle \partial \phi
\cdot \partial \phi \rangle F_{\mu \nu }F^{\mu \nu }-\frac{1}{2\Lambda ^{4}}%
\langle \partial _{\mu }\phi \partial _{\nu }\phi \rangle F_{~~\alpha }^{\mu
}F^{\alpha \nu }  \notag \\
&&-\frac{1}{2M^{4}}\det F+\frac{1}{32M^{4}}\left( F_{\mu \nu }F^{\mu \nu
}\right) ^{2}+O\left( \left( \phi ,F\right) ^{6}\right) .
\end{eqnarray}
For further convenience it is useful to use the spinor notation writing
\begin{equation}
F_{\mu \nu }\overline{\sigma }_{A\overset{\cdot }{A}}^{\mu }\overline{\sigma 
}_{B\overset{\cdot }{B}}^{\nu }=\varepsilon _{\overset{\cdot }{A}\overset{%
\cdot }{B}}\Phi _{AB}+\varepsilon _{AB}\Phi _{\overset{\cdot }{A}\overset{%
\cdot }{B}}\,,
\end{equation}
where in our convention $\overline{\sigma }^{\mu }=\left( 1,-\sigma
^{i}\right) $ and $\varepsilon ^{12}=\varepsilon ^{\overset{\cdot }{1}%
\overset{\cdot }{2}}=1$ . For individual terms in the Lagrangian we get then 
\begin{eqnarray*}
-\frac{2}{3}c\lambda ^{3}\langle \phi \partial _{\mu }\phi \partial \phi
_{\nu }\rangle F^{\mu \nu } &=&\frac{1}{6}c\lambda ^{3}\left[ \langle \phi
\partial _{~~\overset{\cdot }{M}}^{N}\phi \partial _{N\overset{\cdot }{N}%
}\phi\rangle \overline{\Phi }^{\overset{\cdot }{M}\overset{\cdot }{N}}+
\langle \phi \partial _{M}^{~~\overset{\cdot }{N}}\phi \partial _{N\overset{%
\cdot }{N}}\phi\rangle  \Phi ^{MN}\right]  \\
-\frac{1}{8\Lambda ^{4}}\langle \partial \phi \cdot \partial \phi \rangle
F_{\mu \nu }F^{\mu \nu } &=&-\frac{1}{16\Lambda ^{4}}\langle \partial \phi
\cdot \partial \phi \rangle \left( \Phi ^{2}+\overline{\Phi }^{2}\right)  \\
-\frac{1}{2\Lambda ^{4}}\langle \partial _{\mu }\phi \partial _{\nu }\phi
\rangle F_{~~\alpha }^{\mu }F^{\alpha \nu } &=&\frac{1}{16\Lambda ^{4}}\left[
\langle \partial _{A\overset{\cdot }{M}}\phi \partial _{~~~\overset{\cdot }{N%
}}^{A}\phi \rangle \overline{\Phi }^{\overset{\cdot }{A}\overset{\cdot }{N}}%
\overline{\Phi }_{\overset{\cdot }{A}}^{\overset{\cdot }{~~~N}}-\langle
\partial _{M\overset{\cdot }{A}}\phi \partial _{A\overset{\cdot }{N}}\phi
\rangle \overline{\Phi }^{\overset{\cdot }{A}\overset{\cdot }{N}}\Phi
^{AM}\right.  \\
&&\left. -\langle \partial _{A\overset{\cdot }{M}}\phi \partial _{N\overset{%
\cdot }{A}}\phi \rangle \Phi ^{AN}\overline{\Phi }^{\overset{\cdot }{A}%
\overset{\cdot }{M}}+\langle \partial _{N\overset{\cdot }{A}}\phi \partial
_{N}^{~~~\overset{\cdot }{A}}\phi \rangle \Phi ^{AN}\Phi _{A}^{~~~M}\right] .
\end{eqnarray*}%
This leads to the following 4pt seed amplitudes (here we use the Feynman
rules which assign $\sqrt{2}|p]|p]$ and $\sqrt{2}|p\rangle |p\rangle $ to
helicity plus and minus vector external lines respectively)%
\begin{eqnarray}
A\left( 1_{a}^{\phi },2_{b}^{\phi },3_{c}^{\phi },4_{d}^{\phi }\right) 
&=&\lambda ^{2}\sum_{\sigma\in S_4 }\langle T^{\sigma \left( a\right) }T^{\sigma
\left( b\right) }T^{\sigma \left( c\right) }T^{\sigma \left( d\right)
}\rangle \left( \sigma (1) \cdot \sigma \left( 3\right) \right) 
\notag \\
&&-\frac{1}{8\Lambda ^{4}}\sum_{\sigma\in S_4 }\langle T^{\sigma \left( a\right)
}T^{\sigma \left( b\right) }\rangle \langle T^{\sigma \left( c\right)
}T^{\sigma \left( d\right) }\rangle \left( \sigma (1) \cdot
\sigma \left( 2\right) \right) ^{2}  \notag \\
&&+\frac{1}{4\Lambda ^{4}}\sum_{\sigma\in S_4 }\langle T^{\sigma \left( a\right)
}T^{\sigma \left( b\right) }\rangle \langle T^{\sigma \left( c\right)
}T^{\sigma \left( d\right) }\rangle \left( \sigma (1) \cdot
\sigma \left( 3\right) \right) ^{2}
\end{eqnarray}%
\begin{eqnarray}
A\left( 1^{+},2_{a}^{\phi },3_{b}^{\phi },4_{c}^{\phi }\right)  &=&\frac{1}{3%
\sqrt{2}}c\lambda ^{3}\sum_{\sigma\in S_3 }\langle T^{\sigma \left( a\right)
}T^{\sigma \left( b\right) }T^{\sigma \left( c\right) }\rangle \lbrack
1|\sigma \left( 2\right) \sigma \left( 3\right) |1] \\
A\left( 1^{+},2^{+},3_{a}^{\phi },4_{b}^{\phi }\right)  &=&0 \\
A\left( 1^{+},2^{-},3_{a}^{\phi },4_{b}^{\phi }\right)  &=&\frac{1}{2\Lambda
^{4}}\langle T^{a}T^{b}\rangle \lbrack 1|3|2\rangle \lbrack 1|4|2\rangle  \\
A\left( 1^{+},2^{+},3^{-},4^{-}\right)  &=&\frac{1}{8M^{4}}\left[ 1,2\right]
^{2}\langle 3,4\rangle ^{2}\,.
\end{eqnarray}

\section{Basis of the 6pt vertices with scalars and massless vectors \label{sec:basis6pt}}

In this appendix we give a list of the independent  contact terms contributing to the general ansatz for the 6pt amplitudes with triplet of massless scalars and $U(1)$ massless vectors (photons), as discussed in Section \ref{sec:bottomup}.
We  indicate in Tab.~\ref{tab:6ptg1} and~\ref{tab:6ptg2} the number of the contact terms of a given type, the power-counting parameter $\rho$ as well as the number of derivatives of the corresponding terms in the Lagrangian.

\begin{table}[htb]
\centering
 \begin{tabular}{|c | c | c|} 
 \hline
 6pt & $\rho=1/4$ (3der) & $\rho=3/4$ (5der) \\
 \hline
 ${+}{-}000\gamma^+$ & 1 & 5 \\
 ${+}{+}{-}{-}0\gamma^+$ & 1 & 7 \\
 \hline\hline
 & $\rho=1/2$ (4der) & $\rho=1$ (6der) \\
 \hline
 $0000\gamma^+\gamma^-$ & 1 & 3\\ 
 ${+}{-}00\gamma^+\gamma^-$ & 3 & 12 \\
 ${+}{+}{-}{-}\gamma^+\gamma^-$ & 2 & 7 \\
 $0000\gamma^+\gamma^+$ & 1 & 5\\ 
 ${+}{-}00\gamma^+\gamma^+$ & 3 & 17 \\
 ${+}{+}{-}{-}\gamma^+\gamma^+$ & 2 & 11 \\
 \hline
 \end{tabular}
 \caption{Number of independent monomials for 6pt vertices with one or two photons.}
 \label{tab:6ptg1}
\end{table}

\begin{table}[htb]
\centering
  \begin{tabular}{|c | c | c| c|} 
 \hline
  & $\rho=3/4$ (5der) & & $\rho=3/4$ (5der)\\
  \hline
 ${+}{-}0\gamma^+\gamma^+\gamma^+$ & 2
& ${+}{-}0\gamma^+\gamma^+\gamma^-$ & 2\\
\hline\hline
 & $\rho=1$ (6der) & & $\rho=1$ (6der)\\
  \hline
 $00\gamma^+\gamma^+\gamma^+\gamma^+$ & 2
& ${+}{-}\gamma^+\gamma^+\gamma^+\gamma^+$ & 2\\
 $00\gamma^+\gamma^+\gamma^+\gamma^-$ & 3
& ${+}{-}\gamma^+\gamma^+\gamma^+\gamma^-$ & 3\\
$00\gamma^+\gamma^+\gamma^-\gamma^-$ & 2
& ${+}{-}\gamma^+\gamma^+\gamma^-\gamma^-$ & 2\\
 \hline\hline
 & $\rho=1$ (6der) & & $\rho=1$ (6der)\\
  \hline
 $\gamma^+\gamma^+\gamma^+\gamma^+\gamma^+\gamma^+$ & 1
& $\gamma^+\gamma^+\gamma^+\gamma^+\gamma^-\gamma^-$ & 1\\ 
\hline
\end{tabular}
 \caption{Number of independent monomials for 6pt vertices with three and more photons.}
 \label{tab:6ptg2}
\end{table}

\newpage
\bibliography{ref}
\bibliographystyle{JHEP}
\end{document}